\documentclass[aps,prb,twocolumn,superscriptaddress]{revtex4-2}
\usepackage[dvipdfmx]{graphicx}
\usepackage{array}
\usepackage{hhline}
\usepackage{dcolumn}
\usepackage{bm}
\usepackage{braket}
\usepackage{amsmath}
\usepackage{amssymb}
\usepackage[version=4]{mhchem}
\usepackage{color}

\usepackage{comment}

\usepackage{hyperref}
\hypersetup{
colorlinks = true,
urlcolor = blue,
linkcolor = blue,
citecolor = blue
}
\begin{document}
\title{
Optical control of the crystal structure in the bilayer nickelate superconductor \ce{La3Ni2O7} via nonlinear phononics}
\author{Shu Kamiyama}
\affiliation{Department of Physics, The University of Osaka, Toyonaka, Osaka 560-0043, Japan}
\author{Tatsuya Kaneko}
\affiliation{Department of Physics, The University of Osaka, Toyonaka, Osaka 560-0043, Japan}
\author{Kazuhiko Kuroki}
\affiliation{Department of Physics, The University of Osaka, Toyonaka, Osaka 560-0043, Japan}
\author{Masayuki Ochi}
\affiliation{Department of Physics, The University of Osaka, Toyonaka, Osaka 560-0043, Japan}
\affiliation{Forefront Research Center, The University of Osaka, 
Toyonaka, Osaka 560-0043, Japan}

\date{\today}
\begin{abstract}
Superconductivity in the bilayer nickelate \ce{La3Ni2O7} occurs when the interlayer Ni-O-Ni bond angle becomes straight under pressure, suggesting a strong relationship between the crystal structure and the emergence of superconductivity. 
In this study, we theoretically propose a way to control the crystal structure of \ce{La3Ni2O7} toward the tetragonal symmetry via light irradiation instead of pressure using the idea of nonlinear phononics.
Here, resonant optical excitation of an infrared-active (IR) lattice vibration induces a nonlinear Raman-mode displacement through the anharmonic phonon-phonon coupling. 
We calculate the light-induced phonon dynamics on the anharmonic lattice potential determined by first-principles calculation. 
We find that the interlayer Ni-O-Ni bond angle gets slightly closer to straight when an appropriate IR mode is selectively excited.
Our study suggests that light irradiation can be a promising way for structural control of \ce{La3Ni2O7}. 
\end{abstract}

\maketitle

\section{Introdunction}
A recently discovered unconventional superconductor, bilayer nickel oxide \ce{La3Ni2O7}, with a maximum transition temperature $T_c$ of 80 K at around 15--40 GPa has garnered significant attention~\cite{Sun_Huo,Zhang_Lin_2,Yang_Wang,Lechermann_Gondolf,Sakakibara_Kitamine,Gu_Le,Shen_Qin,Christiansson_Petocchi,Liu_Huo,Wu_Luo,Cao_Yang,Hou_Yang,Liu_Mei,Zhang_Su,Lu_Pan_15,Zhang_Lin_16,Oh_Zhang,Liao_Chen,Qu_Qu_19,Yang_Zhang,Jiang_Wang,Zhang_Lin_22,Tian_Chen,Jiang_Hou,Luo_Lv,Yang_Sun,Zhang_Pei_29,Wang_Wang_39,Kaneko_Sakakibara,Lu_Pan_41,Ryee_Witt,Ouyang_Wang,Zhou_Guo,Qu_Qu_56,Kakoi_Kaneko,T.Kaneko_2025}.
This material is a member of the bilayer Ruddlesden-Popper compounds, where two \ce{NiO2} planes are connected via inner apical oxygens.
As pointed out in an earlier study by one of the present authors~\cite{Nakata}, \ce{La3Ni2O7} is a system where a nearly half-filled Ni-$d_{3z^2-r^2}$ orbital hybridized with a Ni-$d_{x^2-y^2}$ orbital constitutes a bilayer square lattice, which is one of the promising playgrounds of high-temperature superconductivity~\cite{bilayer0,bilayer1,bilayer2,bilayer3,bilayer4,bilayer5,bilayer6,bilayer7,bilayer8,bilayer9,bilayer10,bilayer11}.
While the origin of superconductivity is likely unconventional~\cite{Ouyang_Gao,Li_Cao,Yi_Meng}, a complicated role of the Ni-$d_{3z^2-r^2}$ and $d_{x^2-y^2}$ orbitals with strong hybridization remains under debate.

\begin{figure}[t]
  \centering
  \includegraphics[width=0.48\textwidth]{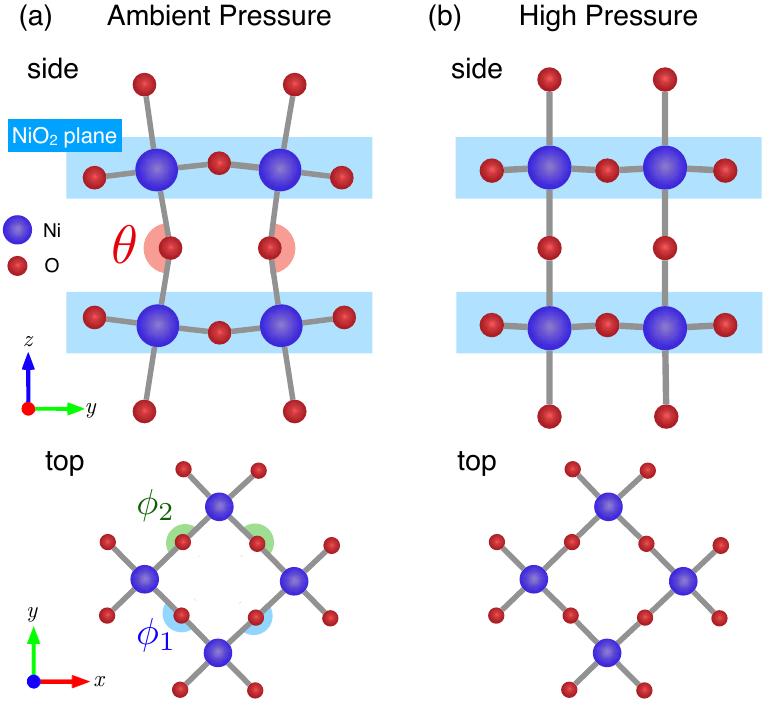}
  \caption{
  Crystal structure of \ce{La3Ni2O7} under (a) ambient pressure and (b) high pressure. 
  }
  \label{fig:intro}
\end{figure}

\ce{La3Ni2O7} undergoes a structural transition depending on temperature and pressure.
Under ambient pressure and room temperature, the Ni-O-Ni interlayer bond angle $\theta$ is around $168^\circ$~\cite{Sun_Huo} [see, Fig.~\ref{fig:intro}(a)] and the system belongs to the orthorhombic $Amam$ phase.
The interlayer bond angle $\theta$ becomes straight under high pressure and low temperature [see, Fig.~\ref{fig:intro}(b)], with the orthorhombic $Fmmm$ phase.
With more pressure, the system becomes the tetragonal $I4/mmm$ phase~\cite{Geisler_Hamlin,Wang_Li,Wang_Wang_50}.
Superconductivity in \ce{La3Ni2O7} emerges under high pressure where the crystal structure becomes tetragonal~\cite{Geisler_Hamlin,Wang_Li,Wang_Wang_50}, which suggests a strong relationship between the tetragonal symmetry of the crystal structure and the emergence of superconductivity.
It is also noteworthy that a trilayer nickelate \ce{La4Ni3O10} also exhibits superconductivity by applying sufficient pressure where the tetragonal phase is realized~\cite{Sakakibara_Ochi,Li_Zhang,Zhu_Peng,Zhang_Pei_49,Li_Chen}.

From this perspective, one important direction in the study of bilayer nickelate superconductors is to realize the tetragonal structure at ambient pressure.
Atomic substitution that stabilizes the tetragonal phase is a possible way. However, rather higher pressure is required to stabilize the tetragonal phase by replacing La with other lanthanoids because La has the largest ionic radius among them~\cite{Zhang_Lin_22,Geisler_Hamlin}.
Therefore, some previous studies have proposed the replacement of La with actinium or alkaline-earth elements such as Sr and Ba, which have a larger ionic radius than La~\cite{Rhodes_Wahl,Wu_Yang}.
Two of the present authors proposed that \ce{Sr3Ni2O5Cl2} is a promising candidate~\cite{3252,3252_yamane} where a larger ionic radius of Sr$^{2+}$ than La$^{3+}$ and a smaller ionic radius of Ni$^{3+}$ than Ni$^{2.5+}$ favor the tetragonal structure even at ambient pressure as shown by first-principles phonon calculation.
Here, while the valence number of nickel is different from \ce{La3Ni2O7}, a crystal field of a NiO$_5$Cl octahedron lowers the $d_{3z^2-r^2}$ orbital energy and thus a nearly half-filled $d_{3z^2-r^2}$ is realized.
It is also noteworthy that recent experiments reported ambient-pressure superconductivity in thin-film \ce{La3Ni2O7}~\cite{E.Ko_2024,G.Zhou_2024,Y.Liu_2025}, where the compressive strain likely reduces the octahedral tilting in the superconducting samples~\cite{Y.Zhao_2024,C.Yue_2025,L.Bhatt_2025,B.Geisler_2025,X.Yi_2025,K.Ushio_2025}.

Another fascinating idea for approaching the tetragonal phase is to manipulate the crystal structure using light instead of applying pressure based on nonlinear phononics~\cite{M.Forst_2011,S.Alaska_2014,R.Mankowsky_2014,M.Fechner_2016,S.Alaska_2015,R.Mankowsky_2017,M.Fechner_2024,C.Paiva_2024,D.M.Juraschek_2017,G.Mingqiang_2018,A.S.Disa_2020,Y.Zeng_2023,R.Tang_2023,T.G.Blank_2023,phono_cavity,phono_review_disa,phono_review_comp}.
In nonlinear phononics, resonant optical excitation of an infrared-active (IR) lattice vibration induces a nonlinear Raman-mode displacement through the anharmonic phonon-phonon coupling. 
While a change in the crystal structure is transient, a sufficiently longer timescale of phonons than electrons allows us to investigate how structural changes affect electronic states of materials.
The generation of intense pulses at mid-infrared and terahertz (THz) frequencies has made nonlinear phononics experimentally possible, and 
the development of probe techniques has enabled us to measure ultrafast dynamics of quantum materials~\cite{M.Forst_2011,phono_review_disa}. 
Nonlinear phononics has been applied to cuprate superconductors~\cite{R.Mankowsky_2014,M.Fechner_2016,S.Alaska_2014}, ferroelectrics~\cite{R.Mankowsky_2017,M.Fechner_2024,S.Alaska_2015,C.Paiva_2024}, magnetic materials~\cite{D.M.Juraschek_2017,A.S.Disa_2020,G.Mingqiang_2018,C.Paiva_2024,Y.Zeng_2023}, and topological materials~\cite{R.Tang_2023,T.G.Blank_2023}.

Based on the idea of nonlinear phononics, in this study, we theoretically investigate the possibility of controlling the crystal structure of \ce{La3Ni2O7} toward the tetragonal phase by light irradiation.
We demonstrate that Raman modes that can make the system closer to the tetragonal phase are activated through the anharmonic coupling with an IR mode resonantly excited by light. 
We calculate the optically driven phonon dynamics using the anharmonic lattice potential determined by first-principles calculation.
We find that the interlayer Ni-O-Ni bond angle gets slightly closer to straight when an appropriate IR mode is selectively excited.
Our study suggests that nonlinear phononics offers a promising pathway to approach the tetragonal crystal structures of the multilayer nickelate superconductors without applying pressure.

This paper is organized as follows.
Methods for first-principles calculations and phonon dynamics are described in Secs.~\ref{sec:method_fp} and \ref{sec:method_model}, respectively.
First-principles construction of the anharmonic lattice potential is described in Sec.~\ref{sec:results_dft}.
In Sec.~\ref{sec:phonon_dynamics}, we present calculation results of phonon dynamics and specify the target IR mode for resonant optical excitation that can effectively modulate the crystal structure toward the tetragonal phase.
The field-amplitude dependence of the structural modulation is discussed in Sec.~\ref{sec:field_amplitude}.
Other possible candidates for the target IR mode are suggested in Sec.~\ref{sec:otherIR}.
Discussions are shown in Sec.~\ref{sec:discussions}.
Section~\ref{sec:summary} summarizes the study.

\section{Methods}
\subsection{First-principles calculation and anharmonic lattice potential\label{sec:method_fp}}
First-principles calculations based on the density functional theory (DFT) are performed by using the PBEsol~\cite{PBEsol_1,PBEsol_2} exchange-correlation functional as implemented in Vienna {\it ab initio} simulation package (VASP)~\cite{vasp1,vasp2,vasp3,vasp,vasp4}.
We use a plane-wave cutoff energy of 600 eV for Kohn-Sham orbitals and an $8\times 8\times 8$ Monkhorst-Pack ${\bm k}$-mesh.
We perform structural optimization until the Hellmann-Feynman force becomes less than 0.01 eV \AA$^{-1}$ for each atom. Both the lattice parameters and atomic coordinates are optimized for the orthorhombic structure (space group: $Amam$). 
The optimized interlayer Ni-O-Ni angle is $170.4^{\circ}$, which is close to the experimental value~\cite{Sun_Huo}, $168^{\circ}$,  and so we expect that PBEsol gives reliable calculation results in the present study~\footnote{
We confirm that the optimized interlayer Ni-O-Ni angle obtained by PBEsol, $170.4^{\circ}$, also agrees well with that obtained by PBE+$U$ ($U=4$ eV)~\cite{ref_PBE,ref_U}, $169.9^{\circ}$.}.
Phonon calculation is performed using the frozen-phonon method as implemented in Phonopy~\cite{phonopy_1, phonopy_2}.
To describe the optical response, we only calculate the $\Gamma$-point phonons~\cite{S.Alaska_2014,R.Mankowsky_2014}.
We analyze irreducible representation of each phonon mode for the point group $mmm$.
The irreducible representations and frequencies of all eigenmodes are presented in Appendix~\ref{sec:phononmode} (Table~\ref{table:eigemmodes_all}).

Next, we construct the anharmonic lattice potential based on our first-principles calculation. The anharmonic lattice potential for a pair of the IR and Raman modes are expressed with their amplitudes $Q_{\rm{IR}}$ and $Q_{\rm{R}}$ as follows:
\begin{align}
    &V(Q_{\rm{IR}},Q_{\rm{R}}) \notag\\
    &=\frac{1}{2}\omega_{\rm{IR}}^2Q_{\rm{IR}}^2 
    + \frac{1}{2}\omega_{\rm{R}}^2Q_{\rm{R}}^2
    -\frac{1}{2}g_{\rm{IR}-\rm{R}} Q_{\rm{IR}}^2Q_{\rm{R}} \label{eq:pot}\\
    &+a_{3;\rm{R}}Q_{\rm{R}}^3
    +a_{4;\rm{R}}Q_{\rm{R}}^4
    +b_{4;\rm{IR}}Q_{\rm{IR}}^4
    -\frac{1}{2} h_{\rm{IR}-\rm{R}} Q_{\rm{IR}}^2Q_{\rm{R}}^2, \notag
\end{align}
where $\omega_{\rm{IR}}$ and $\omega_{\rm{R}}$ are the phonon frequencies of the IR and Raman modes, respectively.
The other coefficients are determined by fitting this equation to the DFT total energies in the way described in Sec.~\ref{sec:results_dft}.
Symmetry constraints on the lattice potential are as follows.
First, due to the space inversion symmetry, an odd-order term of $Q_{\rm{IR}}$ is not allowed in Eq.~(\ref{eq:pot}).
In addition, $g_{\rm{IR}-\rm{R}}$ is non-zero only for the $A_g$ Raman modes because $Q_{\rm{IR}}^2$ for any IR mode belongs to $A_g$ for the present crystal symmetry.
While $h_{\rm{IR}-\rm{R}}$ can be non-zero for the other Raman modes, $-\frac{1}{2} h_{\rm{IR}-\rm{R}} Q_{\rm{IR}}^2Q_{\rm{R}}^2$ does not shift the potential minimum to $Q_{\rm{R}}\neq 0$ unlike $-\frac{1}{2}g_{\rm{IR}-\rm{R}}Q_{\rm{IR}}^2Q_{\rm{R}}$.
Since the modulation of the crystal structure by shifting the potential minimum to $Q_{\rm{R}}\neq 0$ for non-zero $Q_{\rm{IR}}$ is the central purpose of the study of nonlinear phononics, hereafter, we only consider $A_g$ Raman modes, which have non-zero $g_{\rm{IR}-\rm{R}}$.
The phonon frequency and the irreducible representation for all the IR and $A_g$ Raman modes are listed in Table~\ref{table:irr}. 
Eigenmodes for these phonons are presented in Appendix~\ref{sec:phononmode}. 
Several approximations adopted for our lattice potential are discussed in Sec.~\ref{sec:discussions}.

\begin{table}[t]
\begin{tabular}{|c c c||c c c|}
\hline
label & irr.   & $\omega_{\rm{R}}/2\pi$ (THz)  & label & irr. & $\omega_{\rm{IR}}/2\pi$ (THz) \\ \hline
9     & $A_g$ & 2.709              & 29   & $B_{1u}$ & 5.563              \\ \hline
13    & $A_g$ & 3.760              & 30   & $B_{2u}$ & 6.188              \\ \hline
19    & $A_g$ & 4.643              & 34   & $A_u$  & 7.259              \\ \hline
28    & $A_g$ & 5.552              & 36   & $B_{3u}$ & 7.485              \\ \hline
32    & $A_g$ & 6.637              & 37   & $B_{1u}$ & 7.526              \\ \hline
33    & $A_g$ & 6.730              & 40   & $B_{2u}$ & 7.927              \\ \hline
50    & $A_g$ & 9.852              & 42   & $B_{2u}$ & 8.548              \\ \hline
57    & $A_g$ & 11.218             & 43   & $A_u$  & 8.635              \\ \hline
58    & $A_g$ & 11.595             & 44   & $B_{3u}$ & 8.875              \\ \hline
61    & $A_g$ & 13.133             & 45   & $B_{3u}$ & 9.337              \\ 
\hhline{===||---|}
label  & irr.   & $\omega_{\rm{IR}}/2\pi$ (THz) & 46   & $B_{1u}$ & 9.495              \\ \hline
4     & $B_{1u}$   & 1.959              & 48   & $B_{2u}$ & 9.779              \\ \hline
6     & $B_{2u}$   & 2.207              & 53   & $B_{2u}$ & 10.309             \\ \hline
11    & $A_u$    & 3.047              & 54   & $A_u$  & 10.819             \\ \hline
14    & $B_{3u}$   & 3.846              & 55   & $B_{3u}$ & 10.947             \\ \hline
15    & $B_{1u}$   & 3.993              & 59   & $B_{3u}$ & 12.289             \\ \hline
17    & $B_{2u}$   & 4.307              & 62   & $B_{2u}$ & 13.678             \\ \hline
18    & $B_{2u}$   & 4.552              & 63   & $B_{1u}$ & 14.855             \\ \hline
22    & $B_{3u}$   & 4.755              & 66   & $B_{3u}$ & 15.636             \\ \hline
23    & $B_{3u}$   & 4.844              & 67   & $B_{2u}$ & 17.942             \\ \hline
25    & $B_{1u}$   & 4.895              & 70   & $B_{1u}$ & 19.171             \\ \hline
27    & $A_u$    & 5.127              & 72   & $B_{2u}$ & 19.274             \\ \hline
\end{tabular}
\caption{
    Irreducible representation (irr.) and phonon frequency $\omega$ for all the IR and $A_g$ Raman modes at the $\Gamma$ point. A label runs from the lowest frequency mode.
}
\label{table:irr}
\end{table}

\begin{figure}[t]
    \centering
    \includegraphics[width=0.85\linewidth]{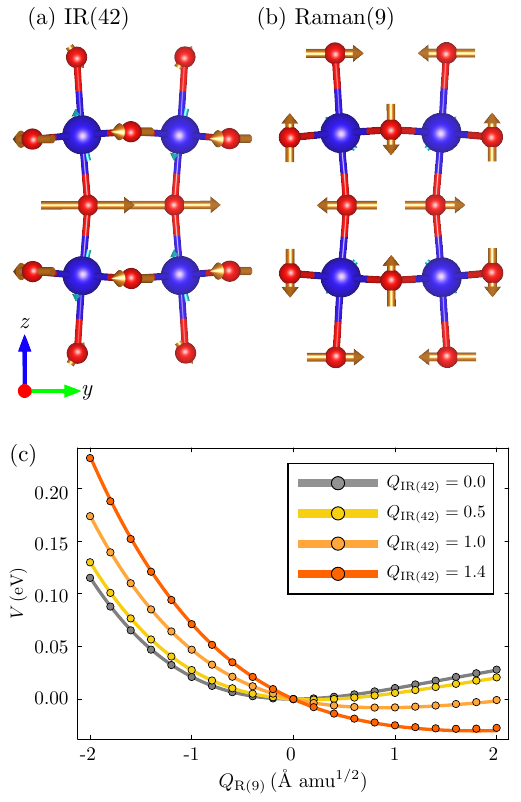}
    \caption{
    Atomic displacements for the (a) IR(42) and (b) Raman(9) modes depicted using VESTA~\cite{vesta}. 
    (c) Anharmonic lattice potential $V(Q_{\rm{IR}},Q_{\rm{R}})$ for the IR(42) and Raman(9) modes, where each circles represent DFT values and curves represent lattice potential fitted by Eq.~(\ref{eq:pot}), respectively.}
    \label{fig:vesta}
\end{figure}

\begin{figure*}[p]
  \centering
  \includegraphics[width=1.0\textwidth]{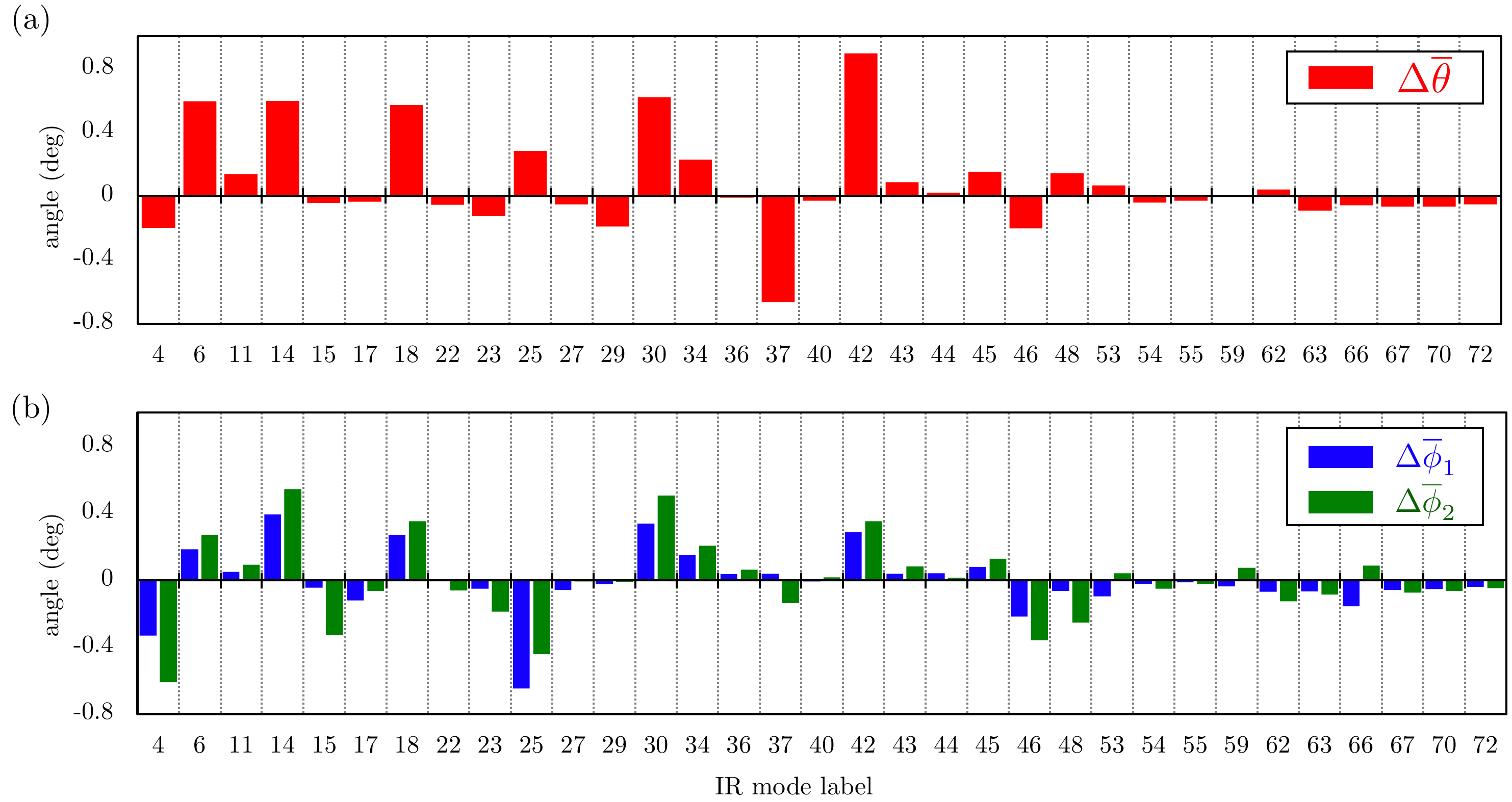}
  \caption{
  Time-averaged change of the bond angles, (a) $\Delta \bar{\theta}$, (b) $\Delta\bar{\phi}_1$ and $\Delta\bar{\phi}_2$, from the original structure after light irradiation.
  The horizontal axis represents the label of the IR mode resonantly excited.
  Time-averaged angles correspond to the dotted lines in Figs.~\ref{fig:IR42_time}(b)--\ref{fig:IR42_time}(d).\label{fig:ave}
  }
  \vspace{7mm}
  \includegraphics[width=0.95\textwidth]{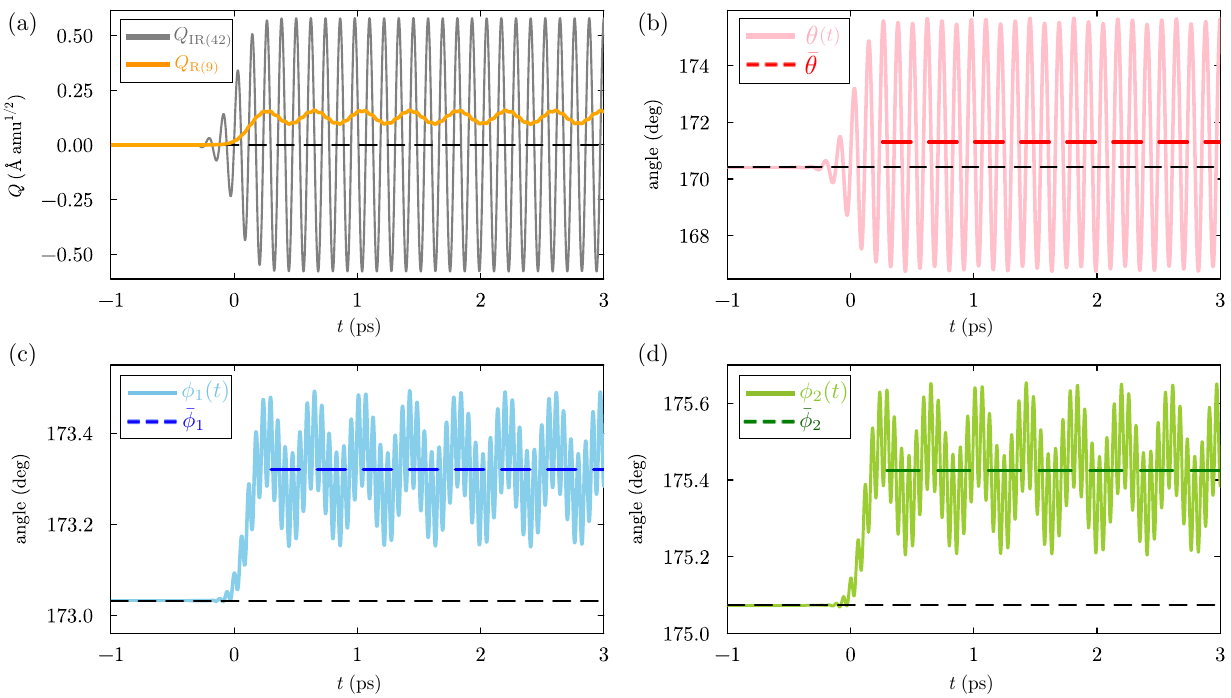}
  \caption{
    Time evolution of (a) phonon amplitudes $Q$ and (b)--(d) bond angles for the case where IR(42) is resonantly excited.
    In (a), $Q_{\rm{IR(42)}}$ and $Q_{\rm{R(9)}}$ are shown as reprensentatives.
    In (b)--(d), the bond angles before light irradiation and the time-averaged angles after light pulse irradiation are shown with dashed lines together with the time evolution of each angle shown with a solid line.
  }
\label{fig:IR42_time}
\end{figure*}

\subsection{Equation of motion for phonon dynamics\label{sec:method_model}}
In this study, we consider the case where a specific IR mode is resonantly excited by the optical pulse field, and Raman oscillations are induced through the IR-Raman coupling. 
Therefore, all the $A_g$ Raman modes $\rm{R}_1, \dots, \rm{R}_n$ are included in the following equation.
We numerically solve the classical equation of motion for the anharmonic lattice potential,
\begin{subequations} 
\label{eqm} 
\begin{align} 
    \ddot{Q}_{\rm{IR}}+\omega_{\rm{IR}}^2Q_{\rm{IR}} 
    &= F(t) - 4b_{4;\rm{IR}} Q_{\rm{IR}}^3 \\
    &+ \sum_{\rm{R}}
    \left(
        g_{\rm{IR}-\rm{R}}Q_{\rm{IR}}Q_{\rm{R}} 
        + h_{\rm{IR}-\rm{R}}Q_{\rm{IR}}Q^2_{\rm{R}}
    \right) \notag \\
    \ddot{Q}_{\rm{R}} + \omega_{\rm{R}}^2 Q_{\rm{R}} 
    &= \frac{1}{2}g_{\rm{IR}-\rm{R}}Q_{\rm{IR}}^2
    + h_{\rm{IR}-\rm{R}}Q_{\rm{IR}}^2 Q_{\rm{R}}
    \label{eq:EOM_R}\\
    &- 3a_{3;\rm{R}} Q_{\rm{R}}^2 - 4a_{4;\rm{R}} Q_{\rm{R}}^3
    \quad(\rm{R} = \rm{R}_1, \dots, \rm{R}_n), \notag
\end{align} 
\end{subequations}
where $F(t) = F_0 e^{-t^2/2\sigma^2}\cos\Omega\,t$ is the driving force generated by the external light pulse. 
The half-width of the Gaussian function is represented as $\tau = 2\sqrt{2\ln 2}\sigma$. We set $F_0 = 20\,\rm{meV} \mbox{\AA}^{-1} \rm{amu}^{-1/2}$ and $\tau = 0.3$ ps unless otherwise noted. The field frequency $\Omega$ is set as $\Omega = \omega_{\rm{IR}}$, where the IR phonon is resonantly excited. 
Note that damping factors are not considered in the equations of motions (see discussions in Sec.~\ref{sec:discussions}). 
A key feature of these equations is that a driving force of the Raman-mode oscillation in Eq.~(\ref{eq:EOM_R}) is proportional to $Q_{\rm{IR}}^2$, resulting in the nonlinear response of the Raman modes and the time-averaged crystal structure with net displacements~\cite{phono_review_disa}.

After the light pulse irradiation (i.e., $F(t) \simeq 0$ for a sufficiently large $t$), the time-averaged crystal structure is modulated by $A_g$ Raman modes while the time average of the IR oscillation becomes zero.
Thus, the time-averaged crystal structure is analyzed based on the original $Amam$ space group since $A_g$ modes fully preserve the crystal symmetry.
We focus on the Ni-O-Ni bond angles: interlayer angle $\theta$ and in-plane angles $\phi_1$ and $\phi_2$ as defined in Fig.~\ref{fig:intro}(a).
Here, there are only two inequivalent in-plane angles due to the reflection symmetry. We note that two NiO$_2$ planes can be transformed each other via the space inversion symmetry, and so there are no additional inequivalent in-plane angles. 
Since the octahedral rotation in the top view of the crystal structure is not allowed for the $Amam$ space group, $\phi_1, \phi_2 \neq \pi$ represents the buckling of the NiO$_2$ plane partially caused by the octahedral tilting. 
The relationship between the phonon amplitude $Q_{\alpha}$ for mode $\alpha$ and the displacement vector $\bm{U}_{\alpha,j}$ for atom $j$ is expressed as
\begin{align}
\label{angle} 
\bm{U}_{\alpha,j} = \frac{Q_{\alpha}}{\sqrt{m_j}} \bm{e}_{\alpha,j},
\end{align} 
where $m_j$ is the mass of atom $j$ and $\bm{e}_{\alpha,j}$ is the dimensionless eigenvector of the phonon mode.

\section{Results}\label{sec:result}

\subsection{First-principles construction of the anharmonic lattice potential\label{sec:results_dft}}

We determine the coefficients in the anharmonic lattice potential, Eq.~(\ref{eq:pot}), by fitting this equation to the DFT total energies in the following way.
First, we use the phonon frequencies $\omega_{\rm{IR}}$ and $\omega_{\rm{R}}$ obtained by first-principles phonon calculation, which are listed in Table~\ref{table:irr}.
Next, $a_{3;\rm{R}}$ and $a_{4;\rm{R}}$ are determined by the fourth-order polynomial fitting using the DFT total energies for various $Q_{\rm{R}} \in [-2.0,2.0]$~\AA~$\rm{amu}^{1/2}$ with $Q_{\rm{IR}}=0$.
The same fitting is performed for $b_{4;\rm{IR}}$ using the DFT total energies for various $Q_{\rm{IR}}\in [-2.0,2.0]$~\AA~$\rm{amu}^{1/2}$ with $Q_{\rm{R}}=0$.
Finally, we determine the IR-Raman coupling coefficients $g_{\rm{IR}-\rm{R}}$ and $h_{\rm{IR}-\rm{R}}$ by the second-order polynomial fitting as a function of $Q_{\rm{R}}$ using the DFT total energies of $Q_{\rm{IR}}=1.0$ and various $Q_{\rm{R}}\in [-0.6,0.6]$~\AA~$\rm{amu}^{1/2}$.
We obtain the coefficients in Eq.~(\ref{eq:pot}) for all the IR and Raman modes considered in this study.

For example, the obtained $V(Q_{\rm{IR}},Q_{\rm{R}})$ for IR(42) and Raman(9) eigenmodes, of which are shown in Figs.~\ref{fig:vesta}(a) and \ref{fig:vesta}(b), is presented in Fig.~\ref{fig:vesta}(c) with curves for various $Q_{\rm{IR}}$.
Here, Raman(9) is a phonon mode that brings the crystal structure closer to the tetragonal structure. 
$V(Q_{\rm{IR}},Q_{\rm{R}})$ agrees well with the DFT total energies shown with circles, which demonstrates that the fitting procedure described above works well. 
We also confirm that the fitting works well for other phonon modes.
As is evident in Eq.~(\ref{eq:pot}), the minimum of the potential curve with respect to $Q_{\rm{R}}$ is shifted when $Q_{\rm{IR}}$ is non-zero in Fig.~\ref{fig:vesta}(c).

\subsection{Nonlinear phonon dynamics and modulation of the crystal structure \label{sec:phonon_dynamics}}

\begin{figure}[htbp]
    \centering
    \includegraphics[width=1.0\linewidth]{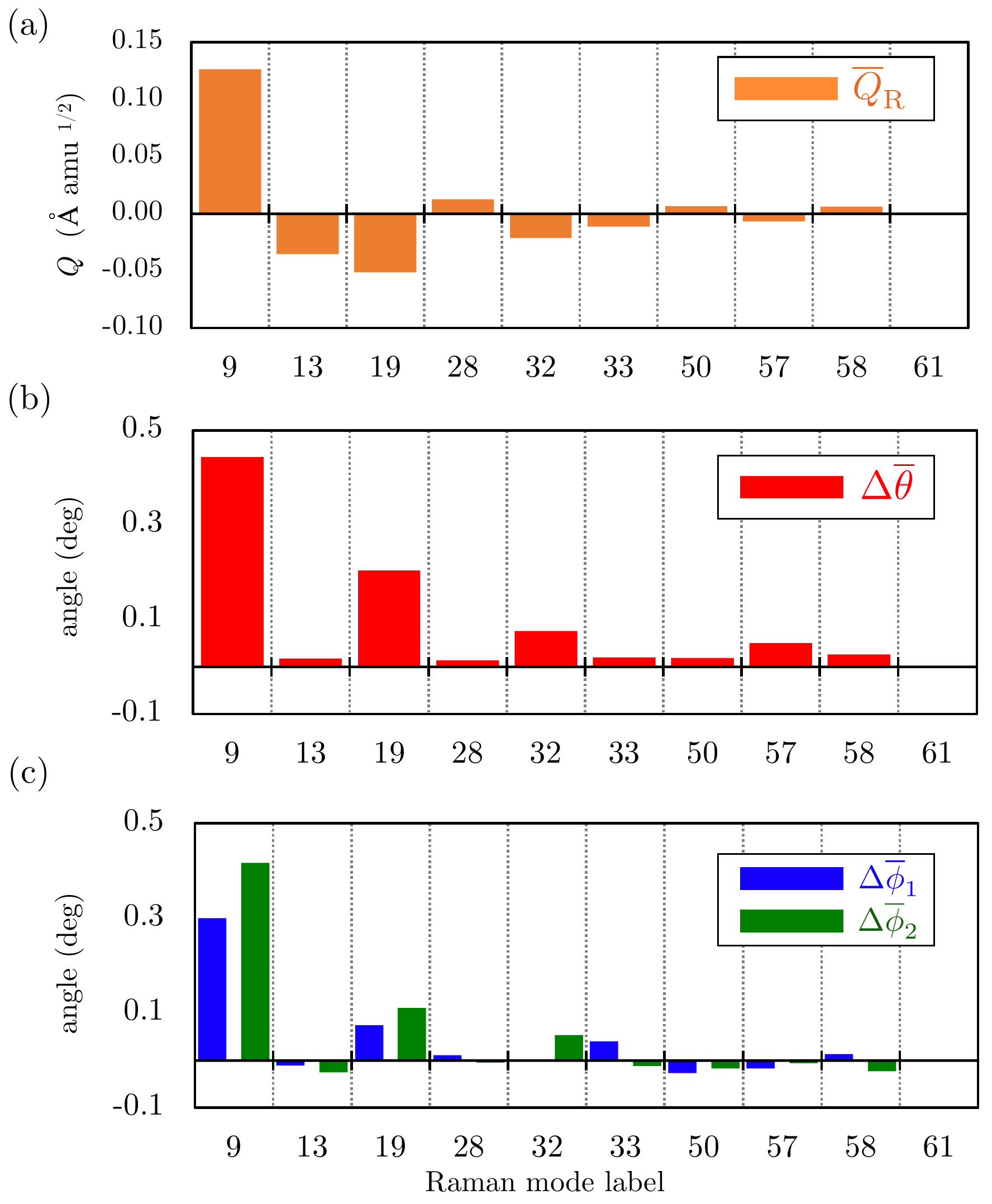}
    \caption{
    (a) Time-averaged phonon amplitude after light irradiation for each Raman mode and
    (b)--(c) changes in the bond angles induced by each Raman mode,
    for the case where IR(42) is optically excited.
    For (b)--(c), bond angles are calculated for the crystal structure with atomic displacement of each $\bar{Q}_{\rm{R}}$.
    }
    \label{fig:IR42_Raman}
\end{figure}

\begin{figure*}[hbt]
    \centering
    \includegraphics[width=1.0\linewidth]{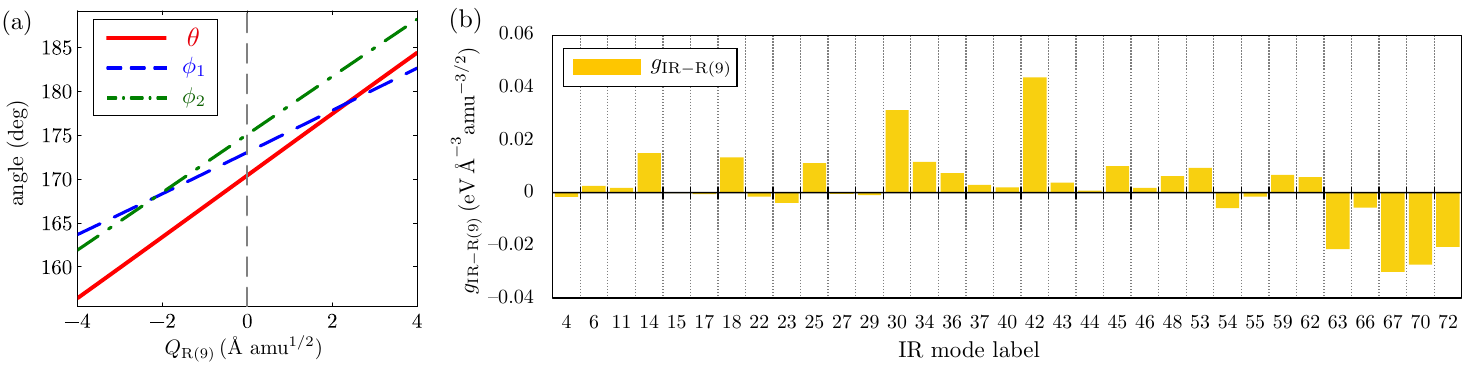}
    \caption{
    (a) Bond angles defined in Fig.~\ref{fig:intro}(a) plotted against the amplitude of Raman(9), $Q_{\rm{R}(9)}$. 
    (b) The IR-Raman coupling coefficient $g_{\rm{IR}-\rm{R}}$ between Raman(9) and each IR mode.
    }
    \label{fig:coupling}
\end{figure*}

A goal of this study is to modulate the crystal structure toward the tetragonal phase at ambient pressure via resonant optical excitation of a specific IR mode and induced Raman oscillations through phonon-phonon coupling.
Therefore, we compare $\bar{\theta}$, $\bar{\phi}_1$, and $\bar{\phi}_2$ among the cases where each IR mode is excited, and identify which IR mode is a good target for optical excitation in the following way.

Figure~\ref{fig:ave} presents the time-averaged change in the bond angles, $\Delta \bar{\theta}$, $\Delta \overline{\phi}_1$, and $\Delta \overline{\phi}_2$, for each case where a specific IR mode is optically excited~\footnote{
We evaluate a time average of $\theta(t)$ after light irradiation in the time domain $[t_{\mathrm{min}}, t_{\mathrm{max}}]$ as follows:
\begin{equation*}
\bar{\theta} = \frac{1}{t_{\mathrm{max}}-t_{\mathrm{min}}}\int_{t_{\mathrm{min}}}^{t_{\mathrm{max}}} dt' \int_{t_{\mathrm{min}}}^{t'}dt\ \frac{\theta(t)}{t'-t_{\mathrm{min}}},
\end{equation*}
where the time average over $t\in[t_{\mathrm{min}}, t']$ is again time-averaged over $t'\in [t_{\mathrm{min}}, t_{\mathrm{max}}]$.
While the first time average, $\int_{t_{\mathrm{min}}}^{t'}dt\ \frac{\theta(t)}{t'-t_{\mathrm{min}}}$, exhibits a decaying oscillation against $t'$, a center of that oscillation is efficiently obtained by the second average.
We set $t_{\mathrm{min}}=0.8\,\rm{ps}$, which is sufficiently larger than $\tau=0.3\,\rm{ps}$ so that $F(t_{\mathrm{min}})\simeq 0$, and $t_{\mathrm{max}}=25\,\rm{ps}$, which is sufficiently large to get a converged value for the time average.}.
In Fig.~\ref{fig:ave}(a), we find that the IR(42) mode with $\omega_{\rm{IR(42)}} = 8.65 \,\rm{THz}$ is the most promising because $\Delta \bar{\theta} = 0.8^{\circ}$ is the largest, i.e., the interlayer Ni-O-Ni angle most effectively becomes close to straight. 
By calculating the band structure for the modulated crystal structure (see, Appendix~\ref{sec:band}), we find that the bilayer splitting energy of Ni-$d_{3z^2-r^2}$ bands increases, i.e., the bilayer coupling becomes strong, in the dynamically modified crystal structure, likely due to the reduced octahedral tilting.
In Fig.~\ref{fig:ave}(b), we also find that the in-plane bond angles driven via IR(42) also increase, which indicates a suppressed buckling of the \ce{NiO2} plane.
Figure~\ref{fig:IR42_time} presents the time evolution of phonon amplitudes $Q$ and the bond angles when IR(42) is optically excited.
As seen in Fig.~\ref{fig:IR42_time}(a), the centers of oscillation for $Q_{\rm{R}}$ and the bond angles are shifted by the optical pulse as the consequence of the nonlinear oscillation of Raman modes.

We further investigate the case where IR(42) is optically excited.
Figure~\ref{fig:IR42_Raman}(a) presents a time-averaged phonon amplitude after light irradiation for each Raman mode. We find that the time-averaged shift, $\overline{Q}_{\rm{R}}$, is the largest for Raman(9) among the $A_g$ Raman modes. 
By calculating the bond angles for the crystal structure with the time-averaged modulation $\overline{Q}_{\rm{R}}$ for each Raman mode, we evaluate the contribution of each mode on the bond angles [see Figs.~\ref{fig:IR42_Raman}(b)--\ref{fig:IR42_Raman}(c)]. 
We find that $\Delta \bar{\theta}$ for Raman(9) is the largest among the $A_g$ Raman modes, indicating that the octahedral tilting is efficiently alleviated by Raman(9). 
In fact, the crystal structure approaches the tetragonal phase by positive $Q_{\rm{R}(9)}$ as shown in Fig.~\ref{fig:vesta}(b) presenting the Raman(9) phonon mode and Fig.~\ref{fig:coupling}(a) presenting the bond angles as a function of $Q_{\rm{R}(9)}$.
As shown in Fig.~\ref{fig:vesta}(b) and Fig.~\ref{fig:IR42_Raman}(c), the buckling of \ce{NiO2} plane partially induced by the octahedral tilting is also alleviated for a positive $Q_{\rm{R}(9)}$.

The IR-Raman(9) coupling coefficient $g_{\rm{IR}-\rm{R}(9)}$ is the largest for IR(42) among all the IR modes as shown in Fig.~\ref{fig:coupling}(b).
This can be naturally understood because atomic displacements for IR(42) and Raman(9) have a large amplitude on the inner apical oxygen atoms as shown in Figs.~\ref{fig:vesta}(a) and \ref{fig:vesta}(b).
The largest $\Delta \bar{\theta}$ for IR(42) in Fig.~\ref{fig:ave} is considered to be due to the strong coupling between IR(42) and Raman(9).

\subsection{Field-amplitude dependence\label{sec:field_amplitude}}

\begin{figure}
    \centering
    \includegraphics[width=0.88\linewidth]{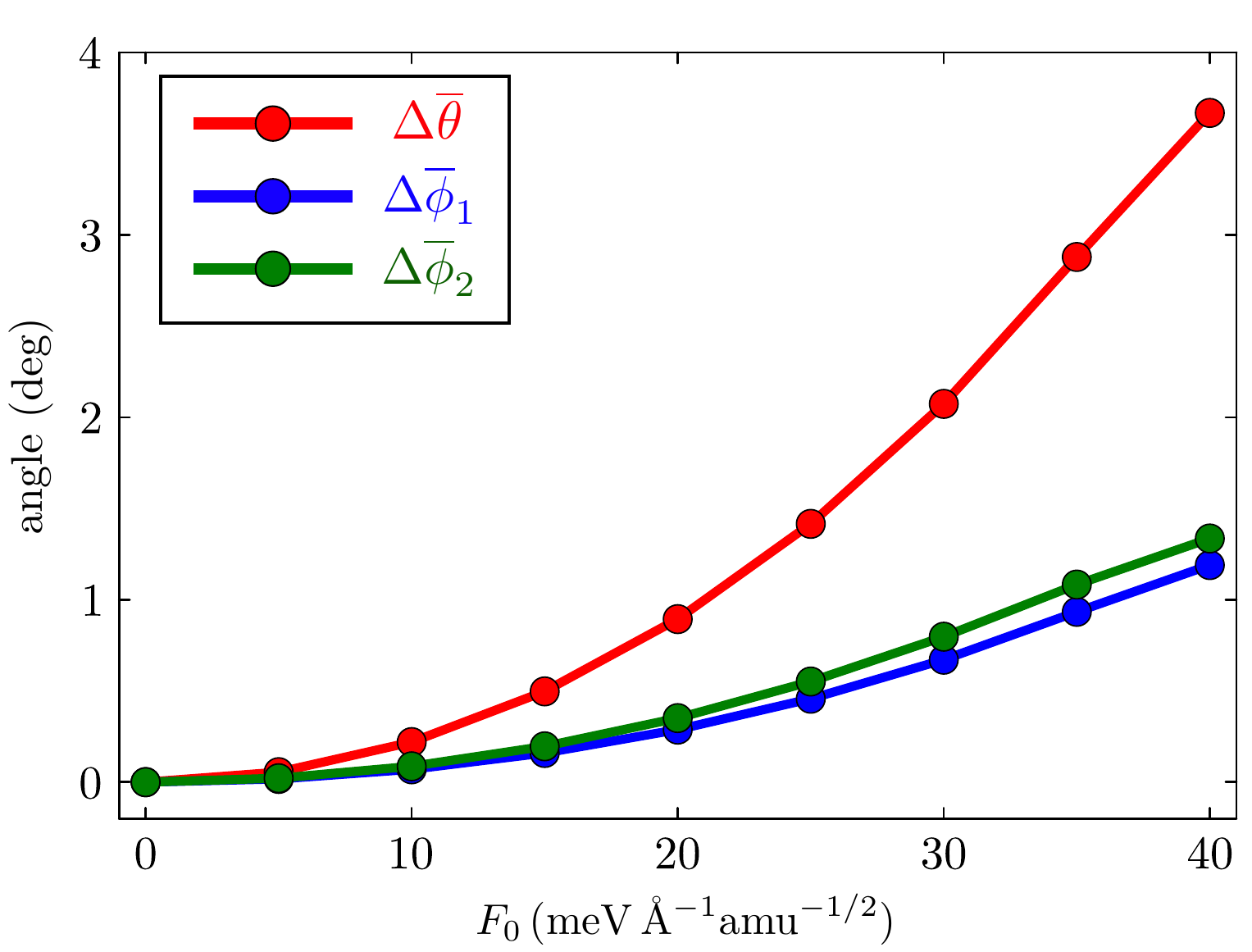}
    \caption{
    Field-amplitude ($F_0$) dependence of $\Delta\bar{\theta}$, $\Delta\bar{\phi}_1$, and $\Delta\bar{\phi}_2$ for the case where IR(42) is optically excited.}
    \label{fig:F0}
\end{figure}

Figure~\ref{fig:F0} shows $\Delta\bar{\theta}$, $\Delta\bar{\phi}_1$, and $\Delta\bar{\phi}_2$ as a function of the field amplitude $F_0$ for the case where IR(42) is optically excited.
We find that the changes in the bond angles are roughly proportional to $F_0^2$.
This originates from approximate relationships, $Q_{\rm{IR}}\propto F_0$ and $Q_{\rm{R}}\propto Q_{\rm{IR}}^2$, which is because the IR and Raman modes are mainly driven by the external field and the phonon-phonon coupling proportional to $Q_{\rm{IR}}^2$, respectively [see, Eq.~(\ref{eqm})].
In other words, we here demonstrate the quadratic $F_0$ dependence of the structural changes as a consequence of the nonlinear Raman oscillation, which is a key aspect of nonlinear phononics for realizing a non-zero displacement of the time-averaged crystal structure after light irradiation.
We also note that a change in the bond angle can be greatly increased by using high-intensity light.

\subsection{Other IR modes\label{sec:otherIR}}
\begin{figure}
    \centering
    \includegraphics[width=1.0\linewidth]{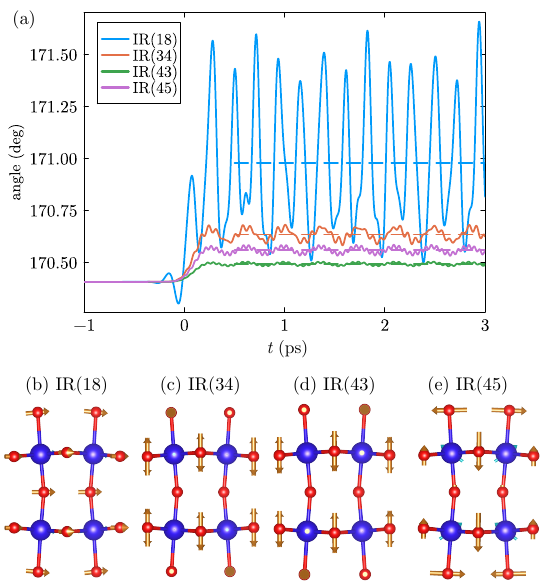}
    \caption{
    (a) Time-evolution of interlayer bond angle $\theta$ for the cases where other IR modes than IR(42) are optically excited. Dashed lines present the time-averaged value of the interlayer bond angle, $\bar{\theta}$.
    (b)--(e) Phonon modes of IR(18), (34), (43), and (45) are shown, respectively.
    }
    \label{fig:otherIR}
\end{figure}

While IR(42) gives the largest increase in $\Delta \bar{\theta}$, $\theta(t)$ is strongly oscillating and $\Delta \theta(t)$ becomes even negative in its oscillation as shown in Fig.~\ref{fig:IR42_time}(b).
This is because the linearly excited IR(42) phonon strongly contributes to the oscillation of the inner apical oxygen as shown in Fig.~\ref{fig:vesta}(a). 
Therefore, we also present the cases where other IR modes are optically excited in Fig.~\ref{fig:otherIR}(a).
In particular, we show the cases where $\Delta \theta(t)$ is always positive, i.e., the interlayer bond angle is always increased after light irradiation.
For these cases, IR(18), (34), (43), and (45) are optically excited, the eigenmodes of which are shown in Figs.~\ref{fig:otherIR}(b)--\ref{fig:otherIR}(e).
Since the inner apical oxygen moves little in these IR modes, $\theta$ is less affected by the IR-mode oscillation, which gives the smaller oscillation of $\theta(t)$ compared to that for IR(42) as shown in Fig.~\ref{fig:IR42_time}(b).
These IR modes can also be good candidates as a target state of light irradiation.

\section{Discussions~\label{sec:discussions}}

In this study, we adopt several approximations for the anharmonic lattice potential, Eq.~(\ref{eq:pot}).
First, we do not consider the Raman-Raman scattering.
Reference~\cite{Raman-Raman} discussed that the role of Raman-Raman coupling terms like $h_{\rm{R}_1 - \rm{R}_2} Q_{\rm{R}_1}Q_{\rm{R}_2}$ is to renormalize the light-induced phonon displacements.
In fact, they evaluated $h_{\rm{R}_1 - \rm{R}_2}$ for an actual material and found that their values are sufficiently smaller than a critical threshold against an unstable deformation of the lattice. 
Thus, we expect that our results are quailtatively kept unchanged even by including the Raman-Raman coupling while the size of the Raman displacements can be somewhat renormalized.

Second, we do not consider ${\bm q}\neq{\bm 0}$ phonons, which are required when one considers higher-order potential coupling terms~\cite{P.G.Klemens_1966,T.Samuel_2018,T.Feng_16}. 
To include such coupling terms, we should consider intractably many phonon modes over the whole Brillouin zone. On the other hand, the nonlinear oscillation, which is an objective of our study, requires $Q_{\rm{IR}}^2$ dependence of the potential term, and so the lowest-order term that can induce the nonlinear oscillation is $Q_{\rm{IR}}^2Q_{\rm{R}}$, where only phonons at the $\Gamma$ point involve due to the translational invariance of the crystal (i.e., the total crystal momentum should be zero for each term).
Thus, in the present study we ignore ${\bm q}\neq{\bm 0}$ phonons for simplicity as many theoretical studies on nonlinear phononics do~\cite{M.Fechner_2016,R.Mankowsky_2014,S.Alaska_2014,R.Tang_2023,T.G.Blank_2023,A.S.Disa_2020,phono_cavity,Y.Zeng_2023,S.Alaska_2015,G.Mingqiang_2018}.

Third, in our simulation of phonon dynamics, we do not explicitly consider the phonon lifetime, which can effectively represent several types of phonon-phonon coupling discarded in this study and electron-phonon coupling.
In fact, when one seeks to see the transient modulated crystal structure after light irradiation theoretically, damping effects are not necessarily included. 
Nonetheless, in some studies, the lifetime was taken as the damping term in the equation of motion phenomenologically~\cite{R.Mankowsky_2014,Y.Zeng_2023}.
We show the phonon dynamics including a damping term in Appendix~\ref{sec:lifetime}. We confirm that a change in the crystal structure is similar to our simulation presented in Sec.~\ref{sec:result} within a short timescale following the field irradiation.

Finally, phonon dynamics is treated by the classical equation of motion rather than by the quantum theory of phonons.
In the quantum treatment of phonon anharmonicity, one needs to consider the energy conservation of each scattering process~\cite{A.Maradudin_62_1,A.Maradudin_62_2,T.Feng_14,T.Feng_16}.
From this perspective, ${\bm q}\neq{\bm 0}$ phonons are necessary to satisfy the energy conservation together with the crystal momentum conservation.
Considering anharmonic phonon dynamics in the quantum manner is important but very challenging, and so is a future objective.
There are various theoretical developments in dealing with anharmonicity with very many degrees of freedom~\cite{L.Monacelli_21, A.Siciliano_23, J.M.Lihm_21}.
These techniques might be useful to develop simulation methods of phonon dynamics.
In addition, we have not considered the dynamics of electronic states associated with lattice motions, which would be one of the future tasks.

Our theoretical proposal of the structural control can be experimentally realizeable by the pump-probe method using a terahertz or mid-infrared laser~\cite{R.Mankowsky_2014,M.Forst_2011,M.Fechner_2024,M.Forst_2011,A.S.Disa_2020,T.G.Blank_2023}. 
By examining the optical response in both equilibrium and pumped nonequilibrium states, one can reveal changes in electronic states~\cite{M.Forst_2011,M.Forst_2011,A.S.Disa_2020,T.G.Blank_2023}. 
In addition, using time-resolved X-rays during laser irradiation, it is possible to observe changes in the crystal structure~\cite{R.Mankowsky_2014,M.Fechner_2024}. 

The external driving force $F(t)$ considered in this study and the actual electric field $\bm{E}(t)$ are related by $F(t) = \bm{E}(t) \cdot \bm{Z}_{\alpha}$, where $\bm{Z}_{\alpha}$ is the mode effective charge of the phonon mode $\alpha$.
Although first-principles evaluation of the effective charge requires the correct description of the metallicity and the screening effect in the material, \ce{La3Ni2O7} hosts the non-trivial density wave resulting in the insulating states~\cite{density_wave_327_326,Sun_Huo,Hou_Yang,Zhang_Su,Zhang_Pei_29,Wang_Wang_39,Chen_Liu,Kakoi_Oi,Xie_Huo,Chen_Choi,Wang_Jiang,Dan_Zhou,Khasanov_Hicken,Yi_Meng,Meng_Yang,LaBollita_Pardo_116,Zhang_Xu,Ni_Ji,Lin_Zhang}, which likely originates from strong electron correlation effects.
While first-principles evaluation of the effective charge for metals has been recently proposed by a dynamical extension of the effective charge~\cite{dynamicalZ1,dynamicalZ2,dynamicalZ3,dynamicalZ4}, it is still challenging to include the strong electron correlation effects into DFT calculation. Therefore, we do not evaluate $\bm{Z}_{\alpha}$ in this study, and hence we cannot show an explicit correspondence between the electric-field strength and $F_0$. Nevertheless, by assuming a typical value of $|\bm{Z}_{\alpha}| \sim 0.1\ e\,{\rm{amu}}^{-1/2}$~\cite{A.S.Disa_2020}, the electric field strength is estimated to be $|\bm{E}| \sim 10$ MV cm$^{-1}$ for $F_0 = $ 10 meV $\mbox{\AA}^{-1} {\rm{amu}}^{-1/2}$.

\section{Summary\label{sec:summary}}
In this study, we have investigated the possibility of controlling the crystal structure of the bilayer nickelate \ce{La3Ni2O7} via light irradiation based on nonlinear phononics.
We have constructed the anharmonic lattice potential using first-principle calculation and calculated phonon dynamics by solving the equation of motion for that potential.
We have found that the interlayer Ni-O-Ni bond angle gets slightly closer to straight when an appropriate IR mode, which is identified in this study, is selectively excited.
If structural control can be achieved through experiments, we expect that it will lead to hints for elucidating the relationship between the crystal structure of multi-layer nickelate superconductors and the mechanisms of superconductivity or density wave appearance. 
We expect our findings stimulate an experimental study on the structural control of \ce{La3Ni2O7} using light.

\begin{acknowledgments}
We thank Dr.~S. Kitou, Dr.~Y. Murotani, Dr.~H. Sakakibara and Dr.~A. Togo for fruitful discussions.
This work was supported by Grants-in-Aid for Scientific Research from JSPS, KAKENHI Grants No.~JP24K06939, No.~JP24H00191 (T.K.), No.~JP22K04907 (K.K.),  and No. JP24K01333. 
\end{acknowledgments}

\section*{Data Availability}
The data that support the findings of this article are openly available~\footnote{\url{https://hdl.handle.net/11094/102582}}.

\appendix
\section{Eigenmodes for the $\Gamma$ point phonons and point group analysis} \label{sec:phononmode}
At the $\Gamma$ point, $A_u$, $B_{1u}$, $B_{2u}$, $B_{3u}$ phonon modes are classified as IR modes, which have an odd parity with respect to the space inversion symmetry.
Since $Q_{\rm{IR}}^2$ for any IR mode belongs to the $A_g$ representation as shown in diagonal elements in Table~\ref{table:product}, $Q_{\rm{IR}}^2Q_{\rm{R}}$ can be included in the Hamiltonian (i.e., $Q_{\rm{IR}}^2Q_{\rm{R}}$ belongs to the $A_g$ representation) only when the Raman mode belongs to the $A_g$ representation, e.g., $B_{1u}\otimes B_{1u}\otimes A_g = A_g\otimes A_g = A_g$. 
Thus, we do not consider the Raman modes that do not belong to the $A_g$ representation.

\begin{figure}[h]
    \centering
    \includegraphics[width=0.95\linewidth]{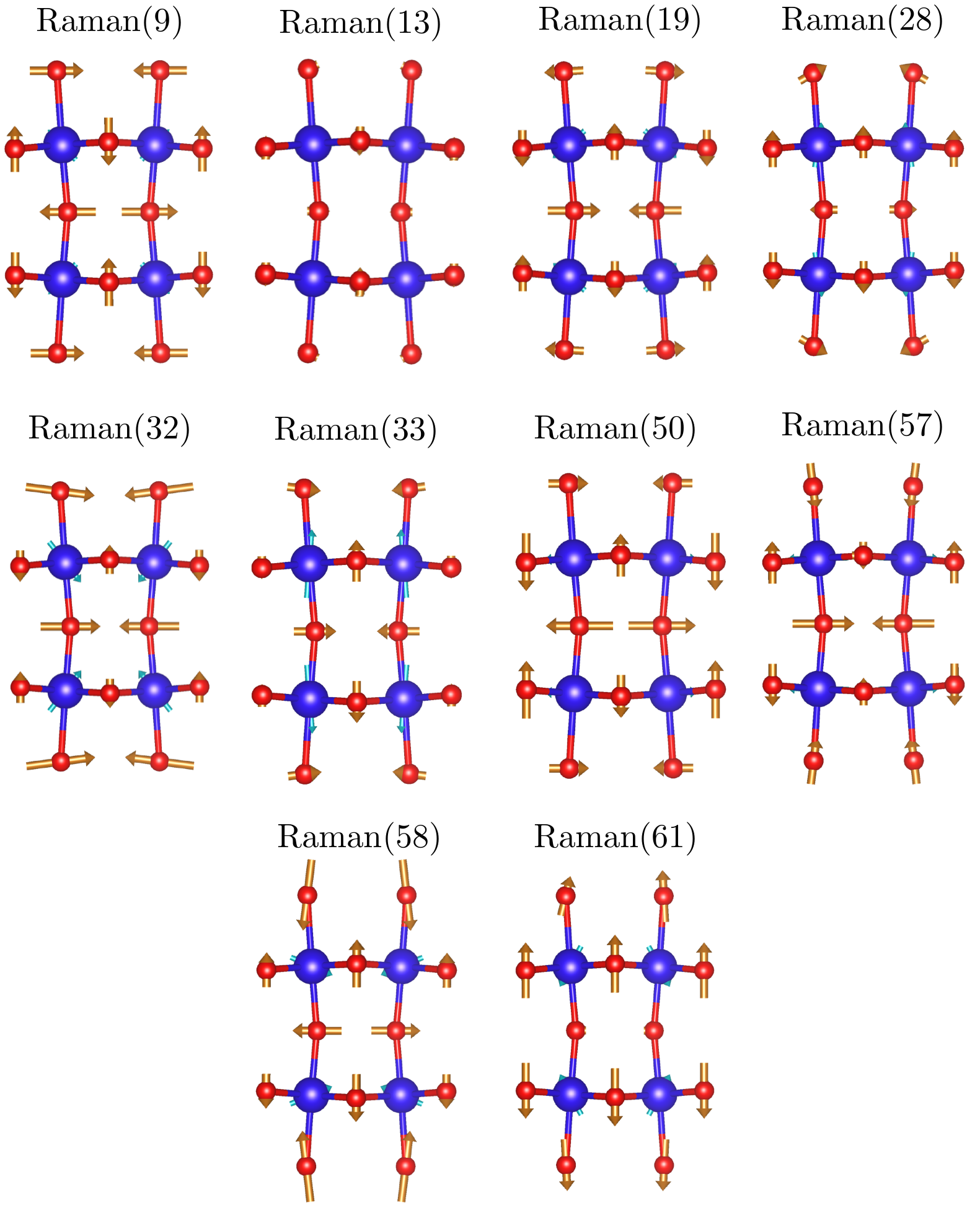}
    \caption{Atomic displacements for all the $A_g$ Raman modes.
    La atoms and their modulated vectors are ommited for simplicity.}
    \label{fig:allAg}
\end{figure}

We illustrate atomic displacements of all the $A_g$ Raman modes at the $\Gamma$ point in Fig.~\ref{fig:allAg}.
\begin{table}[h]
\begin{tabular}{|ccc||ccc|}
\hline
label & irr. & $\omega/2\pi$ (THz) & label & irr. & $\omega/2\pi$ (THz) \\ \hline
1    & acoustic & 0             & 37   & $B_{1u}$ & 7.526         \\ \hline
2    & acoustic & 0             & 38   & $B_{3g}$ & 7.626         \\ \hline
3    & acoustic & 0             & 39   & $B_{1g}$ & 7.875         \\ \hline
4    & $B_{1u}$ & 1.959         & 40   & $B_{2u}$ & 7.927         \\ \hline
5    & $B_{3g}$ & 2.015         & 41   & $B_{3g}$ & 8.546         \\ \hline
6    & $B_{2u}$ & 2.207         & 42   & $B_{2u}$ & 8.548         \\ \hline
7    & $B_{2g}$ & 2.316         & 43   & $A_u$  & 8.635         \\ \hline
8    & $B_{1g}$ & 2.447         & 44   & $B_{3u}$ & 8.875         \\ \hline
9    & $A_g$  & 2.709         & 45   & $B_{3u}$ & 9.337         \\ \hline
10   & $B_{1g}$ & 2.963         & 46   & $B_{1u}$ & 9.495         \\ \hline
11   & $A_u$  & 3.047         & 47   & $B_{2g}$ & 9.557         \\ \hline
12   & $B_{3g}$ & 3.561         & 48   & $B_{2u}$ & 9.779         \\ \hline
13   & $A_g$  & 3.760         & 49   & $B_{1g}$ & 9.810         \\ \hline
14   & $B_{3u}$ & 3.846         & 50   & $A_g$  & 9.852         \\ \hline
15   & $B_{1u}$ & 3.993         & 51   & $B_{3g}$ & 9.880         \\ \hline
16   & $B_{2g}$ & 4.023         & 52   & $B_{1g}$ & 10.169        \\ \hline
17   & $B_{2u}$ & 4.307         & 53   & $B_{2u}$ & 10.309        \\ \hline
18   & $B_{2u}$ & 4.552         & 54   & $A_u$  & 10.819        \\ \hline
19   & $A_g$  & 4.643         & 55   & $B_{3u}$ & 10.947        \\ \hline
20   & $B_{1g}$ & 4.717         & 56   & $B_{3g}$ & 11.137        \\ \hline
21   & $B_{3g}$ & 4.718         & 57   & $A_g$  & 11.218        \\ \hline
22   & $B_{3u}$ & 4.755         & 58   & $A_g$  & 11.595        \\ \hline
23   & $B_{3u}$ & 4.844         & 59   & $B_{3u}$ & 12.289        \\ \hline
24   & $B_{2g}$ & 4.862         & 60   & $B_{1g}$ & 12.908        \\ \hline
25   & $B_{1u}$ & 4.895         & 61   & $A_g$  & 13.133        \\ \hline
26   & $B_{1g}$ & 4.948         & 62   & $B_{2u}$ & 13.678        \\ \hline
27   & $A_u$  & 5.127         & 63   & $B_{1u}$ & 14.855        \\ \hline
28   & $A_g$  & 5.552         & 64   & $B_{1g}$ & 14.859        \\ \hline
29   & $B_{1u}$ & 5.563         & 65   & $B_{2g}$ & 15.099        \\ \hline
30   & $B_{2u}$ & 6.188         & 66   & $B_{3u}$ & 15.636        \\ \hline
31   & $B_{1g}$ & 6.294         & 67   & $B_{2u}$ & 17.942        \\ \hline
32   & $A_g$  & 6.637         & 68   & $B_{1g}$ & 18.527        \\ \hline
33   & $A_g$  & 6.730         & 69   & $B_{2g}$ & 19.066        \\ \hline
34   & $A_u$  & 7.259         & 70   & $B_{1u}$ & 19.171        \\ \hline
35   & $B_{2g}$ & 7.451         & 71   & $B_{1g}$ & 19.205        \\ \hline
36   & $B_{3u}$ & 7.485         & 72   & $B_{2u}$ & 19.274        \\ \hline
\end{tabular}
\caption{All phonon modes at the $\Gamma$ point in \ce{La3Ni2O7}.}
\label{table:eigemmodes_all}
\end{table}

\begin{table}[htbp]
\begin{tabular}{|c||c|c|c|c|c|c|c|c|}
 \hline
$mmm$ & $A_g$ & $A_u$ & $B_{1g}$ & $B_{1u}$ & $B_{2g}$ & $B_{2u}$ & $B_{3g}$ & $B_{3u}$ \\ \hline \hline
$A_g$  & $A_g$ & $A_u$ & $B_{1g}$ & $B_{1u}$ & $B_{2g}$ & $B_{2u}$ & $B_{3g}$ & $B_{3u}$ \\ \hline
$A_u$  &$\cdot$& $A_g$ & $B_{1u}$ & $B_{1g}$ & $B_{2u}$ & $B_{2g}$ & $B_{3u}$ & $B_{3g}$ \\ \hline
$B_{1g}$ &$\cdot$&$\cdot$& $A_g$  & $A_u$  & $B_{3g}$ & $B_{3u}$ & $B_{2g}$ & $B_{2u}$ \\ \hline
$B_{1u}$ &$\cdot$&$\cdot$&$\cdot$ & $A_g$  & $B_{3u}$ & $B_{3g}$ & $B_{2u}$ & $B_{2g}$ \\ \hline
$B_{2g}$ &$\cdot$&$\cdot$&$\cdot$ &$\cdot$ & $A_g$  & $A_u$  & $B_{1g}$ & $B_{1u}$ \\ \hline
$B_{2u}$ &$\cdot$&$\cdot$&$\cdot$ &$\cdot$ &$\cdot$ & $A_g$  & $B_{1u}$ & $B_{1g}$ \\ \hline
$B_{3g}$ &$\cdot$&$\cdot$&$\cdot$ &$\cdot$ &$\cdot$ &$\cdot$ & $A_g$  & $A_u$  \\ \hline
$B_{3u}$ &$\cdot$&$\cdot$&$\cdot$ &$\cdot$ &$\cdot$ &$\cdot$ &$\cdot$ & $A_g$  \\ \hline
\end{tabular}
\caption{
Product table of point group $mmm$~\cite{group1,group2}. Note that the table is symmetric.
}
\label{table:product}
\end{table}

\section{Electronic band dispersion for the modulated crystal structure\label{sec:band}}
\begin{figure}[htbp]
    \centering
    \includegraphics[width=0.95\linewidth]{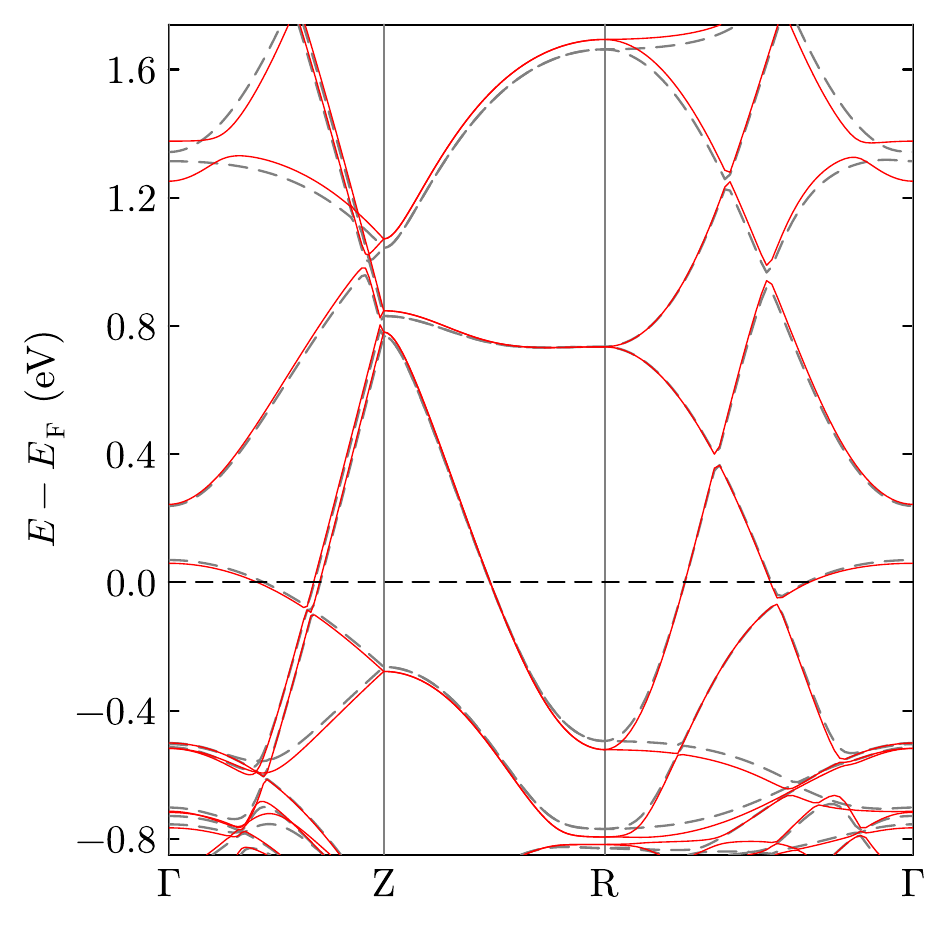}
    \caption{
    Electronic band dispersion for the original crystal structure (gray dashed lines) and the modulated crystal structure for the case where IR(42) is optically excited with $F_0 = 40\,\rm{meV} \mbox{\AA}^{-1} \rm{amu}^{-1/2}$ (red solid lines).
    The $k$-path is set to $\Gamma=(0,0,0) \to {\rm Z}=(\pi/a,0,0) \to {\rm R}=(\pi/a,\pi/2b,\pi/2c) \to \Gamma=(0,0,0)$ with the lattice constants $b$ and $c$. 
    }
    \label{fig:band}
\end{figure}

In Fig.~\ref{fig:band}, we present the electronic band dispersion for the original (unmodulated) crystal structure and that for the modulated crystal structure.
To see how the electronic structure is affected by the structural modulation by light irradiation, we calculate the electronic band dispersion for the time-averaged crystal structure after light irradiation, where IR(42) is optically excited with $F_0 = 40\,\rm{meV} \mbox{\AA}^{-1} \rm{amu}^{-1/2}$.  A $12\times 12 \times 12$ ${\bm k}$-mesh is used for determining the Fermi energy.

We find that the bilayer splitting energy for the $d_{3z^2-r^2}$ orbital is slightly increased. For example, at the Z point, the $d_{3z^2-r^2}$ bonding and anti-bonding bands reside around $E-E_{\mathrm{F}} \simeq -0.3$ and $+1.1$ eV, respectively. At the R point, the $d_{3z^2-r^2}$ bonding and anti-bonding bands reside around $E-E_{\mathrm{F}} \simeq -0.8$ and $+1.7$ eV, respectively. For both points, we can see that the energy difference between the bonding and anti-bonding states becomes larger for the modulated crystal structure. Here, we refer the readers to the previous study, e.g., Ref.~\cite{Sun_Huo}, for the orbital-character assignment of the band dispersion.
The increase in the bilayer splitting energy is likely due to the increase in the interlayer bond angle $\theta$, which can enhance the interlayer hopping between the $d_{3z^2-r^2}$ orbitals. According to previous studies, the interlayer hopping between the $d_{3z^2-r^2}$ orbitals is crucial for superconductivity since it determines the interlayer exchange coupling between spins of the $d_{3z^2-r^2}$ orbital. 
We confirm that the slight enhancement of the bilayer splitting is also obtained when one includes the $+U$ correction in the band structure calculation.

\section{Effect of damping} \label{sec:lifetime}
\begin{figure}[htbp]
\centering
\includegraphics[width = 1.0\linewidth]{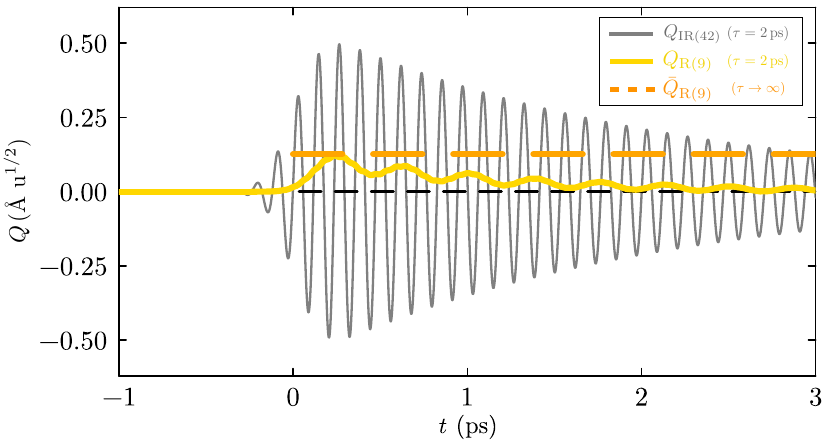}
\caption{
Time evolution of the phonon amplitude of IR(42) and Raman(9) including a damping term. 
The orange dashed line indicates the time-averaged value of $Q_{\rm{R}(9)}$ without a damping term.
}
\label{fig:damp}
\end{figure}

In this section, we present the results of phonon dynamics with finite phonon lifetimes.
The effect of the lifetime is taken into account by phenomenologically introducing a damping term into the equation of motion~\cite{R.Mankowsky_2014,Y.Zeng_2023},
\begin{align}
&\ddot{Q}_{\rm{IR}}+2\gamma_{\rm{IR}}\dot{Q}_{\rm{IR}} +\omega_{\rm{IR}}^2Q_{\rm{IR}} \notag \\
    &= F(t) - 4b_{4;\rm{IR}} Q_{\rm{IR}}^3 
    + \sum_{\rm{R}}
    \left(
        g_{\rm{IR}-\rm{R}}Q_{\rm{IR}}Q_{\rm{R}} 
        + h_{\rm{IR}-\rm{R}}Q_{\rm{IR}}Q^2_{\rm{R}}
    \right)\\
 &\ddot{Q}_{\rm{R}} + 2\gamma_{\rm{R}} \dot{Q}_{\rm{R}} + \omega_{\rm{R}}^2 Q_{\rm{R}} \notag \\
    &= \frac{1}{2}g_{\rm{IR}-\rm{R}}Q_{\rm{IR}}^2 + h_{\rm{IR}-\rm{R}}Q_{\rm{IR}}^2 Q_{\rm{R}}
    - 3a_{3;\rm{R}} Q_{\rm{R}}^2 - 4a_{4;\rm{R}} Q_{\rm{R}}^3\notag \\
    &(\rm{R} = \rm{R}_1, \dots, \rm{R}_n),
\end{align}
where $\gamma_{\rm{IR/R}}$ is the damping constant of the IR/Raman mode, which corresponds to the inverse of phonon lifetime $\tau_{\rm{IR/R}}$, i.e., $\gamma_{\rm{IR/R}} = 1/\tau_{\rm{IR/R}}$.
Here, we set $\tau_{\rm{IR/R}} = 2\,\rm{ps}$.
The results of the phonon dynamics for IR(42) and Raman(9) modes are shown in Fig.~\ref{fig:damp}.
The dashed line indicates the average amplitude of the Raman(9) mode without damping, corresponding to Fig.~\ref{fig:IR42_time}(a).
We confirm that the amplitude of the Raman(9) mode does not differ significantly from the undamped case for a while after light irradiation.

\section{$\mathbf{k}$-mesh convergence} 
\begin{figure}[htbp]
\centering
\includegraphics[width = 0.9\linewidth]{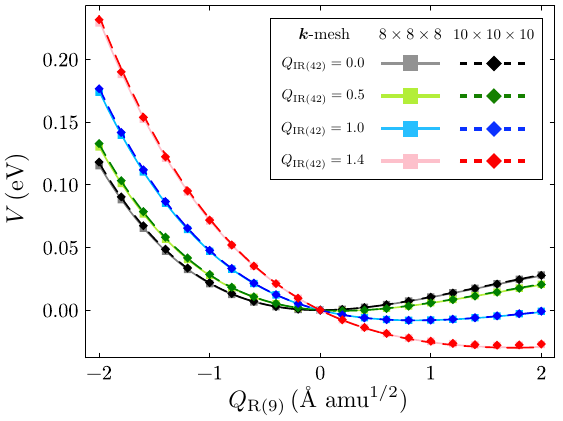}
\caption{
Comparison of the lattice potentials for the IR(42) and Raman(9) modes, obtained by $8\times 8\times 8$ and $10\times 10\times 10$ $\bm{k}$-meshes.
The squares (solid lines) and diamonds (dashed lines) represent the DFT energies (fitting curves) obtained with $8\times 8\times 8$ and $10\times 10\times 10$ $\bm{k}$-meshes, respectively. 
}
\label{fig:kmesh}
\end{figure}

Fig.~\ref{fig:kmesh} presents the comparison of the lattice potentials
for the IR(42) and Raman(9) modes, obtained by $8\times 8 \times 8$ and $10\times 10 \times 10$ $\bm{k}$-meshes, where the former $\bm{k}$-mesh is adopted in the main text. 
The obtained lattice potential agrees very well between two $\bm{k}$-meshes.
In addition, we also verify that the evaluated coupling coefficients in Eq.~(\ref{eq:pot}) barely change: the coupling constant $g_{\rm{IR}(42)-\rm{R}(9)}$ is $4.40\times 10^{-2}$ and $4.46\times 10^{-2}$ [eV \AA$^{-3}$ amu$^{-3/2}$] for $8\times 8 \times 8$ and $10\times 10 \times 10$ $\bm{k}$-meshes, respectively.

\bibliography{main}

\begin{thebibliography}{134}%
\makeatletter
\providecommand \@ifxundefined [1]{%
 \@ifx{#1\undefined}
}%
\providecommand \@ifnum [1]{%
 \ifnum #1\expandafter \@firstoftwo
 \else \expandafter \@secondoftwo
 \fi
}%
\providecommand \@ifx [1]{%
 \ifx #1\expandafter \@firstoftwo
 \else \expandafter \@secondoftwo
 \fi
}%
\providecommand \natexlab [1]{#1}%
\providecommand \enquote  [1]{``#1''}%
\providecommand \bibnamefont  [1]{#1}%
\providecommand \bibfnamefont [1]{#1}%
\providecommand \citenamefont [1]{#1}%
\providecommand \href@noop [0]{\@secondoftwo}%
\providecommand \href [0]{\begingroup \@sanitize@url \@href}%
\providecommand \@href[1]{\@@startlink{#1}\@@href}%
\providecommand \@@href[1]{\endgroup#1\@@endlink}%
\providecommand \@sanitize@url [0]{\catcode `\\12\catcode `\$12\catcode
  `\&12\catcode `\#12\catcode `\^12\catcode `\_12\catcode `\%12\relax}%
\providecommand \@@startlink[1]{}%
\providecommand \@@endlink[0]{}%
\providecommand \url  [0]{\begingroup\@sanitize@url \@url }%
\providecommand \@url [1]{\endgroup\@href {#1}{\urlprefix }}%
\providecommand \urlprefix  [0]{URL }%
\providecommand \Eprint [0]{\href }%
\providecommand \doibase [0]{https://doi.org/}%
\providecommand \selectlanguage [0]{\@gobble}%
\providecommand \bibinfo  [0]{\@secondoftwo}%
\providecommand \bibfield  [0]{\@secondoftwo}%
\providecommand \translation [1]{[#1]}%
\providecommand \BibitemOpen [0]{}%
\providecommand \bibitemStop [0]{}%
\providecommand \bibitemNoStop [0]{.\EOS\space}%
\providecommand \EOS [0]{\spacefactor3000\relax}%
\providecommand \BibitemShut  [1]{\csname bibitem#1\endcsname}%
\let\auto@bib@innerbib\@empty
\bibitem [{\citenamefont {Sun}\ \emph {et~al.}(2023)\citenamefont {Sun},
  \citenamefont {Huo}, \citenamefont {Hu}, \citenamefont {Li}, \citenamefont
  {Liu}, \citenamefont {Han}, \citenamefont {Tang}, \citenamefont {Mao},
  \citenamefont {Yang}, \citenamefont {Wang}, \citenamefont {Cheng},
  \citenamefont {Yao}, \citenamefont {Zhang},\ and\ \citenamefont
  {Wang}}]{Sun_Huo}%
  \BibitemOpen
  \bibfield  {author} {\bibinfo {author} {\bibfnamefont {H.}~\bibnamefont
  {Sun}}, \bibinfo {author} {\bibfnamefont {M.}~\bibnamefont {Huo}}, \bibinfo
  {author} {\bibfnamefont {X.}~\bibnamefont {Hu}}, \bibinfo {author}
  {\bibfnamefont {J.}~\bibnamefont {Li}}, \bibinfo {author} {\bibfnamefont
  {Z.}~\bibnamefont {Liu}}, \bibinfo {author} {\bibfnamefont {Y.}~\bibnamefont
  {Han}}, \bibinfo {author} {\bibfnamefont {L.}~\bibnamefont {Tang}}, \bibinfo
  {author} {\bibfnamefont {Z.}~\bibnamefont {Mao}}, \bibinfo {author}
  {\bibfnamefont {P.}~\bibnamefont {Yang}}, \bibinfo {author} {\bibfnamefont
  {B.}~\bibnamefont {Wang}}, \bibinfo {author} {\bibfnamefont {J.}~\bibnamefont
  {Cheng}}, \bibinfo {author} {\bibfnamefont {D.-X.}\ \bibnamefont {Yao}},
  \bibinfo {author} {\bibfnamefont {G.-M.}\ \bibnamefont {Zhang}},\ and\
  \bibinfo {author} {\bibfnamefont {M.}~\bibnamefont {Wang}},\ }\bibfield
  {title} {\bibinfo {title} {Signatures of superconductivity near 80 {K} in a
  nickelate under high pressure},\ }\href
  {https://doi.org/10.1038/s41586-023-06408-7} {\bibfield  {journal} {\bibinfo
  {journal} {Nature}\ }\textbf {\bibinfo {volume} {621}},\ \bibinfo {pages}
  {493} (\bibinfo {year} {2023})}\BibitemShut {NoStop}%
\bibitem [{\citenamefont {Zhang}\ \emph
  {et~al.}(2023{\natexlab{a}})\citenamefont {Zhang}, \citenamefont {Lin},
  \citenamefont {Moreo},\ and\ \citenamefont {Dagotto}}]{Zhang_Lin_2}%
  \BibitemOpen
  \bibfield  {author} {\bibinfo {author} {\bibfnamefont {Y.}~\bibnamefont
  {Zhang}}, \bibinfo {author} {\bibfnamefont {L.-F.}\ \bibnamefont {Lin}},
  \bibinfo {author} {\bibfnamefont {A.}~\bibnamefont {Moreo}},\ and\ \bibinfo
  {author} {\bibfnamefont {E.}~\bibnamefont {Dagotto}},\ }\bibfield  {title}
  {\bibinfo {title} {Electronic structure, dimer physics, orbital-selective
  behavior, and magnetic tendencies in the bilayer nickelate superconductor
  {La}$_{3}${Ni}$_{2}${O}$_{7}$ under pressure},\ }\href
  {https://doi.org/10.1103/PhysRevB.108.L180510} {\bibfield  {journal}
  {\bibinfo  {journal} {Phys. Rev. B}\ }\textbf {\bibinfo {volume} {108}},\
  \bibinfo {pages} {L180510} (\bibinfo {year}
  {2023}{\natexlab{a}})}\BibitemShut {NoStop}%
\bibitem [{\citenamefont {Yang}\ \emph
  {et~al.}(2023{\natexlab{a}})\citenamefont {Yang}, \citenamefont {Wang},\ and\
  \citenamefont {Wang}}]{Yang_Wang}%
  \BibitemOpen
  \bibfield  {author} {\bibinfo {author} {\bibfnamefont {Q.-G.}\ \bibnamefont
  {Yang}}, \bibinfo {author} {\bibfnamefont {D.}~\bibnamefont {Wang}},\ and\
  \bibinfo {author} {\bibfnamefont {Q.-H.}\ \bibnamefont {Wang}},\ }\bibfield
  {title} {\bibinfo {title} {Possible ${s}_{\pm}$-wave superconductivity in
  {La}$_{3}${Ni}$_{2}${O}$_{7}$},\ }\href
  {https://doi.org/10.1103/PhysRevB.108.L140505} {\bibfield  {journal}
  {\bibinfo  {journal} {Phys. Rev. B}\ }\textbf {\bibinfo {volume} {108}},\
  \bibinfo {pages} {L140505} (\bibinfo {year}
  {2023}{\natexlab{a}})}\BibitemShut {NoStop}%
\bibitem [{\citenamefont {Lechermann}\ \emph {et~al.}(2023)\citenamefont
  {Lechermann}, \citenamefont {Gondolf}, \citenamefont {B\"otzel},\ and\
  \citenamefont {Eremin}}]{Lechermann_Gondolf}%
  \BibitemOpen
  \bibfield  {author} {\bibinfo {author} {\bibfnamefont {F.}~\bibnamefont
  {Lechermann}}, \bibinfo {author} {\bibfnamefont {J.}~\bibnamefont {Gondolf}},
  \bibinfo {author} {\bibfnamefont {S.}~\bibnamefont {B\"otzel}},\ and\
  \bibinfo {author} {\bibfnamefont {I.~M.}\ \bibnamefont {Eremin}},\ }\bibfield
   {title} {\bibinfo {title} {Electronic correlations and superconducting
  instability in {La}$_3${Ni}$_2${O}$_7$ under high pressure},\ }\href
  {https://doi.org/10.1103/PhysRevB.108.L201121} {\bibfield  {journal}
  {\bibinfo  {journal} {Phys. Rev. B}\ }\textbf {\bibinfo {volume} {108}},\
  \bibinfo {pages} {L201121} (\bibinfo {year} {2023})}\BibitemShut {NoStop}%
\bibitem [{\citenamefont {Sakakibara}\ \emph
  {et~al.}(2024{\natexlab{a}})\citenamefont {Sakakibara}, \citenamefont
  {Kitamine}, \citenamefont {Ochi},\ and\ \citenamefont
  {Kuroki}}]{Sakakibara_Kitamine}%
  \BibitemOpen
  \bibfield  {author} {\bibinfo {author} {\bibfnamefont {H.}~\bibnamefont
  {Sakakibara}}, \bibinfo {author} {\bibfnamefont {N.}~\bibnamefont
  {Kitamine}}, \bibinfo {author} {\bibfnamefont {M.}~\bibnamefont {Ochi}},\
  and\ \bibinfo {author} {\bibfnamefont {K.}~\bibnamefont {Kuroki}},\
  }\bibfield  {title} {\bibinfo {title} {Possible high ${T}_{c}$
  superconductivity in {La}$_{3}${Ni}$_{2}${O}$_{7}$ under high pressure
  through manifestation of a nearly half-filled bilayer {H}ubbard model},\
  }\href {https://doi.org/10.1103/PhysRevLett.132.106002} {\bibfield  {journal}
  {\bibinfo  {journal} {Phys. Rev. Lett.}\ }\textbf {\bibinfo {volume} {132}},\
  \bibinfo {pages} {106002} (\bibinfo {year} {2024}{\natexlab{a}})}\BibitemShut
  {NoStop}%
\bibitem [{\citenamefont {Gu}\ \emph {et~al.}(2025)\citenamefont {Gu},
  \citenamefont {Le}, \citenamefont {Yang}, \citenamefont {Wu},\ and\
  \citenamefont {Hu}}]{Gu_Le}%
  \BibitemOpen
  \bibfield  {author} {\bibinfo {author} {\bibfnamefont {Y.}~\bibnamefont
  {Gu}}, \bibinfo {author} {\bibfnamefont {C.}~\bibnamefont {Le}}, \bibinfo
  {author} {\bibfnamefont {Z.}~\bibnamefont {Yang}}, \bibinfo {author}
  {\bibfnamefont {X.}~\bibnamefont {Wu}},\ and\ \bibinfo {author}
  {\bibfnamefont {J.}~\bibnamefont {Hu}},\ }\bibfield  {title} {\bibinfo
  {title} {Effective model and pairing tendency in the bilayer ni-based
  superconductor ${\mathrm{la}}_{3}{\mathrm{ni}}_{2}{\mathrm{o}}_{7}$},\ }\href
  {https://link.aps.org/doi/10.1103/PhysRevB.111.174506} {\bibfield  {journal}
  {\bibinfo  {journal} {Phys. Rev. B}\ }\textbf {\bibinfo {volume} {111}},\
  \bibinfo {pages} {174506} (\bibinfo {year} {2025})}\BibitemShut {NoStop}%
\bibitem [{\citenamefont {Shen}\ \emph {et~al.}(2023)\citenamefont {Shen},
  \citenamefont {Qin},\ and\ \citenamefont {Zhang}}]{Shen_Qin}%
  \BibitemOpen
  \bibfield  {author} {\bibinfo {author} {\bibfnamefont {Y.}~\bibnamefont
  {Shen}}, \bibinfo {author} {\bibfnamefont {M.}~\bibnamefont {Qin}},\ and\
  \bibinfo {author} {\bibfnamefont {G.-M.}\ \bibnamefont {Zhang}},\ }\bibfield
  {title} {\bibinfo {title} {Effective bi-layer model hamiltonian and
  density-matrix renormalization group study for the high-${T}_{c}$
  superconductivity in {La}$_3${Ni}$_2${O}$_7$ under high pressure},\ }\href
  {https://doi.org/10.1088/0256-307X/40/12/127401} {\bibfield  {journal}
  {\bibinfo  {journal} {Chin. Phys. Lett.}\ }\textbf {\bibinfo {volume} {40}},\
  \bibinfo {pages} {127401} (\bibinfo {year} {2023})}\BibitemShut {NoStop}%
\bibitem [{\citenamefont {Christiansson}\ \emph {et~al.}(2023)\citenamefont
  {Christiansson}, \citenamefont {Petocchi},\ and\ \citenamefont
  {Werner}}]{Christiansson_Petocchi}%
  \BibitemOpen
  \bibfield  {author} {\bibinfo {author} {\bibfnamefont {V.}~\bibnamefont
  {Christiansson}}, \bibinfo {author} {\bibfnamefont {F.}~\bibnamefont
  {Petocchi}},\ and\ \bibinfo {author} {\bibfnamefont {P.}~\bibnamefont
  {Werner}},\ }\bibfield  {title} {\bibinfo {title} {Correlated electronic
  structure of {La}$_{3}${Ni}$_{2}${O}$_{7}$ under pressure},\ }\href
  {https://doi.org/10.1103/PhysRevLett.131.206501} {\bibfield  {journal}
  {\bibinfo  {journal} {Phys. Rev. Lett.}\ }\textbf {\bibinfo {volume} {131}},\
  \bibinfo {pages} {206501} (\bibinfo {year} {2023})}\BibitemShut {NoStop}%
\bibitem [{\citenamefont {Liu}\ \emph {et~al.}(2024)\citenamefont {Liu},
  \citenamefont {Huo}, \citenamefont {Li}, \citenamefont {Li}, \citenamefont
  {Liu}, \citenamefont {Dai}, \citenamefont {Zhou}, \citenamefont {Hao},
  \citenamefont {Lu}, \citenamefont {Wang},\ and\ \citenamefont
  {Wen}}]{Liu_Huo}%
  \BibitemOpen
  \bibfield  {author} {\bibinfo {author} {\bibfnamefont {Z.}~\bibnamefont
  {Liu}}, \bibinfo {author} {\bibfnamefont {M.}~\bibnamefont {Huo}}, \bibinfo
  {author} {\bibfnamefont {J.}~\bibnamefont {Li}}, \bibinfo {author}
  {\bibfnamefont {Q.}~\bibnamefont {Li}}, \bibinfo {author} {\bibfnamefont
  {Y.}~\bibnamefont {Liu}}, \bibinfo {author} {\bibfnamefont {Y.}~\bibnamefont
  {Dai}}, \bibinfo {author} {\bibfnamefont {X.}~\bibnamefont {Zhou}}, \bibinfo
  {author} {\bibfnamefont {J.}~\bibnamefont {Hao}}, \bibinfo {author}
  {\bibfnamefont {Y.}~\bibnamefont {Lu}}, \bibinfo {author} {\bibfnamefont
  {M.}~\bibnamefont {Wang}},\ and\ \bibinfo {author} {\bibfnamefont {H.-H.}\
  \bibnamefont {Wen}},\ }\bibfield  {title} {\bibinfo {title} {Electronic
  correlations and partial gap in the bilayer nickelate
  {La}$_{3}${Ni}$_{2}${O}$_{7}$},\ }\href
  {https://doi.org/10.1038/s41467-024-52001-5} {\bibfield  {journal} {\bibinfo
  {journal} {Nat. Commun.}\ }\textbf {\bibinfo {volume} {15}},\ \bibinfo
  {pages} {7570} (\bibinfo {year} {2024})}\BibitemShut {NoStop}%
\bibitem [{\citenamefont {W\'{u}}\ \emph {et~al.}(2024)\citenamefont {W\'{u}},
  \citenamefont {Luo}, \citenamefont {Yao},\ and\ \citenamefont
  {Wang}}]{Wu_Luo}%
  \BibitemOpen
  \bibfield  {author} {\bibinfo {author} {\bibfnamefont {W.}~\bibnamefont
  {W\'{u}}}, \bibinfo {author} {\bibfnamefont {Z.}~\bibnamefont {Luo}},
  \bibinfo {author} {\bibfnamefont {D.-X.}\ \bibnamefont {Yao}},\ and\ \bibinfo
  {author} {\bibfnamefont {M.}~\bibnamefont {Wang}},\ }\bibfield  {title}
  {\bibinfo {title} {Superexchange and charge transfer in the nickelate
  superconductor {La}$_{3}${Ni}$_{2}${O}$_{7}$ under pressure},\ }\href
  {https://doi.org/10.1007/s11433-023-2300-4} {\bibfield  {journal} {\bibinfo
  {journal} {Sci. China Phys. Mech. Astron.}\ }\textbf {\bibinfo {volume}
  {67}},\ \bibinfo {pages} {117402} (\bibinfo {year} {2024})}\BibitemShut
  {NoStop}%
\bibitem [{\citenamefont {Cao}\ and\ \citenamefont {Yang}(2024)}]{Cao_Yang}%
  \BibitemOpen
  \bibfield  {author} {\bibinfo {author} {\bibfnamefont {Y.}~\bibnamefont
  {Cao}}\ and\ \bibinfo {author} {\bibfnamefont {Y.-f.}\ \bibnamefont {Yang}},\
  }\bibfield  {title} {\bibinfo {title} {Flat bands promoted by hund's rule
  coupling in the candidate double-layer high-temperature superconductor
  {La}$_{3}${Ni}$_{2}${O}$_{7}$ under high pressure},\ }\href
  {https://doi.org/10.1103/PhysRevB.109.L081105} {\bibfield  {journal}
  {\bibinfo  {journal} {Phys. Rev. B}\ }\textbf {\bibinfo {volume} {109}},\
  \bibinfo {pages} {L081105} (\bibinfo {year} {2024})}\BibitemShut {NoStop}%
\bibitem [{\citenamefont {Hou}\ \emph {et~al.}(2023)\citenamefont {Hou},
  \citenamefont {Yang}, \citenamefont {Liu}, \citenamefont {Li}, \citenamefont
  {Shan}, \citenamefont {Ma}, \citenamefont {Wang}, \citenamefont {Wang},
  \citenamefont {Guo}, \citenamefont {Sun}, \citenamefont {Uwatoko},
  \citenamefont {Wang}, \citenamefont {Zhang}, \citenamefont {Wang},\ and\
  \citenamefont {Cheng}}]{Hou_Yang}%
  \BibitemOpen
  \bibfield  {author} {\bibinfo {author} {\bibfnamefont {J.}~\bibnamefont
  {Hou}}, \bibinfo {author} {\bibfnamefont {P.-T.}\ \bibnamefont {Yang}},
  \bibinfo {author} {\bibfnamefont {Z.-Y.}\ \bibnamefont {Liu}}, \bibinfo
  {author} {\bibfnamefont {J.-Y.}\ \bibnamefont {Li}}, \bibinfo {author}
  {\bibfnamefont {P.-F.}\ \bibnamefont {Shan}}, \bibinfo {author}
  {\bibfnamefont {L.}~\bibnamefont {Ma}}, \bibinfo {author} {\bibfnamefont
  {G.}~\bibnamefont {Wang}}, \bibinfo {author} {\bibfnamefont {N.-N.}\
  \bibnamefont {Wang}}, \bibinfo {author} {\bibfnamefont {H.-Z.}\ \bibnamefont
  {Guo}}, \bibinfo {author} {\bibfnamefont {J.-P.}\ \bibnamefont {Sun}},
  \bibinfo {author} {\bibfnamefont {Y.}~\bibnamefont {Uwatoko}}, \bibinfo
  {author} {\bibfnamefont {M.}~\bibnamefont {Wang}}, \bibinfo {author}
  {\bibfnamefont {G.-M.}\ \bibnamefont {Zhang}}, \bibinfo {author}
  {\bibfnamefont {B.-S.}\ \bibnamefont {Wang}},\ and\ \bibinfo {author}
  {\bibfnamefont {J.-G.}\ \bibnamefont {Cheng}},\ }\bibfield  {title} {\bibinfo
  {title} {Emergence of high-temperature superconducting phase in pressurized
  {La}$_{3}${Ni}$_{2}${O}$_{7}$ crystals},\ }\href
  {https://doi.org/10.1088/0256-307X/40/11/117302} {\bibfield  {journal}
  {\bibinfo  {journal} {Chin. Phys. Lett.}\ }\textbf {\bibinfo {volume} {40}},\
  \bibinfo {pages} {117302} (\bibinfo {year} {2023})}\BibitemShut {NoStop}%
\bibitem [{\citenamefont {Liu}\ \emph {et~al.}(2023{\natexlab{a}})\citenamefont
  {Liu}, \citenamefont {Mei}, \citenamefont {Ye}, \citenamefont {Chen},\ and\
  \citenamefont {Yang}}]{Liu_Mei}%
  \BibitemOpen
  \bibfield  {author} {\bibinfo {author} {\bibfnamefont {Y.-B.}\ \bibnamefont
  {Liu}}, \bibinfo {author} {\bibfnamefont {J.-W.}\ \bibnamefont {Mei}},
  \bibinfo {author} {\bibfnamefont {F.}~\bibnamefont {Ye}}, \bibinfo {author}
  {\bibfnamefont {W.-Q.}\ \bibnamefont {Chen}},\ and\ \bibinfo {author}
  {\bibfnamefont {F.}~\bibnamefont {Yang}},\ }\bibfield  {title} {\bibinfo
  {title} {${s}^{\pm}$-wave pairing and the destructive role of apical-oxygen
  deficiencies in {La}$_{3}${Ni}$_{2}${O}$_{7}$ under pressure},\ }\href
  {https://doi.org/10.1103/PhysRevLett.131.236002} {\bibfield  {journal}
  {\bibinfo  {journal} {Phys. Rev. Lett.}\ }\textbf {\bibinfo {volume} {131}},\
  \bibinfo {pages} {236002} (\bibinfo {year} {2023}{\natexlab{a}})}\BibitemShut
  {NoStop}%
\bibitem [{\citenamefont {Zhang}\ \emph
  {et~al.}(2024{\natexlab{a}})\citenamefont {Zhang}, \citenamefont {Su},
  \citenamefont {Huang}, \citenamefont {Shan}, \citenamefont {Sun},
  \citenamefont {Huo}, \citenamefont {Ye}, \citenamefont {Zhang}, \citenamefont
  {Yang}, \citenamefont {Xu}, \citenamefont {Su}, \citenamefont {Li},
  \citenamefont {Smidman}, \citenamefont {Wang}, \citenamefont {Jiao},\ and\
  \citenamefont {Yuan}}]{Zhang_Su}%
  \BibitemOpen
  \bibfield  {author} {\bibinfo {author} {\bibfnamefont {Y.}~\bibnamefont
  {Zhang}}, \bibinfo {author} {\bibfnamefont {D.}~\bibnamefont {Su}}, \bibinfo
  {author} {\bibfnamefont {Y.}~\bibnamefont {Huang}}, \bibinfo {author}
  {\bibfnamefont {Z.}~\bibnamefont {Shan}}, \bibinfo {author} {\bibfnamefont
  {H.}~\bibnamefont {Sun}}, \bibinfo {author} {\bibfnamefont {M.}~\bibnamefont
  {Huo}}, \bibinfo {author} {\bibfnamefont {K.}~\bibnamefont {Ye}}, \bibinfo
  {author} {\bibfnamefont {J.}~\bibnamefont {Zhang}}, \bibinfo {author}
  {\bibfnamefont {Z.}~\bibnamefont {Yang}}, \bibinfo {author} {\bibfnamefont
  {Y.}~\bibnamefont {Xu}}, \bibinfo {author} {\bibfnamefont {Y.}~\bibnamefont
  {Su}}, \bibinfo {author} {\bibfnamefont {R.}~\bibnamefont {Li}}, \bibinfo
  {author} {\bibfnamefont {M.}~\bibnamefont {Smidman}}, \bibinfo {author}
  {\bibfnamefont {M.}~\bibnamefont {Wang}}, \bibinfo {author} {\bibfnamefont
  {L.}~\bibnamefont {Jiao}},\ and\ \bibinfo {author} {\bibfnamefont
  {H.}~\bibnamefont {Yuan}},\ }\bibfield  {title} {\bibinfo {title}
  {High-temperature superconductivity with zero resistance and strange-metal
  behaviour in {La}$_{3}${Ni}$_{2}${O}$_{7-\delta}$},\ }\href
  {https://doi.org/10.1038/s41567-024-02515-y} {\bibfield  {journal} {\bibinfo
  {journal} {Nat. Phys.}\ }\textbf {\bibinfo {volume} {20}},\ \bibinfo {pages}
  {1269} (\bibinfo {year} {2024}{\natexlab{a}})}\BibitemShut {NoStop}%
\bibitem [{\citenamefont {Lu}\ \emph {et~al.}(2024{\natexlab{a}})\citenamefont
  {Lu}, \citenamefont {Pan}, \citenamefont {Yang},\ and\ \citenamefont
  {Wu}}]{Lu_Pan_15}%
  \BibitemOpen
  \bibfield  {author} {\bibinfo {author} {\bibfnamefont {C.}~\bibnamefont
  {Lu}}, \bibinfo {author} {\bibfnamefont {Z.}~\bibnamefont {Pan}}, \bibinfo
  {author} {\bibfnamefont {F.}~\bibnamefont {Yang}},\ and\ \bibinfo {author}
  {\bibfnamefont {C.}~\bibnamefont {Wu}},\ }\bibfield  {title} {\bibinfo
  {title} {Interlayer-coupling-driven high-temperature superconductivity in
  {La}$_{3}${Ni}$_{2}${O}$_{7}$ under pressure},\ }\href
  {https://doi.org/10.1103/PhysRevLett.132.146002} {\bibfield  {journal}
  {\bibinfo  {journal} {Phys. Rev. Lett.}\ }\textbf {\bibinfo {volume} {132}},\
  \bibinfo {pages} {146002} (\bibinfo {year} {2024}{\natexlab{a}})}\BibitemShut
  {NoStop}%
\bibitem [{\citenamefont {Zhang}\ \emph
  {et~al.}(2024{\natexlab{b}})\citenamefont {Zhang}, \citenamefont {Lin},
  \citenamefont {Moreo}, \citenamefont {Maier},\ and\ \citenamefont
  {Dagotto}}]{Zhang_Lin_16}%
  \BibitemOpen
  \bibfield  {author} {\bibinfo {author} {\bibfnamefont {Y.}~\bibnamefont
  {Zhang}}, \bibinfo {author} {\bibfnamefont {L.-F.}\ \bibnamefont {Lin}},
  \bibinfo {author} {\bibfnamefont {A.}~\bibnamefont {Moreo}}, \bibinfo
  {author} {\bibfnamefont {T.~A.}\ \bibnamefont {Maier}},\ and\ \bibinfo
  {author} {\bibfnamefont {E.}~\bibnamefont {Dagotto}},\ }\bibfield  {title}
  {\bibinfo {title} {Structural phase transition, $s_{\pm}$-wave pairing, and
  magnetic stripe order in bilayered superconductor
  {La}$_{3}${Ni}$_{2}${O}$_{7}$ under pressure},\ }\href
  {https://doi.org/10.1038/s41467-024-46622-z} {\bibfield  {journal} {\bibinfo
  {journal} {Nat. Commun.}\ }\textbf {\bibinfo {volume} {15}},\ \bibinfo
  {pages} {2470} (\bibinfo {year} {2024}{\natexlab{b}})}\BibitemShut {NoStop}%
\bibitem [{\citenamefont {Oh}\ and\ \citenamefont {Zhang}(2023)}]{Oh_Zhang}%
  \BibitemOpen
  \bibfield  {author} {\bibinfo {author} {\bibfnamefont {H.}~\bibnamefont
  {Oh}}\ and\ \bibinfo {author} {\bibfnamefont {Y.-H.}\ \bibnamefont {Zhang}},\
  }\bibfield  {title} {\bibinfo {title} {Type-{II} $t\ensuremath{-}{J}$ model
  and shared superexchange coupling from {H}und's rule in superconducting
  {La}$_{3}${Ni}$_{2}${O}$_{7}$},\ }\href
  {https://doi.org/10.1103/PhysRevB.108.174511} {\bibfield  {journal} {\bibinfo
   {journal} {Phys. Rev. B}\ }\textbf {\bibinfo {volume} {108}},\ \bibinfo
  {pages} {174511} (\bibinfo {year} {2023})}\BibitemShut {NoStop}%
\bibitem [{\citenamefont {Liao}\ \emph {et~al.}(2023)\citenamefont {Liao},
  \citenamefont {Chen}, \citenamefont {Duan}, \citenamefont {Wang},
  \citenamefont {Liu}, \citenamefont {Yu},\ and\ \citenamefont
  {Si}}]{Liao_Chen}%
  \BibitemOpen
  \bibfield  {author} {\bibinfo {author} {\bibfnamefont {Z.}~\bibnamefont
  {Liao}}, \bibinfo {author} {\bibfnamefont {L.}~\bibnamefont {Chen}}, \bibinfo
  {author} {\bibfnamefont {G.}~\bibnamefont {Duan}}, \bibinfo {author}
  {\bibfnamefont {Y.}~\bibnamefont {Wang}}, \bibinfo {author} {\bibfnamefont
  {C.}~\bibnamefont {Liu}}, \bibinfo {author} {\bibfnamefont {R.}~\bibnamefont
  {Yu}},\ and\ \bibinfo {author} {\bibfnamefont {Q.}~\bibnamefont {Si}},\
  }\bibfield  {title} {\bibinfo {title} {Electron correlations and
  superconductivity in {La}$_{3}${Ni}$_{2}${O}$_{7}$ under pressure tuning},\
  }\href {https://doi.org/10.1103/PhysRevB.108.214522} {\bibfield  {journal}
  {\bibinfo  {journal} {Phys. Rev. B}\ }\textbf {\bibinfo {volume} {108}},\
  \bibinfo {pages} {214522} (\bibinfo {year} {2023})}\BibitemShut {NoStop}%
\bibitem [{\citenamefont {Qu}\ \emph {et~al.}(2024)\citenamefont {Qu},
  \citenamefont {Qu}, \citenamefont {Chen}, \citenamefont {Wu}, \citenamefont
  {Yang}, \citenamefont {Li},\ and\ \citenamefont {Su}}]{Qu_Qu_19}%
  \BibitemOpen
  \bibfield  {author} {\bibinfo {author} {\bibfnamefont {X.-Z.}\ \bibnamefont
  {Qu}}, \bibinfo {author} {\bibfnamefont {D.-W.}\ \bibnamefont {Qu}}, \bibinfo
  {author} {\bibfnamefont {J.}~\bibnamefont {Chen}}, \bibinfo {author}
  {\bibfnamefont {C.}~\bibnamefont {Wu}}, \bibinfo {author} {\bibfnamefont
  {F.}~\bibnamefont {Yang}}, \bibinfo {author} {\bibfnamefont {W.}~\bibnamefont
  {Li}},\ and\ \bibinfo {author} {\bibfnamefont {G.}~\bibnamefont {Su}},\
  }\bibfield  {title} {\bibinfo {title} {Bilayer
  ${t\text{\ensuremath{-}}{J}\text{\ensuremath{-}}J}_{\ensuremath{\perp}}$
  model and magnetically mediated pairing in the pressurized nickelate
  {La}$_{3}${Ni}$_{2}${O}$_{7}$},\ }\href
  {https://doi.org/10.1103/PhysRevLett.132.036502} {\bibfield  {journal}
  {\bibinfo  {journal} {Phys. Rev. Lett.}\ }\textbf {\bibinfo {volume} {132}},\
  \bibinfo {pages} {036502} (\bibinfo {year} {2024})}\BibitemShut {NoStop}%
\bibitem [{\citenamefont {Yang}\ \emph
  {et~al.}(2023{\natexlab{b}})\citenamefont {Yang}, \citenamefont {Zhang},\
  and\ \citenamefont {Zhang}}]{Yang_Zhang}%
  \BibitemOpen
  \bibfield  {author} {\bibinfo {author} {\bibfnamefont {Y.-f.}\ \bibnamefont
  {Yang}}, \bibinfo {author} {\bibfnamefont {G.-M.}\ \bibnamefont {Zhang}},\
  and\ \bibinfo {author} {\bibfnamefont {F.-C.}\ \bibnamefont {Zhang}},\
  }\bibfield  {title} {\bibinfo {title} {Interlayer valence bonds and
  two-component theory for high-${T}_{c}$ superconductivity of
  {La}$_{3}${Ni}$_{2}${O}$_{7}$ under pressure},\ }\href
  {https://doi.org/10.1103/PhysRevB.108.L201108} {\bibfield  {journal}
  {\bibinfo  {journal} {Phys. Rev. B}\ }\textbf {\bibinfo {volume} {108}},\
  \bibinfo {pages} {L201108} (\bibinfo {year}
  {2023}{\natexlab{b}})}\BibitemShut {NoStop}%
\bibitem [{\citenamefont {Jiang}\ \emph
  {et~al.}(2024{\natexlab{a}})\citenamefont {Jiang}, \citenamefont {Wang},\
  and\ \citenamefont {Zhang}}]{Jiang_Wang}%
  \BibitemOpen
  \bibfield  {author} {\bibinfo {author} {\bibfnamefont {K.}~\bibnamefont
  {Jiang}}, \bibinfo {author} {\bibfnamefont {Z.}~\bibnamefont {Wang}},\ and\
  \bibinfo {author} {\bibfnamefont {F.-C.}\ \bibnamefont {Zhang}},\ }\bibfield
  {title} {\bibinfo {title} {High-temperature superconductivity in
  {La}$_{3}${Ni}$_{2}${O}$_{7}$},\ }\href
  {https://doi.org/10.1088/0256-307X/41/1/017402} {\bibfield  {journal}
  {\bibinfo  {journal} {Chin. Phys. Lett.}\ }\textbf {\bibinfo {volume} {41}},\
  \bibinfo {pages} {017402} (\bibinfo {year} {2024}{\natexlab{a}})}\BibitemShut
  {NoStop}%
\bibitem [{\citenamefont {Zhang}\ \emph
  {et~al.}(2023{\natexlab{b}})\citenamefont {Zhang}, \citenamefont {Lin},
  \citenamefont {Moreo}, \citenamefont {Maier},\ and\ \citenamefont
  {Dagotto}}]{Zhang_Lin_22}%
  \BibitemOpen
  \bibfield  {author} {\bibinfo {author} {\bibfnamefont {Y.}~\bibnamefont
  {Zhang}}, \bibinfo {author} {\bibfnamefont {L.-F.}\ \bibnamefont {Lin}},
  \bibinfo {author} {\bibfnamefont {A.}~\bibnamefont {Moreo}}, \bibinfo
  {author} {\bibfnamefont {T.~A.}\ \bibnamefont {Maier}},\ and\ \bibinfo
  {author} {\bibfnamefont {E.}~\bibnamefont {Dagotto}},\ }\bibfield  {title}
  {\bibinfo {title} {Trends in electronic structures and ${s}_{\pm}$-wave
  pairing for the rare-earth series in bilayer nickelate superconductor
  ${R}_{3}${Ni}$_{2}${O}$_{7}$},\ }\href
  {https://doi.org/10.1103/PhysRevB.108.165141} {\bibfield  {journal} {\bibinfo
   {journal} {Phys. Rev. B}\ }\textbf {\bibinfo {volume} {108}},\ \bibinfo
  {pages} {165141} (\bibinfo {year} {2023}{\natexlab{b}})}\BibitemShut
  {NoStop}%
\bibitem [{\citenamefont {Tian}\ \emph {et~al.}(2024)\citenamefont {Tian},
  \citenamefont {Chen}, \citenamefont {Wang}, \citenamefont {He},\ and\
  \citenamefont {Lu}}]{Tian_Chen}%
  \BibitemOpen
  \bibfield  {author} {\bibinfo {author} {\bibfnamefont {Y.-H.}\ \bibnamefont
  {Tian}}, \bibinfo {author} {\bibfnamefont {Y.}~\bibnamefont {Chen}}, \bibinfo
  {author} {\bibfnamefont {J.-M.}\ \bibnamefont {Wang}}, \bibinfo {author}
  {\bibfnamefont {R.-Q.}\ \bibnamefont {He}},\ and\ \bibinfo {author}
  {\bibfnamefont {Z.-Y.}\ \bibnamefont {Lu}},\ }\bibfield  {title} {\bibinfo
  {title} {Correlation effects and concomitant two-orbital ${s}_{\pm}$-wave
  superconductivity in {La}$_{3}${Ni}$_{2}${O}$_{7}$ under high pressure},\
  }\href {https://doi.org/10.1103/PhysRevB.109.165154} {\bibfield  {journal}
  {\bibinfo  {journal} {Phys. Rev. B}\ }\textbf {\bibinfo {volume} {109}},\
  \bibinfo {pages} {165154} (\bibinfo {year} {2024})}\BibitemShut {NoStop}%
\bibitem [{\citenamefont {Jiang}\ \emph
  {et~al.}(2024{\natexlab{b}})\citenamefont {Jiang}, \citenamefont {Hou},
  \citenamefont {Fan}, \citenamefont {Lang},\ and\ \citenamefont
  {Ku}}]{Jiang_Hou}%
  \BibitemOpen
  \bibfield  {author} {\bibinfo {author} {\bibfnamefont {R.}~\bibnamefont
  {Jiang}}, \bibinfo {author} {\bibfnamefont {J.}~\bibnamefont {Hou}}, \bibinfo
  {author} {\bibfnamefont {Z.}~\bibnamefont {Fan}}, \bibinfo {author}
  {\bibfnamefont {Z.-J.}\ \bibnamefont {Lang}},\ and\ \bibinfo {author}
  {\bibfnamefont {W.}~\bibnamefont {Ku}},\ }\bibfield  {title} {\bibinfo
  {title} {Pressure driven fractionalization of ionic spins results in
  cupratelike high-${T}_{c}$ superconductivity in
  {La}$_{3}${Ni}$_{2}${O}$_{7}$},\ }\href
  {https://doi.org/10.1103/PhysRevLett.132.126503} {\bibfield  {journal}
  {\bibinfo  {journal} {Phys. Rev. Lett.}\ }\textbf {\bibinfo {volume} {132}},\
  \bibinfo {pages} {126503} (\bibinfo {year} {2024}{\natexlab{b}})}\BibitemShut
  {NoStop}%
\bibitem [{\citenamefont {Luo}\ \emph {et~al.}(2024)\citenamefont {Luo},
  \citenamefont {Lv}, \citenamefont {Wang}, \citenamefont {W{\'{u}}},\ and\
  \citenamefont {Yao}}]{Luo_Lv}%
  \BibitemOpen
  \bibfield  {author} {\bibinfo {author} {\bibfnamefont {Z.}~\bibnamefont
  {Luo}}, \bibinfo {author} {\bibfnamefont {B.}~\bibnamefont {Lv}}, \bibinfo
  {author} {\bibfnamefont {M.}~\bibnamefont {Wang}}, \bibinfo {author}
  {\bibfnamefont {W.}~\bibnamefont {W{\'{u}}}},\ and\ \bibinfo {author}
  {\bibfnamefont {D.-X.}\ \bibnamefont {Yao}},\ }\bibfield  {title} {\bibinfo
  {title} {High-${T}_{C}$ superconductivity in {La}$_{3}${Ni}$_{2}${O}$_{7}$
  based on the bilayer two-orbital $t$-${J}$ model},\ }\href
  {https://doi.org/10.1038/s41535-024-00668-w} {\bibfield  {journal} {\bibinfo
  {journal} {npj Quantum Mater.}\ }\textbf {\bibinfo {volume} {9}},\ \bibinfo
  {pages} {61} (\bibinfo {year} {2024})}\BibitemShut {NoStop}%
\bibitem [{\citenamefont {Yang}\ \emph {et~al.}(2024)\citenamefont {Yang},
  \citenamefont {Sun}, \citenamefont {Hu}, \citenamefont {Xie}, \citenamefont
  {Miao}, \citenamefont {Luo}, \citenamefont {Chen}, \citenamefont {Liang},
  \citenamefont {Zhu}, \citenamefont {Qu}, \citenamefont {Chen}, \citenamefont
  {Huo}, \citenamefont {Huang}, \citenamefont {Zhang}, \citenamefont {Zhang},
  \citenamefont {Yang}, \citenamefont {Wang}, \citenamefont {Peng},
  \citenamefont {Mao}, \citenamefont {Liu}, \citenamefont {Xu}, \citenamefont
  {Qian}, \citenamefont {Yao}, \citenamefont {Wang}, \citenamefont {Zhao},\
  and\ \citenamefont {Zhou}}]{Yang_Sun}%
  \BibitemOpen
  \bibfield  {author} {\bibinfo {author} {\bibfnamefont {J.}~\bibnamefont
  {Yang}}, \bibinfo {author} {\bibfnamefont {H.}~\bibnamefont {Sun}}, \bibinfo
  {author} {\bibfnamefont {X.}~\bibnamefont {Hu}}, \bibinfo {author}
  {\bibfnamefont {Y.}~\bibnamefont {Xie}}, \bibinfo {author} {\bibfnamefont
  {T.}~\bibnamefont {Miao}}, \bibinfo {author} {\bibfnamefont {H.}~\bibnamefont
  {Luo}}, \bibinfo {author} {\bibfnamefont {H.}~\bibnamefont {Chen}}, \bibinfo
  {author} {\bibfnamefont {B.}~\bibnamefont {Liang}}, \bibinfo {author}
  {\bibfnamefont {W.}~\bibnamefont {Zhu}}, \bibinfo {author} {\bibfnamefont
  {G.}~\bibnamefont {Qu}}, \bibinfo {author} {\bibfnamefont {C.-Q.}\
  \bibnamefont {Chen}}, \bibinfo {author} {\bibfnamefont {M.}~\bibnamefont
  {Huo}}, \bibinfo {author} {\bibfnamefont {Y.}~\bibnamefont {Huang}}, \bibinfo
  {author} {\bibfnamefont {S.}~\bibnamefont {Zhang}}, \bibinfo {author}
  {\bibfnamefont {F.}~\bibnamefont {Zhang}}, \bibinfo {author} {\bibfnamefont
  {F.}~\bibnamefont {Yang}}, \bibinfo {author} {\bibfnamefont {Z.}~\bibnamefont
  {Wang}}, \bibinfo {author} {\bibfnamefont {Q.}~\bibnamefont {Peng}}, \bibinfo
  {author} {\bibfnamefont {H.}~\bibnamefont {Mao}}, \bibinfo {author}
  {\bibfnamefont {G.}~\bibnamefont {Liu}}, \bibinfo {author} {\bibfnamefont
  {Z.}~\bibnamefont {Xu}}, \bibinfo {author} {\bibfnamefont {T.}~\bibnamefont
  {Qian}}, \bibinfo {author} {\bibfnamefont {D.-X.}\ \bibnamefont {Yao}},
  \bibinfo {author} {\bibfnamefont {M.}~\bibnamefont {Wang}}, \bibinfo {author}
  {\bibfnamefont {L.}~\bibnamefont {Zhao}},\ and\ \bibinfo {author}
  {\bibfnamefont {X.~J.}\ \bibnamefont {Zhou}},\ }\bibfield  {title} {\bibinfo
  {title} {Orbital-dependent electron correlation in double-layer nickelate
  {La}$_{3}${Ni}$_{2}${O}$_{7}$},\ }\href
  {https://doi.org/10.1038/s41467-024-48701-7} {\bibfield  {journal} {\bibinfo
  {journal} {Nat. Commun.}\ }\textbf {\bibinfo {volume} {15}},\ \bibinfo
  {pages} {4373} (\bibinfo {year} {2024})}\BibitemShut {NoStop}%
\bibitem [{\citenamefont {Zhang}\ \emph
  {et~al.}(2024{\natexlab{c}})\citenamefont {Zhang}, \citenamefont {Pei},
  \citenamefont {Wang}, \citenamefont {Zhao}, \citenamefont {Li}, \citenamefont
  {Cao}, \citenamefont {Zhu}, \citenamefont {Wu},\ and\ \citenamefont
  {Qi}}]{Zhang_Pei_29}%
  \BibitemOpen
  \bibfield  {author} {\bibinfo {author} {\bibfnamefont {M.}~\bibnamefont
  {Zhang}}, \bibinfo {author} {\bibfnamefont {C.}~\bibnamefont {Pei}}, \bibinfo
  {author} {\bibfnamefont {Q.}~\bibnamefont {Wang}}, \bibinfo {author}
  {\bibfnamefont {Y.}~\bibnamefont {Zhao}}, \bibinfo {author} {\bibfnamefont
  {C.}~\bibnamefont {Li}}, \bibinfo {author} {\bibfnamefont {W.}~\bibnamefont
  {Cao}}, \bibinfo {author} {\bibfnamefont {S.}~\bibnamefont {Zhu}}, \bibinfo
  {author} {\bibfnamefont {J.}~\bibnamefont {Wu}},\ and\ \bibinfo {author}
  {\bibfnamefont {Y.}~\bibnamefont {Qi}},\ }\bibfield  {title} {\bibinfo
  {title} {Effects of pressure and doping on {R}uddlesden-{P}opper phases
  {La}$_{n+1}${Ni}$_{n}${O}$_{3n+1}$},\ }\href
  {https://doi.org/https://doi.org/10.1016/j.jmst.2023.11.011} {\bibfield
  {journal} {\bibinfo  {journal} {J. Mater. Sci. Technol.}\ }\textbf {\bibinfo
  {volume} {185}},\ \bibinfo {pages} {147} (\bibinfo {year}
  {2024}{\natexlab{c}})}\BibitemShut {NoStop}%
\bibitem [{\citenamefont {Wang}\ \emph
  {et~al.}(2024{\natexlab{a}})\citenamefont {Wang}, \citenamefont {Wang},
  \citenamefont {Shen}, \citenamefont {Hou}, \citenamefont {Ma}, \citenamefont
  {Shi}, \citenamefont {Ren}, \citenamefont {Gu}, \citenamefont {Ma},
  \citenamefont {Yang}, \citenamefont {Liu}, \citenamefont {Guo}, \citenamefont
  {Sun}, \citenamefont {Zhang}, \citenamefont {Calder}, \citenamefont {Yan},
  \citenamefont {Wang}, \citenamefont {Uwatoko},\ and\ \citenamefont
  {Cheng}}]{Wang_Wang_39}%
  \BibitemOpen
  \bibfield  {author} {\bibinfo {author} {\bibfnamefont {G.}~\bibnamefont
  {Wang}}, \bibinfo {author} {\bibfnamefont {N.~N.}\ \bibnamefont {Wang}},
  \bibinfo {author} {\bibfnamefont {X.~L.}\ \bibnamefont {Shen}}, \bibinfo
  {author} {\bibfnamefont {J.}~\bibnamefont {Hou}}, \bibinfo {author}
  {\bibfnamefont {L.}~\bibnamefont {Ma}}, \bibinfo {author} {\bibfnamefont
  {L.~F.}\ \bibnamefont {Shi}}, \bibinfo {author} {\bibfnamefont {Z.~A.}\
  \bibnamefont {Ren}}, \bibinfo {author} {\bibfnamefont {Y.~D.}\ \bibnamefont
  {Gu}}, \bibinfo {author} {\bibfnamefont {H.~M.}\ \bibnamefont {Ma}}, \bibinfo
  {author} {\bibfnamefont {P.~T.}\ \bibnamefont {Yang}}, \bibinfo {author}
  {\bibfnamefont {Z.~Y.}\ \bibnamefont {Liu}}, \bibinfo {author} {\bibfnamefont
  {H.~Z.}\ \bibnamefont {Guo}}, \bibinfo {author} {\bibfnamefont {J.~P.}\
  \bibnamefont {Sun}}, \bibinfo {author} {\bibfnamefont {G.~M.}\ \bibnamefont
  {Zhang}}, \bibinfo {author} {\bibfnamefont {S.}~\bibnamefont {Calder}},
  \bibinfo {author} {\bibfnamefont {J.-Q.}\ \bibnamefont {Yan}}, \bibinfo
  {author} {\bibfnamefont {B.~S.}\ \bibnamefont {Wang}}, \bibinfo {author}
  {\bibfnamefont {Y.}~\bibnamefont {Uwatoko}},\ and\ \bibinfo {author}
  {\bibfnamefont {J.-G.}\ \bibnamefont {Cheng}},\ }\bibfield  {title} {\bibinfo
  {title} {Pressure-induced superconductivity in polycrystalline
  {La}$_{3}${Ni}$_{2}${O}$_{7-\delta}$},\ }\href
  {https://doi.org/10.1103/PhysRevX.14.011040} {\bibfield  {journal} {\bibinfo
  {journal} {Phys. Rev. X}\ }\textbf {\bibinfo {volume} {14}},\ \bibinfo
  {pages} {011040} (\bibinfo {year} {2024}{\natexlab{a}})}\BibitemShut
  {NoStop}%
\bibitem [{\citenamefont {Kaneko}\ \emph {et~al.}(2024)\citenamefont {Kaneko},
  \citenamefont {Sakakibara}, \citenamefont {Ochi},\ and\ \citenamefont
  {Kuroki}}]{Kaneko_Sakakibara}%
  \BibitemOpen
  \bibfield  {author} {\bibinfo {author} {\bibfnamefont {T.}~\bibnamefont
  {Kaneko}}, \bibinfo {author} {\bibfnamefont {H.}~\bibnamefont {Sakakibara}},
  \bibinfo {author} {\bibfnamefont {M.}~\bibnamefont {Ochi}},\ and\ \bibinfo
  {author} {\bibfnamefont {K.}~\bibnamefont {Kuroki}},\ }\bibfield  {title}
  {\bibinfo {title} {Pair correlations in the two-orbital {H}ubbard ladder:
  Implications for superconductivity in the bilayer nickelate
  {La}$_{3}${Ni}$_{2}${O}$_{7}$},\ }\href
  {https://doi.org/10.1103/PhysRevB.109.045154} {\bibfield  {journal} {\bibinfo
   {journal} {Phys. Rev. B}\ }\textbf {\bibinfo {volume} {109}},\ \bibinfo
  {pages} {045154} (\bibinfo {year} {2024})}\BibitemShut {NoStop}%
\bibitem [{\citenamefont {Lu}\ \emph {et~al.}(2024{\natexlab{b}})\citenamefont
  {Lu}, \citenamefont {Pan}, \citenamefont {Yang},\ and\ \citenamefont
  {Wu}}]{Lu_Pan_41}%
  \BibitemOpen
  \bibfield  {author} {\bibinfo {author} {\bibfnamefont {C.}~\bibnamefont
  {Lu}}, \bibinfo {author} {\bibfnamefont {Z.}~\bibnamefont {Pan}}, \bibinfo
  {author} {\bibfnamefont {F.}~\bibnamefont {Yang}},\ and\ \bibinfo {author}
  {\bibfnamefont {C.}~\bibnamefont {Wu}},\ }\bibfield  {title} {\bibinfo
  {title} {Interplay of two ${E}_{g}$ orbitals in superconducting
  {La}$_{3}${Ni}$_{2}${O}$_{7}$ under pressure},\ }\href
  {https://doi.org/10.1103/PhysRevB.110.094509} {\bibfield  {journal} {\bibinfo
   {journal} {Phys. Rev. B}\ }\textbf {\bibinfo {volume} {110}},\ \bibinfo
  {pages} {094509} (\bibinfo {year} {2024}{\natexlab{b}})}\BibitemShut
  {NoStop}%
\bibitem [{\citenamefont {Ryee}\ \emph {et~al.}(2024)\citenamefont {Ryee},
  \citenamefont {Witt},\ and\ \citenamefont {Wehling}}]{Ryee_Witt}%
  \BibitemOpen
  \bibfield  {author} {\bibinfo {author} {\bibfnamefont {S.}~\bibnamefont
  {Ryee}}, \bibinfo {author} {\bibfnamefont {N.}~\bibnamefont {Witt}},\ and\
  \bibinfo {author} {\bibfnamefont {T.~O.}\ \bibnamefont {Wehling}},\
  }\bibfield  {title} {\bibinfo {title} {Quenched pair breaking by interlayer
  correlations as a key to superconductivity in
  {La}$_{3}${Ni}$_{2}${O}$_{7}$},\ }\href
  {https://doi.org/10.1103/PhysRevLett.133.096002} {\bibfield  {journal}
  {\bibinfo  {journal} {Phys. Rev. Lett.}\ }\textbf {\bibinfo {volume} {133}},\
  \bibinfo {pages} {096002} (\bibinfo {year} {2024})}\BibitemShut {NoStop}%
\bibitem [{\citenamefont {Ouyang}\ \emph
  {et~al.}(2024{\natexlab{a}})\citenamefont {Ouyang}, \citenamefont {Wang},
  \citenamefont {Wang}, \citenamefont {He}, \citenamefont {Huang},\ and\
  \citenamefont {Lu}}]{Ouyang_Wang}%
  \BibitemOpen
  \bibfield  {author} {\bibinfo {author} {\bibfnamefont {Z.}~\bibnamefont
  {Ouyang}}, \bibinfo {author} {\bibfnamefont {J.-M.}\ \bibnamefont {Wang}},
  \bibinfo {author} {\bibfnamefont {J.-X.}\ \bibnamefont {Wang}}, \bibinfo
  {author} {\bibfnamefont {R.-Q.}\ \bibnamefont {He}}, \bibinfo {author}
  {\bibfnamefont {L.}~\bibnamefont {Huang}},\ and\ \bibinfo {author}
  {\bibfnamefont {Z.-Y.}\ \bibnamefont {Lu}},\ }\bibfield  {title} {\bibinfo
  {title} {Hund electronic correlation in {La}$_{3}${Ni}$_{2}${O}$_{7}$ under
  high pressure},\ }\href {https://doi.org/10.1103/PhysRevB.109.115114}
  {\bibfield  {journal} {\bibinfo  {journal} {Phys. Rev. B}\ }\textbf {\bibinfo
  {volume} {109}},\ \bibinfo {pages} {115114} (\bibinfo {year}
  {2024}{\natexlab{a}})}\BibitemShut {NoStop}%
\bibitem [{\citenamefont {Zhou}\ \emph
  {et~al.}(2025{\natexlab{a}})\citenamefont {Zhou}, \citenamefont {Guo},
  \citenamefont {Cai}, \citenamefont {Sun}, \citenamefont {Li}, \citenamefont
  {Zhao}, \citenamefont {Wang}, \citenamefont {Han}, \citenamefont {Chen},
  \citenamefont {Chen}, \citenamefont {Wu}, \citenamefont {Ding}, \citenamefont
  {Xiang}, \citenamefont {Mao},\ and\ \citenamefont {Sun}}]{Zhou_Guo}%
  \BibitemOpen
  \bibfield  {author} {\bibinfo {author} {\bibfnamefont {Y.}~\bibnamefont
  {Zhou}}, \bibinfo {author} {\bibfnamefont {J.}~\bibnamefont {Guo}}, \bibinfo
  {author} {\bibfnamefont {S.}~\bibnamefont {Cai}}, \bibinfo {author}
  {\bibfnamefont {H.}~\bibnamefont {Sun}}, \bibinfo {author} {\bibfnamefont
  {C.}~\bibnamefont {Li}}, \bibinfo {author} {\bibfnamefont {J.}~\bibnamefont
  {Zhao}}, \bibinfo {author} {\bibfnamefont {P.}~\bibnamefont {Wang}}, \bibinfo
  {author} {\bibfnamefont {J.}~\bibnamefont {Han}}, \bibinfo {author}
  {\bibfnamefont {X.}~\bibnamefont {Chen}}, \bibinfo {author} {\bibfnamefont
  {Y.}~\bibnamefont {Chen}}, \bibinfo {author} {\bibfnamefont {Q.}~\bibnamefont
  {Wu}}, \bibinfo {author} {\bibfnamefont {Y.}~\bibnamefont {Ding}}, \bibinfo
  {author} {\bibfnamefont {T.}~\bibnamefont {Xiang}}, \bibinfo {author}
  {\bibfnamefont {H.-k.}\ \bibnamefont {Mao}},\ and\ \bibinfo {author}
  {\bibfnamefont {L.}~\bibnamefont {Sun}},\ }\bibfield  {title} {\bibinfo
  {title} {Investigations of key issues on the reproducibility of high-tc
  superconductivity emerging from compressed \ce{La3Ni2O7}},\ }\href
  {https://doi.org/10.1063/5.0247684} {\bibfield  {journal} {\bibinfo
  {journal} {Matter Radiat. Extremes}\ }\textbf {\bibinfo {volume} {10}},\
  \bibinfo {pages} {027801} (\bibinfo {year} {2025}{\natexlab{a}})}\BibitemShut
  {NoStop}%
\bibitem [{\citenamefont {Qu}\ \emph {et~al.}(2023)\citenamefont {Qu},
  \citenamefont {Qu}, \citenamefont {Yi}, \citenamefont {Li},\ and\
  \citenamefont {Su}}]{Qu_Qu_56}%
  \BibitemOpen
  \bibfield  {author} {\bibinfo {author} {\bibfnamefont {X.-Z.}\ \bibnamefont
  {Qu}}, \bibinfo {author} {\bibfnamefont {D.-W.}\ \bibnamefont {Qu}}, \bibinfo
  {author} {\bibfnamefont {X.-W.}\ \bibnamefont {Yi}}, \bibinfo {author}
  {\bibfnamefont {W.}~\bibnamefont {Li}},\ and\ \bibinfo {author}
  {\bibfnamefont {G.}~\bibnamefont {Su}},\ }\href@noop {} {\bibinfo {title}
  {Roles of hund's rule and hybridization in the two-orbital model for
  high-{$T_c$} superconductivity in the bilayer nickelate}} (\bibinfo {year}
  {2023}),\ \Eprint {https://arxiv.org/abs/2311.12769} {arXiv:2311.12769}
  \BibitemShut {NoStop}%
\bibitem [{\citenamefont {Kakoi}\ \emph
  {et~al.}(2024{\natexlab{a}})\citenamefont {Kakoi}, \citenamefont {Kaneko},
  \citenamefont {Sakakibara}, \citenamefont {Ochi},\ and\ \citenamefont
  {Kuroki}}]{Kakoi_Kaneko}%
  \BibitemOpen
  \bibfield  {author} {\bibinfo {author} {\bibfnamefont {M.}~\bibnamefont
  {Kakoi}}, \bibinfo {author} {\bibfnamefont {T.}~\bibnamefont {Kaneko}},
  \bibinfo {author} {\bibfnamefont {H.}~\bibnamefont {Sakakibara}}, \bibinfo
  {author} {\bibfnamefont {M.}~\bibnamefont {Ochi}},\ and\ \bibinfo {author}
  {\bibfnamefont {K.}~\bibnamefont {Kuroki}},\ }\bibfield  {title} {\bibinfo
  {title} {Pair correlations of the hybridized orbitals in a ladder model for
  the bilayer nickelate {La}$_{3}${Ni}$_2${O}$_7$},\ }\href
  {https://doi.org/10.1103/PhysRevB.109.L201124} {\bibfield  {journal}
  {\bibinfo  {journal} {Phys. Rev. B}\ }\textbf {\bibinfo {volume} {109}},\
  \bibinfo {pages} {L201124} (\bibinfo {year}
  {2024}{\natexlab{a}})}\BibitemShut {NoStop}%
\bibitem [{\citenamefont {Kaneko}\ \emph {et~al.}(2025)\citenamefont {Kaneko},
  \citenamefont {Kakoi},\ and\ \citenamefont {Kuroki}}]{T.Kaneko_2025}%
  \BibitemOpen
  \bibfield  {author} {\bibinfo {author} {\bibfnamefont {T.}~\bibnamefont
  {Kaneko}}, \bibinfo {author} {\bibfnamefont {M.}~\bibnamefont {Kakoi}},\ and\
  \bibinfo {author} {\bibfnamefont {K.}~\bibnamefont {Kuroki}},\ }\bibfield
  {title} {\bibinfo {title} {$t$-$j$ model for strongly correlated two-orbital
  systems: Application to bilayer nickelate superconductors},\ }\href
  {https://link.aps.org/doi/10.1103/bsgt-sg2s} {\bibfield  {journal} {\bibinfo
  {journal} {Phys. Rev. B}\ }\textbf {\bibinfo {volume} {112}},\ \bibinfo
  {pages} {075143} (\bibinfo {year} {2025})}\BibitemShut {NoStop}%
\bibitem [{\citenamefont {Nakata}\ \emph {et~al.}(2017)\citenamefont {Nakata},
  \citenamefont {Ogura}, \citenamefont {Usui},\ and\ \citenamefont
  {Kuroki}}]{Nakata}%
  \BibitemOpen
  \bibfield  {author} {\bibinfo {author} {\bibfnamefont {M.}~\bibnamefont
  {Nakata}}, \bibinfo {author} {\bibfnamefont {D.}~\bibnamefont {Ogura}},
  \bibinfo {author} {\bibfnamefont {H.}~\bibnamefont {Usui}},\ and\ \bibinfo
  {author} {\bibfnamefont {K.}~\bibnamefont {Kuroki}},\ }\bibfield  {title}
  {\bibinfo {title} {Finite-energy spin fluctuations as a pairing glue in
  systems with coexisting electron and hole bands},\ }\href
  {https://doi.org/10.1103/PhysRevB.95.214509} {\bibfield  {journal} {\bibinfo
  {journal} {Phys. Rev. B}\ }\textbf {\bibinfo {volume} {95}},\ \bibinfo
  {pages} {214509} (\bibinfo {year} {2017})}\BibitemShut {NoStop}%
\bibitem [{\citenamefont {Dagotto}\ \emph {et~al.}(1992)\citenamefont
  {Dagotto}, \citenamefont {Riera},\ and\ \citenamefont
  {Scalapino}}]{bilayer0}%
  \BibitemOpen
  \bibfield  {author} {\bibinfo {author} {\bibfnamefont {E.}~\bibnamefont
  {Dagotto}}, \bibinfo {author} {\bibfnamefont {J.}~\bibnamefont {Riera}},\
  and\ \bibinfo {author} {\bibfnamefont {D.}~\bibnamefont {Scalapino}},\
  }\bibfield  {title} {\bibinfo {title} {Superconductivity in ladders and
  coupled planes},\ }\href {https://doi.org/10.1103/PhysRevB.45.5744}
  {\bibfield  {journal} {\bibinfo  {journal} {Phys. Rev. B}\ }\textbf {\bibinfo
  {volume} {45}},\ \bibinfo {pages} {5744} (\bibinfo {year}
  {1992})}\BibitemShut {NoStop}%
\bibitem [{\citenamefont {Bulut}\ \emph {et~al.}(1992)\citenamefont {Bulut},
  \citenamefont {Scalapino},\ and\ \citenamefont {Scalettar}}]{bilayer1}%
  \BibitemOpen
  \bibfield  {author} {\bibinfo {author} {\bibfnamefont {N.}~\bibnamefont
  {Bulut}}, \bibinfo {author} {\bibfnamefont {D.~J.}\ \bibnamefont
  {Scalapino}},\ and\ \bibinfo {author} {\bibfnamefont {R.~T.}\ \bibnamefont
  {Scalettar}},\ }\bibfield  {title} {\bibinfo {title} {Nodeless d-wave pairing
  in a two-layer hubbard model},\ }\href
  {https://doi.org/10.1103/PhysRevB.45.5577} {\bibfield  {journal} {\bibinfo
  {journal} {Phys. Rev. B}\ }\textbf {\bibinfo {volume} {45}},\ \bibinfo
  {pages} {5577} (\bibinfo {year} {1992})}\BibitemShut {NoStop}%
\bibitem [{\citenamefont {Scalettar}\ \emph {et~al.}(1994)\citenamefont
  {Scalettar}, \citenamefont {Cannon}, \citenamefont {Scalapino},\ and\
  \citenamefont {Sugar}}]{bilayer2}%
  \BibitemOpen
  \bibfield  {author} {\bibinfo {author} {\bibfnamefont {R.~T.}\ \bibnamefont
  {Scalettar}}, \bibinfo {author} {\bibfnamefont {J.~W.}\ \bibnamefont
  {Cannon}}, \bibinfo {author} {\bibfnamefont {D.~J.}\ \bibnamefont
  {Scalapino}},\ and\ \bibinfo {author} {\bibfnamefont {R.~L.}\ \bibnamefont
  {Sugar}},\ }\bibfield  {title} {\bibinfo {title} {Magnetic and pairing
  correlations in coupled hubbard planes},\ }\href
  {https://doi.org/10.1103/PhysRevB.50.13419} {\bibfield  {journal} {\bibinfo
  {journal} {Phys. Rev. B}\ }\textbf {\bibinfo {volume} {50}},\ \bibinfo
  {pages} {13419} (\bibinfo {year} {1994})}\BibitemShut {NoStop}%
\bibitem [{\citenamefont {Hetzel}\ \emph {et~al.}(1994)\citenamefont {Hetzel},
  \citenamefont {von~der Linden},\ and\ \citenamefont {Hanke}}]{bilayer3}%
  \BibitemOpen
  \bibfield  {author} {\bibinfo {author} {\bibfnamefont {R.~E.}\ \bibnamefont
  {Hetzel}}, \bibinfo {author} {\bibfnamefont {W.}~\bibnamefont {von~der
  Linden}},\ and\ \bibinfo {author} {\bibfnamefont {W.}~\bibnamefont {Hanke}},\
  }\bibfield  {title} {\bibinfo {title} {Pairing correlations in a two-layer
  hubbard model},\ }\href {https://doi.org/10.1103/PhysRevB.50.4159} {\bibfield
   {journal} {\bibinfo  {journal} {Phys. Rev. B}\ }\textbf {\bibinfo {volume}
  {50}},\ \bibinfo {pages} {4159} (\bibinfo {year} {1994})}\BibitemShut
  {NoStop}%
\bibitem [{\citenamefont {dos Santos}(1995)}]{bilayer4}%
  \BibitemOpen
  \bibfield  {author} {\bibinfo {author} {\bibfnamefont {R.~R.}\ \bibnamefont
  {dos Santos}},\ }\bibfield  {title} {\bibinfo {title} {Magnetism and pairing
  in hubbard bilayers},\ }\href {https://doi.org/10.1103/PhysRevB.51.15540}
  {\bibfield  {journal} {\bibinfo  {journal} {Phys. Rev. B}\ }\textbf {\bibinfo
  {volume} {51}},\ \bibinfo {pages} {15540} (\bibinfo {year}
  {1995})}\BibitemShut {NoStop}%
\bibitem [{\citenamefont {Liechtenstein}\ \emph {et~al.}(1995)\citenamefont
  {Liechtenstein}, \citenamefont {Mazin},\ and\ \citenamefont
  {Andersen}}]{bilayer5}%
  \BibitemOpen
  \bibfield  {author} {\bibinfo {author} {\bibfnamefont {A.~I.}\ \bibnamefont
  {Liechtenstein}}, \bibinfo {author} {\bibfnamefont {I.~I.}\ \bibnamefont
  {Mazin}},\ and\ \bibinfo {author} {\bibfnamefont {O.~K.}\ \bibnamefont
  {Andersen}},\ }\bibfield  {title} {\bibinfo {title} {$\mathit{s}$-wave
  superconductivity from an antiferromagnetic spin-fluctuation model for
  bilayer materials},\ }\href {https://doi.org/10.1103/PhysRevLett.74.2303}
  {\bibfield  {journal} {\bibinfo  {journal} {Phys. Rev. Lett.}\ }\textbf
  {\bibinfo {volume} {74}},\ \bibinfo {pages} {2303} (\bibinfo {year}
  {1995})}\BibitemShut {NoStop}%
\bibitem [{\citenamefont {Kuroki}\ \emph {et~al.}(2002)\citenamefont {Kuroki},
  \citenamefont {Kimura},\ and\ \citenamefont {Arita}}]{bilayer6}%
  \BibitemOpen
  \bibfield  {author} {\bibinfo {author} {\bibfnamefont {K.}~\bibnamefont
  {Kuroki}}, \bibinfo {author} {\bibfnamefont {T.}~\bibnamefont {Kimura}},\
  and\ \bibinfo {author} {\bibfnamefont {R.}~\bibnamefont {Arita}},\ }\bibfield
   {title} {\bibinfo {title} {High-temperature superconductivity in dimer array
  systems},\ }\href {https://doi.org/10.1103/PhysRevB.66.184508} {\bibfield
  {journal} {\bibinfo  {journal} {Phys. Rev. B}\ }\textbf {\bibinfo {volume}
  {66}},\ \bibinfo {pages} {184508} (\bibinfo {year} {2002})}\BibitemShut
  {NoStop}%
\bibitem [{\citenamefont {Kancharla}\ and\ \citenamefont
  {Okamoto}(2007)}]{bilayer7}%
  \BibitemOpen
  \bibfield  {author} {\bibinfo {author} {\bibfnamefont {S.~S.}\ \bibnamefont
  {Kancharla}}\ and\ \bibinfo {author} {\bibfnamefont {S.}~\bibnamefont
  {Okamoto}},\ }\bibfield  {title} {\bibinfo {title} {Band insulator to mott
  insulator transition in a bilayer hubbard model},\ }\href
  {https://doi.org/10.1103/PhysRevB.75.193103} {\bibfield  {journal} {\bibinfo
  {journal} {Phys. Rev. B}\ }\textbf {\bibinfo {volume} {75}},\ \bibinfo
  {pages} {193103} (\bibinfo {year} {2007})}\BibitemShut {NoStop}%
\bibitem [{\citenamefont {Bouadim}\ \emph {et~al.}(2008)\citenamefont
  {Bouadim}, \citenamefont {Batrouni}, \citenamefont {H\'{e}bert},\ and\
  \citenamefont {Scalettar}}]{bilayer8}%
  \BibitemOpen
  \bibfield  {author} {\bibinfo {author} {\bibfnamefont {K.}~\bibnamefont
  {Bouadim}}, \bibinfo {author} {\bibfnamefont {G.~G.}\ \bibnamefont
  {Batrouni}}, \bibinfo {author} {\bibfnamefont {F.}~\bibnamefont
  {H\'{e}bert}},\ and\ \bibinfo {author} {\bibfnamefont {R.~T.}\ \bibnamefont
  {Scalettar}},\ }\bibfield  {title} {\bibinfo {title} {Magnetic and transport
  properties of a coupled hubbard bilayer with electron and hole doping},\
  }\href {https://doi.org/10.1103/PhysRevB.77.144527} {\bibfield  {journal}
  {\bibinfo  {journal} {Phys. Rev. B}\ }\textbf {\bibinfo {volume} {77}},\
  \bibinfo {pages} {144527} (\bibinfo {year} {2008})}\BibitemShut {NoStop}%
\bibitem [{\citenamefont {Lanat\`a}\ \emph {et~al.}(2009)\citenamefont
  {Lanat\`a}, \citenamefont {Barone},\ and\ \citenamefont
  {Fabrizio}}]{bilayer9}%
  \BibitemOpen
  \bibfield  {author} {\bibinfo {author} {\bibfnamefont {N.}~\bibnamefont
  {Lanat\`a}}, \bibinfo {author} {\bibfnamefont {P.}~\bibnamefont {Barone}},\
  and\ \bibinfo {author} {\bibfnamefont {M.}~\bibnamefont {Fabrizio}},\
  }\bibfield  {title} {\bibinfo {title} {Superconductivity in the doped bilayer
  hubbard model},\ }\href {https://doi.org/10.1103/PhysRevB.80.224524}
  {\bibfield  {journal} {\bibinfo  {journal} {Phys. Rev. B}\ }\textbf {\bibinfo
  {volume} {80}},\ \bibinfo {pages} {224524} (\bibinfo {year}
  {2009})}\BibitemShut {NoStop}%
\bibitem [{\citenamefont {Zhai}\ \emph {et~al.}(2009)\citenamefont {Zhai},
  \citenamefont {Wang},\ and\ \citenamefont {Lee}}]{bilayer10}%
  \BibitemOpen
  \bibfield  {author} {\bibinfo {author} {\bibfnamefont {H.}~\bibnamefont
  {Zhai}}, \bibinfo {author} {\bibfnamefont {F.}~\bibnamefont {Wang}},\ and\
  \bibinfo {author} {\bibfnamefont {D.-H.}\ \bibnamefont {Lee}},\ }\bibfield
  {title} {\bibinfo {title} {Antiferromagnetically driven electronic
  correlations in iron pnictides and cuprates},\ }\href
  {https://doi.org/10.1103/PhysRevB.80.064517} {\bibfield  {journal} {\bibinfo
  {journal} {Phys. Rev. B}\ }\textbf {\bibinfo {volume} {80}},\ \bibinfo
  {pages} {064517} (\bibinfo {year} {2009})}\BibitemShut {NoStop}%
\bibitem [{\citenamefont {Maier}\ and\ \citenamefont
  {Scalapino}(2011)}]{bilayer11}%
  \BibitemOpen
  \bibfield  {author} {\bibinfo {author} {\bibfnamefont {T.~A.}\ \bibnamefont
  {Maier}}\ and\ \bibinfo {author} {\bibfnamefont {D.~J.}\ \bibnamefont
  {Scalapino}},\ }\bibfield  {title} {\bibinfo {title} {Pair structure and the
  pairing interaction in a bilayer hubbard model for unconventional
  superconductivity},\ }\href {https://doi.org/10.1103/PhysRevB.84.180513}
  {\bibfield  {journal} {\bibinfo  {journal} {Phys. Rev. B}\ }\textbf {\bibinfo
  {volume} {84}},\ \bibinfo {pages} {180513} (\bibinfo {year}
  {2011})}\BibitemShut {NoStop}%
\bibitem [{\citenamefont {Ouyang}\ \emph
  {et~al.}(2024{\natexlab{b}})\citenamefont {Ouyang}, \citenamefont {Gao},\
  and\ \citenamefont {Lu}}]{Ouyang_Gao}%
  \BibitemOpen
  \bibfield  {author} {\bibinfo {author} {\bibfnamefont {Z.}~\bibnamefont
  {Ouyang}}, \bibinfo {author} {\bibfnamefont {M.}~\bibnamefont {Gao}},\ and\
  \bibinfo {author} {\bibfnamefont {Z.-Y.}\ \bibnamefont {Lu}},\ }\bibfield
  {title} {\bibinfo {title} {Absence of electron-phonon coupling
  superconductivity in the bilayer phase of {La}$_{3}${Ni}$_2${O}$_{7}$ under
  pressure},\ }\href {https://www.nature.com/articles/s41535-024-00689-5}
  {\bibfield  {journal} {\bibinfo  {journal} {npj Quantum Mater.}\ }\textbf
  {\bibinfo {volume} {9}},\ \bibinfo {pages} {80} (\bibinfo {year}
  {2024}{\natexlab{b}})}\BibitemShut {NoStop}%
\bibitem [{\citenamefont {Li}\ \emph {et~al.}(2025)\citenamefont {Li},
  \citenamefont {Cao}, \citenamefont {Liu}, \citenamefont {Peng}, \citenamefont
  {Lin}, \citenamefont {Pei}, \citenamefont {Zhang}, \citenamefont {Wu},
  \citenamefont {Du}, \citenamefont {Zhao}, \citenamefont {Zhai}, \citenamefont
  {Zhang}, \citenamefont {Zhao}, \citenamefont {Lin}, \citenamefont {Tan},
  \citenamefont {Qi}, \citenamefont {Li}, \citenamefont {Guo}, \citenamefont
  {Yang},\ and\ \citenamefont {Yang}}]{Li_Cao}%
  \BibitemOpen
  \bibfield  {author} {\bibinfo {author} {\bibfnamefont {Y.}~\bibnamefont
  {Li}}, \bibinfo {author} {\bibfnamefont {Y.}~\bibnamefont {Cao}}, \bibinfo
  {author} {\bibfnamefont {L.}~\bibnamefont {Liu}}, \bibinfo {author}
  {\bibfnamefont {P.}~\bibnamefont {Peng}}, \bibinfo {author} {\bibfnamefont
  {H.}~\bibnamefont {Lin}}, \bibinfo {author} {\bibfnamefont {C.}~\bibnamefont
  {Pei}}, \bibinfo {author} {\bibfnamefont {M.}~\bibnamefont {Zhang}}, \bibinfo
  {author} {\bibfnamefont {H.}~\bibnamefont {Wu}}, \bibinfo {author}
  {\bibfnamefont {X.}~\bibnamefont {Du}}, \bibinfo {author} {\bibfnamefont
  {W.}~\bibnamefont {Zhao}}, \bibinfo {author} {\bibfnamefont {K.}~\bibnamefont
  {Zhai}}, \bibinfo {author} {\bibfnamefont {X.}~\bibnamefont {Zhang}},
  \bibinfo {author} {\bibfnamefont {J.}~\bibnamefont {Zhao}}, \bibinfo {author}
  {\bibfnamefont {M.}~\bibnamefont {Lin}}, \bibinfo {author} {\bibfnamefont
  {P.}~\bibnamefont {Tan}}, \bibinfo {author} {\bibfnamefont {Y.}~\bibnamefont
  {Qi}}, \bibinfo {author} {\bibfnamefont {G.}~\bibnamefont {Li}}, \bibinfo
  {author} {\bibfnamefont {H.}~\bibnamefont {Guo}}, \bibinfo {author}
  {\bibfnamefont {L.}~\bibnamefont {Yang}},\ and\ \bibinfo {author}
  {\bibfnamefont {L.}~\bibnamefont {Yang}},\ }\bibfield  {title} {\bibinfo
  {title} {Distinct ultrafast dynamics of bilayer and trilayer nickelate
  superconductors regarding the density-wave-like transitions},\ }\href
  {https://www.sciencedirect.com/science/article/pii/S2095927324007503}
  {\bibfield  {journal} {\bibinfo  {journal} {Science Bulletin}\ }\textbf
  {\bibinfo {volume} {70}},\ \bibinfo {pages} {180} (\bibinfo {year}
  {2025})}\BibitemShut {NoStop}%
\bibitem [{\citenamefont {Yi}\ \emph {et~al.}(2024)\citenamefont {Yi},
  \citenamefont {Meng}, \citenamefont {Li}, \citenamefont {Liao}, \citenamefont
  {Li}, \citenamefont {You}, \citenamefont {Gu},\ and\ \citenamefont
  {Su}}]{Yi_Meng}%
  \BibitemOpen
  \bibfield  {author} {\bibinfo {author} {\bibfnamefont {X.-W.}\ \bibnamefont
  {Yi}}, \bibinfo {author} {\bibfnamefont {Y.}~\bibnamefont {Meng}}, \bibinfo
  {author} {\bibfnamefont {J.-W.}\ \bibnamefont {Li}}, \bibinfo {author}
  {\bibfnamefont {Z.-W.}\ \bibnamefont {Liao}}, \bibinfo {author}
  {\bibfnamefont {W.}~\bibnamefont {Li}}, \bibinfo {author} {\bibfnamefont
  {J.-Y.}\ \bibnamefont {You}}, \bibinfo {author} {\bibfnamefont
  {B.}~\bibnamefont {Gu}},\ and\ \bibinfo {author} {\bibfnamefont
  {G.}~\bibnamefont {Su}},\ }\bibfield  {title} {\bibinfo {title} {Nature of
  charge density waves and metal-insulator transition in pressurized
  {La}$_{3}${Ni}$_2${O}$_{7}$},\ }\href
  {https://doi.org/10.1103/PhysRevB.110.L140508} {\bibfield  {journal}
  {\bibinfo  {journal} {Phys. Rev. B}\ }\textbf {\bibinfo {volume} {110}},\
  \bibinfo {pages} {L140508} (\bibinfo {year} {2024})}\BibitemShut {NoStop}%
\bibitem [{\citenamefont {Geisler}\ \emph {et~al.}(2024)\citenamefont
  {Geisler}, \citenamefont {Hamlin}, \citenamefont {Stewart}, \citenamefont
  {Hennig},\ and\ \citenamefont {Hirschfeld}}]{Geisler_Hamlin}%
  \BibitemOpen
  \bibfield  {author} {\bibinfo {author} {\bibfnamefont {B.}~\bibnamefont
  {Geisler}}, \bibinfo {author} {\bibfnamefont {J.~J.}\ \bibnamefont {Hamlin}},
  \bibinfo {author} {\bibfnamefont {G.~R.}\ \bibnamefont {Stewart}}, \bibinfo
  {author} {\bibfnamefont {R.~G.}\ \bibnamefont {Hennig}},\ and\ \bibinfo
  {author} {\bibfnamefont {P.~J.}\ \bibnamefont {Hirschfeld}},\ }\bibfield
  {title} {\bibinfo {title} {Structural transitions, octahedral rotations, and
  electronic properties of ${A}_{3}${Ni}$_{2}${O}$_{7}$ rare-earth nickelates
  under high pressure},\ }\href {https://doi.org/10.1038/s41535-024-00648-0}
  {\bibfield  {journal} {\bibinfo  {journal} {npj Quantum Mater.}\ }\textbf
  {\bibinfo {volume} {9}},\ \bibinfo {pages} {38} (\bibinfo {year}
  {2024})}\BibitemShut {NoStop}%
\bibitem [{\citenamefont {Wang}\ \emph
  {et~al.}(2024{\natexlab{b}})\citenamefont {Wang}, \citenamefont {Li},
  \citenamefont {Xie}, \citenamefont {Liu}, \citenamefont {Sun}, \citenamefont
  {Huang}, \citenamefont {Gao}, \citenamefont {Nakagawa}, \citenamefont {Fu},
  \citenamefont {Dong}, \citenamefont {Cao}, \citenamefont {Yu}, \citenamefont
  {Kawaguchi}, \citenamefont {Kadobayashi}, \citenamefont {Wang}, \citenamefont
  {Jin}, \citenamefont {Mao},\ and\ \citenamefont {Liu}}]{Wang_Li}%
  \BibitemOpen
  \bibfield  {author} {\bibinfo {author} {\bibfnamefont {L.}~\bibnamefont
  {Wang}}, \bibinfo {author} {\bibfnamefont {Y.}~\bibnamefont {Li}}, \bibinfo
  {author} {\bibfnamefont {S.-Y.}\ \bibnamefont {Xie}}, \bibinfo {author}
  {\bibfnamefont {F.}~\bibnamefont {Liu}}, \bibinfo {author} {\bibfnamefont
  {H.}~\bibnamefont {Sun}}, \bibinfo {author} {\bibfnamefont {C.}~\bibnamefont
  {Huang}}, \bibinfo {author} {\bibfnamefont {Y.}~\bibnamefont {Gao}}, \bibinfo
  {author} {\bibfnamefont {T.}~\bibnamefont {Nakagawa}}, \bibinfo {author}
  {\bibfnamefont {B.}~\bibnamefont {Fu}}, \bibinfo {author} {\bibfnamefont
  {B.}~\bibnamefont {Dong}}, \bibinfo {author} {\bibfnamefont {Z.}~\bibnamefont
  {Cao}}, \bibinfo {author} {\bibfnamefont {R.}~\bibnamefont {Yu}}, \bibinfo
  {author} {\bibfnamefont {S.~I.}\ \bibnamefont {Kawaguchi}}, \bibinfo {author}
  {\bibfnamefont {H.}~\bibnamefont {Kadobayashi}}, \bibinfo {author}
  {\bibfnamefont {M.}~\bibnamefont {Wang}}, \bibinfo {author} {\bibfnamefont
  {C.}~\bibnamefont {Jin}}, \bibinfo {author} {\bibfnamefont {H.-k.}\
  \bibnamefont {Mao}},\ and\ \bibinfo {author} {\bibfnamefont {H.}~\bibnamefont
  {Liu}},\ }\bibfield  {title} {\bibinfo {title} {Structure responsible for the
  superconducting state in {La}$_{3}${Ni}$_{2}${O}$_{7}$ at high-pressure and
  low-temperature conditions},\ }\href {https://doi.org/10.1021/jacs.3c13094}
  {\bibfield  {journal} {\bibinfo  {journal} {J. Am. Chem. Soc.}\ }\textbf
  {\bibinfo {volume} {146}},\ \bibinfo {pages} {7506} (\bibinfo {year}
  {2024}{\natexlab{b}})}\BibitemShut {NoStop}%
\bibitem [{\citenamefont {Wang}\ \emph
  {et~al.}(2024{\natexlab{c}})\citenamefont {Wang}, \citenamefont {Wang},
  \citenamefont {Shen}, \citenamefont {Hou}, \citenamefont {Luo}, \citenamefont
  {Ma}, \citenamefont {Yang}, \citenamefont {Shi}, \citenamefont {Dou},
  \citenamefont {Feng}, \citenamefont {Yang}, \citenamefont {Shi},
  \citenamefont {Ren}, \citenamefont {Ma}, \citenamefont {Yang}, \citenamefont
  {Liu}, \citenamefont {Liu}, \citenamefont {Zhang}, \citenamefont {Dong},
  \citenamefont {Wang}, \citenamefont {Jiang}, \citenamefont {Hu},
  \citenamefont {Nagasaki}, \citenamefont {Kitagawa}, \citenamefont {Calder},
  \citenamefont {Yan}, \citenamefont {Sun}, \citenamefont {Wang}, \citenamefont
  {Zhou}, \citenamefont {Uwatoko},\ and\ \citenamefont {Cheng}}]{Wang_Wang_50}%
  \BibitemOpen
  \bibfield  {author} {\bibinfo {author} {\bibfnamefont {N.}~\bibnamefont
  {Wang}}, \bibinfo {author} {\bibfnamefont {G.}~\bibnamefont {Wang}}, \bibinfo
  {author} {\bibfnamefont {X.}~\bibnamefont {Shen}}, \bibinfo {author}
  {\bibfnamefont {J.}~\bibnamefont {Hou}}, \bibinfo {author} {\bibfnamefont
  {J.}~\bibnamefont {Luo}}, \bibinfo {author} {\bibfnamefont {X.}~\bibnamefont
  {Ma}}, \bibinfo {author} {\bibfnamefont {H.}~\bibnamefont {Yang}}, \bibinfo
  {author} {\bibfnamefont {L.}~\bibnamefont {Shi}}, \bibinfo {author}
  {\bibfnamefont {J.}~\bibnamefont {Dou}}, \bibinfo {author} {\bibfnamefont
  {J.}~\bibnamefont {Feng}}, \bibinfo {author} {\bibfnamefont {J.}~\bibnamefont
  {Yang}}, \bibinfo {author} {\bibfnamefont {Y.}~\bibnamefont {Shi}}, \bibinfo
  {author} {\bibfnamefont {Z.}~\bibnamefont {Ren}}, \bibinfo {author}
  {\bibfnamefont {H.}~\bibnamefont {Ma}}, \bibinfo {author} {\bibfnamefont
  {P.}~\bibnamefont {Yang}}, \bibinfo {author} {\bibfnamefont {Z.}~\bibnamefont
  {Liu}}, \bibinfo {author} {\bibfnamefont {Y.}~\bibnamefont {Liu}}, \bibinfo
  {author} {\bibfnamefont {H.}~\bibnamefont {Zhang}}, \bibinfo {author}
  {\bibfnamefont {X.}~\bibnamefont {Dong}}, \bibinfo {author} {\bibfnamefont
  {Y.}~\bibnamefont {Wang}}, \bibinfo {author} {\bibfnamefont {K.}~\bibnamefont
  {Jiang}}, \bibinfo {author} {\bibfnamefont {J.}~\bibnamefont {Hu}}, \bibinfo
  {author} {\bibfnamefont {S.}~\bibnamefont {Nagasaki}}, \bibinfo {author}
  {\bibfnamefont {K.}~\bibnamefont {Kitagawa}}, \bibinfo {author}
  {\bibfnamefont {S.}~\bibnamefont {Calder}}, \bibinfo {author} {\bibfnamefont
  {J.}~\bibnamefont {Yan}}, \bibinfo {author} {\bibfnamefont {J.}~\bibnamefont
  {Sun}}, \bibinfo {author} {\bibfnamefont {B.}~\bibnamefont {Wang}}, \bibinfo
  {author} {\bibfnamefont {R.}~\bibnamefont {Zhou}}, \bibinfo {author}
  {\bibfnamefont {Y.}~\bibnamefont {Uwatoko}},\ and\ \bibinfo {author}
  {\bibfnamefont {J.}~\bibnamefont {Cheng}},\ }\bibfield  {title} {\bibinfo
  {title} {Bulk high-temperature superconductivity in pressurized tetragonal
  \ce{La2PrNi2O7}},\ }\href {https://doi.org/10.1038/s41586-024-07996-8}
  {\bibfield  {journal} {\bibinfo  {journal} {Nature}\ }\textbf {\bibinfo
  {volume} {634}},\ \bibinfo {pages} {579} (\bibinfo {year}
  {2024}{\natexlab{c}})}\BibitemShut {NoStop}%
\bibitem [{\citenamefont {Sakakibara}\ \emph
  {et~al.}(2024{\natexlab{b}})\citenamefont {Sakakibara}, \citenamefont {Ochi},
  \citenamefont {Nagata}, \citenamefont {Ueki}, \citenamefont {Sakurai},
  \citenamefont {Matsumoto}, \citenamefont {Terashima}, \citenamefont {Hirose},
  \citenamefont {Ohta}, \citenamefont {Kato}, \citenamefont {Takano},\ and\
  \citenamefont {Kuroki}}]{Sakakibara_Ochi}%
  \BibitemOpen
  \bibfield  {author} {\bibinfo {author} {\bibfnamefont {H.}~\bibnamefont
  {Sakakibara}}, \bibinfo {author} {\bibfnamefont {M.}~\bibnamefont {Ochi}},
  \bibinfo {author} {\bibfnamefont {H.}~\bibnamefont {Nagata}}, \bibinfo
  {author} {\bibfnamefont {Y.}~\bibnamefont {Ueki}}, \bibinfo {author}
  {\bibfnamefont {H.}~\bibnamefont {Sakurai}}, \bibinfo {author} {\bibfnamefont
  {R.}~\bibnamefont {Matsumoto}}, \bibinfo {author} {\bibfnamefont
  {K.}~\bibnamefont {Terashima}}, \bibinfo {author} {\bibfnamefont
  {K.}~\bibnamefont {Hirose}}, \bibinfo {author} {\bibfnamefont
  {H.}~\bibnamefont {Ohta}}, \bibinfo {author} {\bibfnamefont {M.}~\bibnamefont
  {Kato}}, \bibinfo {author} {\bibfnamefont {Y.}~\bibnamefont {Takano}},\ and\
  \bibinfo {author} {\bibfnamefont {K.}~\bibnamefont {Kuroki}},\ }\bibfield
  {title} {\bibinfo {title} {Theoretical analysis on the possibility of
  superconductivity in the trilayer {R}uddlesden-{P}opper nickelate
  {La}$_{4}${Ni}$_{3}${O}$_{10}$ under pressure and its experimental
  examination: Comparison with {La}$_{3}${Ni}$_{2}${O}$_{7}$},\ }\href
  {https://doi.org/10.1103/PhysRevB.109.144511} {\bibfield  {journal} {\bibinfo
   {journal} {Phys. Rev. B}\ }\textbf {\bibinfo {volume} {109}},\ \bibinfo
  {pages} {144511} (\bibinfo {year} {2024}{\natexlab{b}})}\BibitemShut
  {NoStop}%
\bibitem [{\citenamefont {Li}\ \emph {et~al.}(2024{\natexlab{a}})\citenamefont
  {Li}, \citenamefont {Zhang}, \citenamefont {Xiang}, \citenamefont {Zhang},
  \citenamefont {Zhu},\ and\ \citenamefont {Wen}}]{Li_Zhang}%
  \BibitemOpen
  \bibfield  {author} {\bibinfo {author} {\bibfnamefont {Q.}~\bibnamefont
  {Li}}, \bibinfo {author} {\bibfnamefont {Y.-J.}\ \bibnamefont {Zhang}},
  \bibinfo {author} {\bibfnamefont {Z.-N.}\ \bibnamefont {Xiang}}, \bibinfo
  {author} {\bibfnamefont {Y.}~\bibnamefont {Zhang}}, \bibinfo {author}
  {\bibfnamefont {X.}~\bibnamefont {Zhu}},\ and\ \bibinfo {author}
  {\bibfnamefont {H.-H.}\ \bibnamefont {Wen}},\ }\bibfield  {title} {\bibinfo
  {title} {Signature of superconductivity in pressurized
  {La}$_{4}${Ni}$_{3}${O}$_{10}$},\ }\href
  {https://doi.org/10.1088/0256-307X/41/1/017401} {\bibfield  {journal}
  {\bibinfo  {journal} {Chin. Phys. Lett.}\ }\textbf {\bibinfo {volume} {41}},\
  \bibinfo {pages} {017401} (\bibinfo {year} {2024}{\natexlab{a}})}\BibitemShut
  {NoStop}%
\bibitem [{\citenamefont {Zhu}\ \emph {et~al.}(2024)\citenamefont {Zhu},
  \citenamefont {Peng}, \citenamefont {Zhang}, \citenamefont {Pan},
  \citenamefont {Chen}, \citenamefont {Chen}, \citenamefont {Ren},
  \citenamefont {Liu}, \citenamefont {Hao}, \citenamefont {Li}, \citenamefont
  {Xing}, \citenamefont {Lan}, \citenamefont {Han}, \citenamefont {Wang},
  \citenamefont {Jia}, \citenamefont {Wo}, \citenamefont {Gu}, \citenamefont
  {Gu}, \citenamefont {Ji}, \citenamefont {Wang}, \citenamefont {Gou},
  \citenamefont {Shen}, \citenamefont {Ying}, \citenamefont {Chen},
  \citenamefont {Yang}, \citenamefont {Cao}, \citenamefont {Zheng},
  \citenamefont {Zeng}, \citenamefont {Guo},\ and\ \citenamefont
  {Zhao}}]{Zhu_Peng}%
  \BibitemOpen
  \bibfield  {author} {\bibinfo {author} {\bibfnamefont {Y.}~\bibnamefont
  {Zhu}}, \bibinfo {author} {\bibfnamefont {D.}~\bibnamefont {Peng}}, \bibinfo
  {author} {\bibfnamefont {E.}~\bibnamefont {Zhang}}, \bibinfo {author}
  {\bibfnamefont {B.}~\bibnamefont {Pan}}, \bibinfo {author} {\bibfnamefont
  {X.}~\bibnamefont {Chen}}, \bibinfo {author} {\bibfnamefont {L.}~\bibnamefont
  {Chen}}, \bibinfo {author} {\bibfnamefont {H.}~\bibnamefont {Ren}}, \bibinfo
  {author} {\bibfnamefont {F.}~\bibnamefont {Liu}}, \bibinfo {author}
  {\bibfnamefont {Y.}~\bibnamefont {Hao}}, \bibinfo {author} {\bibfnamefont
  {N.}~\bibnamefont {Li}}, \bibinfo {author} {\bibfnamefont {Z.}~\bibnamefont
  {Xing}}, \bibinfo {author} {\bibfnamefont {F.}~\bibnamefont {Lan}}, \bibinfo
  {author} {\bibfnamefont {J.}~\bibnamefont {Han}}, \bibinfo {author}
  {\bibfnamefont {J.}~\bibnamefont {Wang}}, \bibinfo {author} {\bibfnamefont
  {D.}~\bibnamefont {Jia}}, \bibinfo {author} {\bibfnamefont {H.}~\bibnamefont
  {Wo}}, \bibinfo {author} {\bibfnamefont {Y.}~\bibnamefont {Gu}}, \bibinfo
  {author} {\bibfnamefont {Y.}~\bibnamefont {Gu}}, \bibinfo {author}
  {\bibfnamefont {L.}~\bibnamefont {Ji}}, \bibinfo {author} {\bibfnamefont
  {W.}~\bibnamefont {Wang}}, \bibinfo {author} {\bibfnamefont {H.}~\bibnamefont
  {Gou}}, \bibinfo {author} {\bibfnamefont {Y.}~\bibnamefont {Shen}}, \bibinfo
  {author} {\bibfnamefont {T.}~\bibnamefont {Ying}}, \bibinfo {author}
  {\bibfnamefont {X.}~\bibnamefont {Chen}}, \bibinfo {author} {\bibfnamefont
  {W.}~\bibnamefont {Yang}}, \bibinfo {author} {\bibfnamefont {H.}~\bibnamefont
  {Cao}}, \bibinfo {author} {\bibfnamefont {C.}~\bibnamefont {Zheng}}, \bibinfo
  {author} {\bibfnamefont {Q.}~\bibnamefont {Zeng}}, \bibinfo {author}
  {\bibfnamefont {J.-G.}\ \bibnamefont {Guo}},\ and\ \bibinfo {author}
  {\bibfnamefont {J.}~\bibnamefont {Zhao}},\ }\bibfield  {title} {\bibinfo
  {title} {Superconductivity in pressurized trilayer
  {La}$_{4}${Ni}$_{3}${O}$_{10-\delta}$ single crystals},\ }\href
  {https://doi.org/10.1038/s41586-024-07553-3} {\bibfield  {journal} {\bibinfo
  {journal} {Nature}\ }\textbf {\bibinfo {volume} {631}},\ \bibinfo {pages}
  {531} (\bibinfo {year} {2024})}\BibitemShut {NoStop}%
\bibitem [{\citenamefont {Zhang}\ \emph {et~al.}(2025)\citenamefont {Zhang},
  \citenamefont {Pei}, \citenamefont {Peng}, \citenamefont {Du}, \citenamefont
  {Hu}, \citenamefont {Cao}, \citenamefont {Wang}, \citenamefont {Wu},
  \citenamefont {Li}, \citenamefont {Liu}, \citenamefont {Wen}, \citenamefont
  {Song}, \citenamefont {Zhao}, \citenamefont {Li}, \citenamefont {Cao},
  \citenamefont {Zhu}, \citenamefont {Zhang}, \citenamefont {Yu}, \citenamefont
  {Cheng}, \citenamefont {Zhang}, \citenamefont {Li}, \citenamefont {Zhao},
  \citenamefont {Chen}, \citenamefont {Jin}, \citenamefont {Guo}, \citenamefont
  {Wu}, \citenamefont {Yang}, \citenamefont {Zeng}, \citenamefont {Yan},
  \citenamefont {Yang},\ and\ \citenamefont {Qi}}]{Zhang_Pei_49}%
  \BibitemOpen
  \bibfield  {author} {\bibinfo {author} {\bibfnamefont {M.}~\bibnamefont
  {Zhang}}, \bibinfo {author} {\bibfnamefont {C.}~\bibnamefont {Pei}}, \bibinfo
  {author} {\bibfnamefont {D.}~\bibnamefont {Peng}}, \bibinfo {author}
  {\bibfnamefont {X.}~\bibnamefont {Du}}, \bibinfo {author} {\bibfnamefont
  {W.}~\bibnamefont {Hu}}, \bibinfo {author} {\bibfnamefont {Y.}~\bibnamefont
  {Cao}}, \bibinfo {author} {\bibfnamefont {Q.}~\bibnamefont {Wang}}, \bibinfo
  {author} {\bibfnamefont {J.}~\bibnamefont {Wu}}, \bibinfo {author}
  {\bibfnamefont {Y.}~\bibnamefont {Li}}, \bibinfo {author} {\bibfnamefont
  {H.}~\bibnamefont {Liu}}, \bibinfo {author} {\bibfnamefont {C.}~\bibnamefont
  {Wen}}, \bibinfo {author} {\bibfnamefont {J.}~\bibnamefont {Song}}, \bibinfo
  {author} {\bibfnamefont {Y.}~\bibnamefont {Zhao}}, \bibinfo {author}
  {\bibfnamefont {C.}~\bibnamefont {Li}}, \bibinfo {author} {\bibfnamefont
  {W.}~\bibnamefont {Cao}}, \bibinfo {author} {\bibfnamefont {S.}~\bibnamefont
  {Zhu}}, \bibinfo {author} {\bibfnamefont {Q.}~\bibnamefont {Zhang}}, \bibinfo
  {author} {\bibfnamefont {N.}~\bibnamefont {Yu}}, \bibinfo {author}
  {\bibfnamefont {P.}~\bibnamefont {Cheng}}, \bibinfo {author} {\bibfnamefont
  {L.}~\bibnamefont {Zhang}}, \bibinfo {author} {\bibfnamefont
  {Z.}~\bibnamefont {Li}}, \bibinfo {author} {\bibfnamefont {J.}~\bibnamefont
  {Zhao}}, \bibinfo {author} {\bibfnamefont {Y.}~\bibnamefont {Chen}}, \bibinfo
  {author} {\bibfnamefont {C.}~\bibnamefont {Jin}}, \bibinfo {author}
  {\bibfnamefont {H.}~\bibnamefont {Guo}}, \bibinfo {author} {\bibfnamefont
  {C.}~\bibnamefont {Wu}}, \bibinfo {author} {\bibfnamefont {F.}~\bibnamefont
  {Yang}}, \bibinfo {author} {\bibfnamefont {Q.}~\bibnamefont {Zeng}}, \bibinfo
  {author} {\bibfnamefont {S.}~\bibnamefont {Yan}}, \bibinfo {author}
  {\bibfnamefont {L.}~\bibnamefont {Yang}},\ and\ \bibinfo {author}
  {\bibfnamefont {Y.}~\bibnamefont {Qi}},\ }\bibfield  {title} {\bibinfo
  {title} {Superconductivity in trilayer nickelate \ce{La4Ni3O10} under
  pressure},\ }\href {https://link.aps.org/doi/10.1103/PhysRevX.15.021005}
  {\bibfield  {journal} {\bibinfo  {journal} {Phys. Rev. X}\ }\textbf {\bibinfo
  {volume} {15}},\ \bibinfo {pages} {021005} (\bibinfo {year}
  {2025})}\BibitemShut {NoStop}%
\bibitem [{\citenamefont {Li}\ \emph {et~al.}(2024{\natexlab{b}})\citenamefont
  {Li}, \citenamefont {Chen}, \citenamefont {Huang}, \citenamefont {Han},
  \citenamefont {Huo}, \citenamefont {Huang}, \citenamefont {Ma}, \citenamefont
  {Qiu}, \citenamefont {Chen}, \citenamefont {Hu}, \citenamefont {Chen},
  \citenamefont {Xie}, \citenamefont {Shen}, \citenamefont {Sun}, \citenamefont
  {Yao},\ and\ \citenamefont {Wang}}]{Li_Chen}%
  \BibitemOpen
  \bibfield  {author} {\bibinfo {author} {\bibfnamefont {J.}~\bibnamefont
  {Li}}, \bibinfo {author} {\bibfnamefont {C.-Q.}\ \bibnamefont {Chen}},
  \bibinfo {author} {\bibfnamefont {C.}~\bibnamefont {Huang}}, \bibinfo
  {author} {\bibfnamefont {Y.}~\bibnamefont {Han}}, \bibinfo {author}
  {\bibfnamefont {M.}~\bibnamefont {Huo}}, \bibinfo {author} {\bibfnamefont
  {X.}~\bibnamefont {Huang}}, \bibinfo {author} {\bibfnamefont
  {P.}~\bibnamefont {Ma}}, \bibinfo {author} {\bibfnamefont {Z.}~\bibnamefont
  {Qiu}}, \bibinfo {author} {\bibfnamefont {J.}~\bibnamefont {Chen}}, \bibinfo
  {author} {\bibfnamefont {X.}~\bibnamefont {Hu}}, \bibinfo {author}
  {\bibfnamefont {L.}~\bibnamefont {Chen}}, \bibinfo {author} {\bibfnamefont
  {T.}~\bibnamefont {Xie}}, \bibinfo {author} {\bibfnamefont {B.}~\bibnamefont
  {Shen}}, \bibinfo {author} {\bibfnamefont {H.}~\bibnamefont {Sun}}, \bibinfo
  {author} {\bibfnamefont {D.-X.}\ \bibnamefont {Yao}},\ and\ \bibinfo {author}
  {\bibfnamefont {M.}~\bibnamefont {Wang}},\ }\bibfield  {title} {\bibinfo
  {title} {Structural transition, electric transport, and electronic structures
  in the compressed trilayer nickelate {La}$_{4}${Ni}$_{3}${O}$_{10}$},\ }\href
  {https://doi.org/10.1007/s11433-023-2329-x} {\bibfield  {journal} {\bibinfo
  {journal} {Sci. China Phys. Mech. Astron.}\ }\textbf {\bibinfo {volume}
  {67}},\ \bibinfo {pages} {117403} (\bibinfo {year}
  {2024}{\natexlab{b}})}\BibitemShut {NoStop}%
\bibitem [{\citenamefont {Rhodes}\ and\ \citenamefont
  {Wahl}(2024)}]{Rhodes_Wahl}%
  \BibitemOpen
  \bibfield  {author} {\bibinfo {author} {\bibfnamefont {L.~C.}\ \bibnamefont
  {Rhodes}}\ and\ \bibinfo {author} {\bibfnamefont {P.}~\bibnamefont {Wahl}},\
  }\bibfield  {title} {\bibinfo {title} {Structural routes to stabilize
  superconducting {La}$_{3}${Ni}$_{2}${O}$_{7}$ at ambient pressure},\ }\href
  {https://doi.org/10.1103/PhysRevMaterials.8.044801} {\bibfield  {journal}
  {\bibinfo  {journal} {Phys. Rev. Mater.}\ }\textbf {\bibinfo {volume} {8}},\
  \bibinfo {pages} {044801} (\bibinfo {year} {2024})}\BibitemShut {NoStop}%
\bibitem [{\citenamefont {Wu}\ \emph {et~al.}(2024)\citenamefont {Wu},
  \citenamefont {Yang}, \citenamefont {Ma}, \citenamefont {Dai}, \citenamefont
  {Shi}, \citenamefont {Yuan}, \citenamefont {Lin},\ and\ \citenamefont
  {Cao}}]{Wu_Yang}%
  \BibitemOpen
  \bibfield  {author} {\bibinfo {author} {\bibfnamefont {S.}~\bibnamefont
  {Wu}}, \bibinfo {author} {\bibfnamefont {Z.}~\bibnamefont {Yang}}, \bibinfo
  {author} {\bibfnamefont {X.}~\bibnamefont {Ma}}, \bibinfo {author}
  {\bibfnamefont {J.}~\bibnamefont {Dai}}, \bibinfo {author} {\bibfnamefont
  {M.}~\bibnamefont {Shi}}, \bibinfo {author} {\bibfnamefont {H.-Q.}\
  \bibnamefont {Yuan}}, \bibinfo {author} {\bibfnamefont {H.-Q.}\ \bibnamefont
  {Lin}},\ and\ \bibinfo {author} {\bibfnamefont {C.}~\bibnamefont {Cao}},\
  }\href@noop {} {\bibinfo {title} {\ce{Ac3Ni2O7} and \ce{La2}{$Ae$}\ce{Ni2O6F}
  ({$Ae$} = \ce{Sr}, \ce{Ba}): Benchmark materials for bilayer nickelate
  superconductivity}} (\bibinfo {year} {2024}),\ \Eprint
  {https://arxiv.org/abs/2403.11713} {arXiv:2403.11713} \BibitemShut {NoStop}%
\bibitem [{\citenamefont {Ochi}\ \emph {et~al.}(2025)\citenamefont {Ochi},
  \citenamefont {Sakakibara}, \citenamefont {Usui},\ and\ \citenamefont
  {Kuroki}}]{3252}%
  \BibitemOpen
  \bibfield  {author} {\bibinfo {author} {\bibfnamefont {M.}~\bibnamefont
  {Ochi}}, \bibinfo {author} {\bibfnamefont {H.}~\bibnamefont {Sakakibara}},
  \bibinfo {author} {\bibfnamefont {H.}~\bibnamefont {Usui}},\ and\ \bibinfo
  {author} {\bibfnamefont {K.}~\bibnamefont {Kuroki}},\ }\bibfield  {title}
  {\bibinfo {title} {Theoretical study on the crystal structure of a bilayer
  nickel-oxychloride \ce{Sr3Ni2O5Cl2} and analysis on the occurrence of
  possible unconventional superconductivity},\ }\href
  {https://journals.aps.org/prb/abstract/10.1103/PhysRevB.111.064511}
  {\bibfield  {journal} {\bibinfo  {journal} {Phys. Rev. B}\ }\textbf {\bibinfo
  {volume} {111}},\ \bibinfo {pages} {064511} (\bibinfo {year}
  {2025})}\BibitemShut {NoStop}%
\bibitem [{\citenamefont {Yamane}\ \emph {et~al.}(2025)\citenamefont {Yamane},
  \citenamefont {Matsushita}, \citenamefont {Adachi}, \citenamefont
  {Matsumoto}, \citenamefont {Terashima}, \citenamefont {Hiroto}, \citenamefont
  {Sakurai},\ and\ \citenamefont {Takano}}]{3252_yamane}%
  \BibitemOpen
  \bibfield  {author} {\bibinfo {author} {\bibfnamefont {K.}~\bibnamefont
  {Yamane}}, \bibinfo {author} {\bibfnamefont {Y.}~\bibnamefont {Matsushita}},
  \bibinfo {author} {\bibfnamefont {S.}~\bibnamefont {Adachi}}, \bibinfo
  {author} {\bibfnamefont {R.}~\bibnamefont {Matsumoto}}, \bibinfo {author}
  {\bibfnamefont {K.}~\bibnamefont {Terashima}}, \bibinfo {author}
  {\bibfnamefont {T.}~\bibnamefont {Hiroto}}, \bibinfo {author} {\bibfnamefont
  {H.}~\bibnamefont {Sakurai}},\ and\ \bibinfo {author} {\bibfnamefont
  {Y.}~\bibnamefont {Takano}},\ }\bibfield  {title} {\bibinfo {title}
  {High-pressure synthesis of bilayer nickelate \ce{Sr3Ni2O5Cl2} with
  tetragonal crystal structure},\ }\href
  {https://doi.org/10.1107/S2053229625002281} {\bibfield  {journal} {\bibinfo
  {journal} {Acta Cryst.}\ }\textbf {\bibinfo {volume} {C81}},\ \bibinfo
  {pages} {259} (\bibinfo {year} {2025})}\BibitemShut {NoStop}%
\bibitem [{\citenamefont {Ko}\ \emph {et~al.}(2025)\citenamefont {Ko},
  \citenamefont {Yu}, \citenamefont {Liu}, \citenamefont {Bhatt}, \citenamefont
  {Li}, \citenamefont {Thampy}, \citenamefont {Kuo}, \citenamefont {Wang},
  \citenamefont {Lee}, \citenamefont {Lee}, \citenamefont {Lee}, \citenamefont
  {Goodge}, \citenamefont {Muller},\ and\ \citenamefont {Hwang}}]{E.Ko_2024}%
  \BibitemOpen
  \bibfield  {author} {\bibinfo {author} {\bibfnamefont {E.~K.}\ \bibnamefont
  {Ko}}, \bibinfo {author} {\bibfnamefont {Y.}~\bibnamefont {Yu}}, \bibinfo
  {author} {\bibfnamefont {Y.}~\bibnamefont {Liu}}, \bibinfo {author}
  {\bibfnamefont {L.}~\bibnamefont {Bhatt}}, \bibinfo {author} {\bibfnamefont
  {J.}~\bibnamefont {Li}}, \bibinfo {author} {\bibfnamefont {V.}~\bibnamefont
  {Thampy}}, \bibinfo {author} {\bibfnamefont {C.-T.}\ \bibnamefont {Kuo}},
  \bibinfo {author} {\bibfnamefont {B.~Y.}\ \bibnamefont {Wang}}, \bibinfo
  {author} {\bibfnamefont {Y.}~\bibnamefont {Lee}}, \bibinfo {author}
  {\bibfnamefont {K.}~\bibnamefont {Lee}}, \bibinfo {author} {\bibfnamefont
  {J.-S.}\ \bibnamefont {Lee}}, \bibinfo {author} {\bibfnamefont {B.~H.}\
  \bibnamefont {Goodge}}, \bibinfo {author} {\bibfnamefont {D.~A.}\
  \bibnamefont {Muller}},\ and\ \bibinfo {author} {\bibfnamefont {H.~Y.}\
  \bibnamefont {Hwang}},\ }\bibfield  {title} {\bibinfo {title} {Signatures of
  ambient pressure superconductivity in thin film {L}a$_3${N}i$_2${O}$_7$},\
  }\href {https://doi.org/10.1038/s41586-024-08525-3} {\bibfield  {journal}
  {\bibinfo  {journal} {Nature}\ }\textbf {\bibinfo {volume} {638}},\ \bibinfo
  {pages} {935} (\bibinfo {year} {2025})}\BibitemShut {NoStop}%
\bibitem [{\citenamefont {Zhou}\ \emph
  {et~al.}(2025{\natexlab{b}})\citenamefont {Zhou}, \citenamefont {Lv},
  \citenamefont {Wang}, \citenamefont {Nie}, \citenamefont {Chen},
  \citenamefont {Li}, \citenamefont {Huang}, \citenamefont {Chen},
  \citenamefont {Sun}, \citenamefont {Xue},\ and\ \citenamefont
  {Chen}}]{G.Zhou_2024}%
  \BibitemOpen
  \bibfield  {author} {\bibinfo {author} {\bibfnamefont {G.}~\bibnamefont
  {Zhou}}, \bibinfo {author} {\bibfnamefont {W.}~\bibnamefont {Lv}}, \bibinfo
  {author} {\bibfnamefont {H.}~\bibnamefont {Wang}}, \bibinfo {author}
  {\bibfnamefont {Z.}~\bibnamefont {Nie}}, \bibinfo {author} {\bibfnamefont
  {Y.}~\bibnamefont {Chen}}, \bibinfo {author} {\bibfnamefont {Y.}~\bibnamefont
  {Li}}, \bibinfo {author} {\bibfnamefont {H.}~\bibnamefont {Huang}}, \bibinfo
  {author} {\bibfnamefont {W.}~\bibnamefont {Chen}}, \bibinfo {author}
  {\bibfnamefont {Y.}~\bibnamefont {Sun}}, \bibinfo {author} {\bibfnamefont
  {Q.-K.}\ \bibnamefont {Xue}},\ and\ \bibinfo {author} {\bibfnamefont
  {Z.}~\bibnamefont {Chen}},\ }\bibfield  {title} {\bibinfo {title}
  {Ambient-pressure superconductivity onset above 40~{K} in \ce{(La,Pr)3Ni2O7}
  films},\ }\href {https://doi.org/10.1038/s41586-025-08755-z} {\bibfield
  {journal} {\bibinfo  {journal} {Nature}\ }\textbf {\bibinfo {volume} {640}},\
  \bibinfo {pages} {641} (\bibinfo {year} {2025}{\natexlab{b}})}\BibitemShut
  {NoStop}%
\bibitem [{\citenamefont {Liu}\ \emph {et~al.}(2025)\citenamefont {Liu},
  \citenamefont {Ko}, \citenamefont {Tarn}, \citenamefont {Bhatt},
  \citenamefont {Goodge}, \citenamefont {Muller}, \citenamefont {Raghu},
  \citenamefont {Yu},\ and\ \citenamefont {Hwang}}]{Y.Liu_2025}%
  \BibitemOpen
  \bibfield  {author} {\bibinfo {author} {\bibfnamefont {Y.}~\bibnamefont
  {Liu}}, \bibinfo {author} {\bibfnamefont {E.~K.}\ \bibnamefont {Ko}},
  \bibinfo {author} {\bibfnamefont {Y.}~\bibnamefont {Tarn}}, \bibinfo {author}
  {\bibfnamefont {L.}~\bibnamefont {Bhatt}}, \bibinfo {author} {\bibfnamefont
  {B.~H.}\ \bibnamefont {Goodge}}, \bibinfo {author} {\bibfnamefont {D.~A.}\
  \bibnamefont {Muller}}, \bibinfo {author} {\bibfnamefont {S.}~\bibnamefont
  {Raghu}}, \bibinfo {author} {\bibfnamefont {Y.}~\bibnamefont {Yu}},\ and\
  \bibinfo {author} {\bibfnamefont {H.~Y.}\ \bibnamefont {Hwang}},\ }\bibfield
  {title} {\bibinfo {title} {Superconductivity and normal-state transport in
  compressively strained \ce{La2PrNi2O7} thin films},\ }\href
  {https://doi.org/10.1038/s41563-025-02258-y} {\bibfield  {journal} {\bibinfo
  {journal} {Nat. Mater.}\ }\textbf {\bibinfo {volume} {24}},\ \bibinfo {pages}
  {1221} (\bibinfo {year} {2025})}\BibitemShut {NoStop}%
\bibitem [{\citenamefont {Zhao}\ and\ \citenamefont
  {Botana}(2025)}]{Y.Zhao_2024}%
  \BibitemOpen
  \bibfield  {author} {\bibinfo {author} {\bibfnamefont {Y.-F.}\ \bibnamefont
  {Zhao}}\ and\ \bibinfo {author} {\bibfnamefont {A.~S.}\ \bibnamefont
  {Botana}},\ }\bibfield  {title} {\bibinfo {title} {Electronic structure of
  ruddlesden-popper nickelates: Strain to mimic the effects of pressure},\
  }\href {https://link.aps.org/doi/10.1103/PhysRevB.111.115154} {\bibfield
  {journal} {\bibinfo  {journal} {Phys. Rev. B}\ }\textbf {\bibinfo {volume}
  {111}},\ \bibinfo {pages} {115154} (\bibinfo {year} {2025})}\BibitemShut
  {NoStop}%
\bibitem [{\citenamefont {Yue}\ \emph {et~al.}(2025)\citenamefont {Yue},
  \citenamefont {Miao}, \citenamefont {Huang}, \citenamefont {Hua},
  \citenamefont {Li}, \citenamefont {Li}, \citenamefont {Zhou}, \citenamefont
  {Lv}, \citenamefont {Yang}, \citenamefont {Sun}, \citenamefont {Sun},
  \citenamefont {Lin}, \citenamefont {Xue}, \citenamefont {Chen},\ and\
  \citenamefont {Chen}}]{C.Yue_2025}%
  \BibitemOpen
  \bibfield  {author} {\bibinfo {author} {\bibfnamefont {C.}~\bibnamefont
  {Yue}}, \bibinfo {author} {\bibfnamefont {J.-J.}\ \bibnamefont {Miao}},
  \bibinfo {author} {\bibfnamefont {H.}~\bibnamefont {Huang}}, \bibinfo
  {author} {\bibfnamefont {Y.}~\bibnamefont {Hua}}, \bibinfo {author}
  {\bibfnamefont {P.}~\bibnamefont {Li}}, \bibinfo {author} {\bibfnamefont
  {Y.}~\bibnamefont {Li}}, \bibinfo {author} {\bibfnamefont {G.}~\bibnamefont
  {Zhou}}, \bibinfo {author} {\bibfnamefont {W.}~\bibnamefont {Lv}}, \bibinfo
  {author} {\bibfnamefont {Q.}~\bibnamefont {Yang}}, \bibinfo {author}
  {\bibfnamefont {H.}~\bibnamefont {Sun}}, \bibinfo {author} {\bibfnamefont
  {Y.-J.}\ \bibnamefont {Sun}}, \bibinfo {author} {\bibfnamefont
  {J.}~\bibnamefont {Lin}}, \bibinfo {author} {\bibfnamefont {Q.-K.}\
  \bibnamefont {Xue}}, \bibinfo {author} {\bibfnamefont {Z.}~\bibnamefont
  {Chen}},\ and\ \bibinfo {author} {\bibfnamefont {W.-Q.}\ \bibnamefont
  {Chen}},\ }\bibfield  {title} {\bibinfo {title} {Correlated electronic
  structures and unconventional superconductivity in bilayer nickelate
  heterostructures},\ }\href {https://doi.org/10.1093/nsr/nwaf253} {\bibfield
  {journal} {\bibinfo  {journal} {National Science Review}\ ,\ \bibinfo {pages}
  {nwaf253}} (\bibinfo {year} {2025})}\BibitemShut {NoStop}%
\bibitem [{\citenamefont {Bhatt}\ \emph {et~al.}(2025)\citenamefont {Bhatt},
  \citenamefont {Jiang}, \citenamefont {Ko}, \citenamefont {Schnitzer},
  \citenamefont {Pan}, \citenamefont {Segedin}, \citenamefont {Liu},
  \citenamefont {Yu}, \citenamefont {Zhao}, \citenamefont {Morales},
  \citenamefont {Brooks}, \citenamefont {Botana}, \citenamefont {Hwang},
  \citenamefont {Mundy}, \citenamefont {Muller},\ and\ \citenamefont
  {Goodge}}]{L.Bhatt_2025}%
  \BibitemOpen
  \bibfield  {author} {\bibinfo {author} {\bibfnamefont {L.}~\bibnamefont
  {Bhatt}}, \bibinfo {author} {\bibfnamefont {A.~Y.}\ \bibnamefont {Jiang}},
  \bibinfo {author} {\bibfnamefont {E.~K.}\ \bibnamefont {Ko}}, \bibinfo
  {author} {\bibfnamefont {N.}~\bibnamefont {Schnitzer}}, \bibinfo {author}
  {\bibfnamefont {G.~A.}\ \bibnamefont {Pan}}, \bibinfo {author} {\bibfnamefont
  {D.~F.}\ \bibnamefont {Segedin}}, \bibinfo {author} {\bibfnamefont
  {Y.}~\bibnamefont {Liu}}, \bibinfo {author} {\bibfnamefont {Y.}~\bibnamefont
  {Yu}}, \bibinfo {author} {\bibfnamefont {Y.-F.}\ \bibnamefont {Zhao}},
  \bibinfo {author} {\bibfnamefont {E.~A.}\ \bibnamefont {Morales}}, \bibinfo
  {author} {\bibfnamefont {C.~M.}\ \bibnamefont {Brooks}}, \bibinfo {author}
  {\bibfnamefont {A.~S.}\ \bibnamefont {Botana}}, \bibinfo {author}
  {\bibfnamefont {H.~Y.}\ \bibnamefont {Hwang}}, \bibinfo {author}
  {\bibfnamefont {J.~A.}\ \bibnamefont {Mundy}}, \bibinfo {author}
  {\bibfnamefont {D.~A.}\ \bibnamefont {Muller}},\ and\ \bibinfo {author}
  {\bibfnamefont {B.~H.}\ \bibnamefont {Goodge}},\ }\href@noop {} {\bibinfo
  {title} {Resolving structural origins for superconductivity in
  strain-engineered \ce{La3Ni2O7} thin films}} (\bibinfo {year} {2025}),\
  \Eprint {https://arxiv.org/abs/2501.08204} {arXiv:2501.08204} \BibitemShut
  {NoStop}%
\bibitem [{\citenamefont {Geisler}\ \emph {et~al.}(2025)\citenamefont
  {Geisler}, \citenamefont {Hamlin}, \citenamefont {Stewart}, \citenamefont
  {Hennig},\ and\ \citenamefont {Hirschfeld}}]{B.Geisler_2025}%
  \BibitemOpen
  \bibfield  {author} {\bibinfo {author} {\bibfnamefont {B.}~\bibnamefont
  {Geisler}}, \bibinfo {author} {\bibfnamefont {J.~J.}\ \bibnamefont {Hamlin}},
  \bibinfo {author} {\bibfnamefont {G.~R.}\ \bibnamefont {Stewart}}, \bibinfo
  {author} {\bibfnamefont {R.~G.}\ \bibnamefont {Hennig}},\ and\ \bibinfo
  {author} {\bibfnamefont {P.~J.}\ \bibnamefont {Hirschfeld}},\ }\href@noop {}
  {\bibinfo {title} {Electronic reconstruction and interface engineering of
  emergent spin fluctuations in compressively strained \ce{La3Ni2O7} on
  \ce{SrLaAlO4}(001)}} (\bibinfo {year} {2025}),\ \Eprint
  {https://arxiv.org/abs/2503.10902} {arXiv:2503.10902} \BibitemShut {NoStop}%
\bibitem [{\citenamefont {Yi}\ \emph {et~al.}(2025)\citenamefont {Yi},
  \citenamefont {Li}, \citenamefont {You}, \citenamefont {Gu},\ and\
  \citenamefont {Su}}]{X.Yi_2025}%
  \BibitemOpen
  \bibfield  {author} {\bibinfo {author} {\bibfnamefont {X.-W.}\ \bibnamefont
  {Yi}}, \bibinfo {author} {\bibfnamefont {W.}~\bibnamefont {Li}}, \bibinfo
  {author} {\bibfnamefont {J.-Y.}\ \bibnamefont {You}}, \bibinfo {author}
  {\bibfnamefont {B.}~\bibnamefont {Gu}},\ and\ \bibinfo {author}
  {\bibfnamefont {G.}~\bibnamefont {Su}},\ }\href@noop {} {\bibinfo {title}
  {Unifying strain-driven and pressure-driven superconductivity in
  \ce{La3Ni2O7}: Suppressed charge/spin density waves and enhanced interlayer
  coupling}} (\bibinfo {year} {2025}),\ \Eprint
  {https://arxiv.org/abs/2505.12733} {arXiv:2505.12733} \BibitemShut {NoStop}%
\bibitem [{\citenamefont {Ushio}\ \emph {et~al.}(2025)\citenamefont {Ushio},
  \citenamefont {Kamiyama}, \citenamefont {Hoshi}, \citenamefont {Mizuno},
  \citenamefont {Ochi}, \citenamefont {Kuroki},\ and\ \citenamefont
  {Sakakibara}}]{K.Ushio_2025}%
  \BibitemOpen
  \bibfield  {author} {\bibinfo {author} {\bibfnamefont {K.}~\bibnamefont
  {Ushio}}, \bibinfo {author} {\bibfnamefont {S.}~\bibnamefont {Kamiyama}},
  \bibinfo {author} {\bibfnamefont {Y.}~\bibnamefont {Hoshi}}, \bibinfo
  {author} {\bibfnamefont {R.}~\bibnamefont {Mizuno}}, \bibinfo {author}
  {\bibfnamefont {M.}~\bibnamefont {Ochi}}, \bibinfo {author} {\bibfnamefont
  {K.}~\bibnamefont {Kuroki}},\ and\ \bibinfo {author} {\bibfnamefont
  {H.}~\bibnamefont {Sakakibara}},\ }\href@noop {} {\bibinfo {title}
  {Theoretical study on ambient pressure superconductivity in \ce{La3Ni2O7}
  thin films: structural analysis, model construction, and robustness of
  $s\pm$-wave pairing}} (\bibinfo {year} {2025}),\ \Eprint
  {https://arxiv.org/abs/2506.20497} {arXiv:2506.20497} \BibitemShut {NoStop}%
\bibitem [{\citenamefont {F{\"o}rst}\ \emph {et~al.}(2011)\citenamefont
  {F{\"o}rst}, \citenamefont {Manzoni}, \citenamefont {Kaiser}, \citenamefont
  {Tomioka}, \citenamefont {Tokura}, \citenamefont {Merlin},\ and\
  \citenamefont {Cavalleri}}]{M.Forst_2011}%
  \BibitemOpen
  \bibfield  {author} {\bibinfo {author} {\bibfnamefont {M.}~\bibnamefont
  {F{\"o}rst}}, \bibinfo {author} {\bibfnamefont {C.}~\bibnamefont {Manzoni}},
  \bibinfo {author} {\bibfnamefont {S.}~\bibnamefont {Kaiser}}, \bibinfo
  {author} {\bibfnamefont {Y.}~\bibnamefont {Tomioka}}, \bibinfo {author}
  {\bibfnamefont {Y.}~\bibnamefont {Tokura}}, \bibinfo {author} {\bibfnamefont
  {R.}~\bibnamefont {Merlin}},\ and\ \bibinfo {author} {\bibfnamefont
  {A.}~\bibnamefont {Cavalleri}},\ }\bibfield  {title} {\bibinfo {title}
  {Nonlinear phononics as an ultrafast route to lattice control},\ }\href
  {https://doi.org/10.1038/nphys2055} {\bibfield  {journal} {\bibinfo
  {journal} {Nat. Phys.}\ }\textbf {\bibinfo {volume} {7}},\ \bibinfo {pages}
  {854} (\bibinfo {year} {2011})}\BibitemShut {NoStop}%
\bibitem [{\citenamefont {Subedi}\ \emph {et~al.}(2014)\citenamefont {Subedi},
  \citenamefont {Cavalleri},\ and\ \citenamefont {Georges}}]{S.Alaska_2014}%
  \BibitemOpen
  \bibfield  {author} {\bibinfo {author} {\bibfnamefont {A.}~\bibnamefont
  {Subedi}}, \bibinfo {author} {\bibfnamefont {A.}~\bibnamefont {Cavalleri}},\
  and\ \bibinfo {author} {\bibfnamefont {A.}~\bibnamefont {Georges}},\
  }\bibfield  {title} {\bibinfo {title} {Theory of nonlinear phononics for
  coherent light control of solids},\ }\href
  {https://doi.org/10.1103/PhysRevB.89.220301} {\bibfield  {journal} {\bibinfo
  {journal} {Phys. Rev. B}\ }\textbf {\bibinfo {volume} {89}},\ \bibinfo
  {pages} {220301} (\bibinfo {year} {2014})}\BibitemShut {NoStop}%
\bibitem [{\citenamefont {Mankowsky}\ \emph {et~al.}(2014)\citenamefont
  {Mankowsky}, \citenamefont {Subedi}, \citenamefont {F{\"o}rst}, \citenamefont
  {Mariager}, \citenamefont {Chollet}, \citenamefont {Lemke}, \citenamefont
  {Robinson}, \citenamefont {Glownia}, \citenamefont {Minitti}, \citenamefont
  {Frano}, \citenamefont {Fechner}, \citenamefont {Spaldin}, \citenamefont
  {Loew}, \citenamefont {Keimer}, \citenamefont {Georges},\ and\ \citenamefont
  {Cavalleri}}]{R.Mankowsky_2014}%
  \BibitemOpen
  \bibfield  {author} {\bibinfo {author} {\bibfnamefont {R.}~\bibnamefont
  {Mankowsky}}, \bibinfo {author} {\bibfnamefont {A.}~\bibnamefont {Subedi}},
  \bibinfo {author} {\bibfnamefont {M.}~\bibnamefont {F{\"o}rst}}, \bibinfo
  {author} {\bibfnamefont {S.~O.}\ \bibnamefont {Mariager}}, \bibinfo {author}
  {\bibfnamefont {M.}~\bibnamefont {Chollet}}, \bibinfo {author} {\bibfnamefont
  {H.~T.}\ \bibnamefont {Lemke}}, \bibinfo {author} {\bibfnamefont {J.~S.}\
  \bibnamefont {Robinson}}, \bibinfo {author} {\bibfnamefont {J.~M.}\
  \bibnamefont {Glownia}}, \bibinfo {author} {\bibfnamefont {M.~P.}\
  \bibnamefont {Minitti}}, \bibinfo {author} {\bibfnamefont {A.}~\bibnamefont
  {Frano}}, \bibinfo {author} {\bibfnamefont {M.}~\bibnamefont {Fechner}},
  \bibinfo {author} {\bibfnamefont {N.~A.}\ \bibnamefont {Spaldin}}, \bibinfo
  {author} {\bibfnamefont {T.}~\bibnamefont {Loew}}, \bibinfo {author}
  {\bibfnamefont {B.}~\bibnamefont {Keimer}}, \bibinfo {author} {\bibfnamefont
  {A.}~\bibnamefont {Georges}},\ and\ \bibinfo {author} {\bibfnamefont
  {A.}~\bibnamefont {Cavalleri}},\ }\bibfield  {title} {\bibinfo {title}
  {Nonlinear lattice dynamics as a basis for enhanced superconductivity in
  \ce{YBa2Cu3O_{6.5}}},\ }\href {https://doi.org/10.1038/nature13875}
  {\bibfield  {journal} {\bibinfo  {journal} {Nature}\ }\textbf {\bibinfo
  {volume} {516}},\ \bibinfo {pages} {71} (\bibinfo {year} {2014})}\BibitemShut
  {NoStop}%
\bibitem [{\citenamefont {Fechner}\ and\ \citenamefont
  {Spaldin}(2016)}]{M.Fechner_2016}%
  \BibitemOpen
  \bibfield  {author} {\bibinfo {author} {\bibfnamefont {M.}~\bibnamefont
  {Fechner}}\ and\ \bibinfo {author} {\bibfnamefont {N.~A.}\ \bibnamefont
  {Spaldin}},\ }\bibfield  {title} {\bibinfo {title} {Effects of intense
  optical phonon pumping on the structure and electronic properties of yttrium
  barium copper oxide},\ }\href {https://doi.org/10.1103/PhysRevB.94.134307}
  {\bibfield  {journal} {\bibinfo  {journal} {Phys. Rev. B}\ }\textbf {\bibinfo
  {volume} {94}},\ \bibinfo {pages} {134307} (\bibinfo {year}
  {2016})}\BibitemShut {NoStop}%
\bibitem [{\citenamefont {Subedi}(2015)}]{S.Alaska_2015}%
  \BibitemOpen
  \bibfield  {author} {\bibinfo {author} {\bibfnamefont {A.}~\bibnamefont
  {Subedi}},\ }\bibfield  {title} {\bibinfo {title} {Proposal for ultrafast
  switching of ferroelectrics using midinfrared pulses},\ }\href
  {https://doi.org/10.1103/PhysRevB.92.214303} {\bibfield  {journal} {\bibinfo
  {journal} {Phys. Rev. B}\ }\textbf {\bibinfo {volume} {92}},\ \bibinfo
  {pages} {214303} (\bibinfo {year} {2015})}\BibitemShut {NoStop}%
\bibitem [{\citenamefont {Mankowsky}\ \emph {et~al.}(2017)\citenamefont
  {Mankowsky}, \citenamefont {von Hoegen}, \citenamefont {F\"orst},\ and\
  \citenamefont {Cavalleri}}]{R.Mankowsky_2017}%
  \BibitemOpen
  \bibfield  {author} {\bibinfo {author} {\bibfnamefont {R.}~\bibnamefont
  {Mankowsky}}, \bibinfo {author} {\bibfnamefont {A.}~\bibnamefont {von
  Hoegen}}, \bibinfo {author} {\bibfnamefont {M.}~\bibnamefont {F\"orst}},\
  and\ \bibinfo {author} {\bibfnamefont {A.}~\bibnamefont {Cavalleri}},\
  }\bibfield  {title} {\bibinfo {title} {Ultrafast reversal of the
  ferroelectric polarization},\ }\href
  {https://doi.org/10.1103/PhysRevLett.118.197601} {\bibfield  {journal}
  {\bibinfo  {journal} {Phys. Rev. Lett.}\ }\textbf {\bibinfo {volume} {118}},\
  \bibinfo {pages} {197601} (\bibinfo {year} {2017})}\BibitemShut {NoStop}%
\bibitem [{\citenamefont {Fechner}\ \emph {et~al.}(2024)\citenamefont
  {Fechner}, \citenamefont {F{\"o}rst}, \citenamefont {Orenstein},
  \citenamefont {Krapivin}, \citenamefont {Disa}, \citenamefont {Buzzi},
  \citenamefont {von Hoegen}, \citenamefont {de~la Pena}, \citenamefont
  {Nguyen}, \citenamefont {Mankowsky}, \citenamefont {Sander}, \citenamefont
  {Lemke}, \citenamefont {Deng}, \citenamefont {Trigo},\ and\ \citenamefont
  {Cavalleri}}]{M.Fechner_2024}%
  \BibitemOpen
  \bibfield  {author} {\bibinfo {author} {\bibfnamefont {M.}~\bibnamefont
  {Fechner}}, \bibinfo {author} {\bibfnamefont {M.}~\bibnamefont {F{\"o}rst}},
  \bibinfo {author} {\bibfnamefont {G.}~\bibnamefont {Orenstein}}, \bibinfo
  {author} {\bibfnamefont {V.}~\bibnamefont {Krapivin}}, \bibinfo {author}
  {\bibfnamefont {A.~S.}\ \bibnamefont {Disa}}, \bibinfo {author}
  {\bibfnamefont {M.}~\bibnamefont {Buzzi}}, \bibinfo {author} {\bibfnamefont
  {A.}~\bibnamefont {von Hoegen}}, \bibinfo {author} {\bibfnamefont
  {G.}~\bibnamefont {de~la Pena}}, \bibinfo {author} {\bibfnamefont {Q.~L.}\
  \bibnamefont {Nguyen}}, \bibinfo {author} {\bibfnamefont {R.}~\bibnamefont
  {Mankowsky}}, \bibinfo {author} {\bibfnamefont {M.}~\bibnamefont {Sander}},
  \bibinfo {author} {\bibfnamefont {H.}~\bibnamefont {Lemke}}, \bibinfo
  {author} {\bibfnamefont {Y.}~\bibnamefont {Deng}}, \bibinfo {author}
  {\bibfnamefont {M.}~\bibnamefont {Trigo}},\ and\ \bibinfo {author}
  {\bibfnamefont {A.}~\bibnamefont {Cavalleri}},\ }\bibfield  {title} {\bibinfo
  {title} {Quenched lattice fluctuations in optically driven \ce{SrTiO3}},\
  }\href {https://doi.org/10.1038/s41563-023-01791-y} {\bibfield  {journal}
  {\bibinfo  {journal} {Nat. Mater.}\ }\textbf {\bibinfo {volume} {23}},\
  \bibinfo {pages} {363} (\bibinfo {year} {2024})}\BibitemShut {NoStop}%
\bibitem [{\citenamefont {Paiva}\ \emph {et~al.}(2025)\citenamefont {Paiva},
  \citenamefont {Fechner},\ and\ \citenamefont {Juraschek}}]{C.Paiva_2024}%
  \BibitemOpen
  \bibfield  {author} {\bibinfo {author} {\bibfnamefont {C.}~\bibnamefont
  {Paiva}}, \bibinfo {author} {\bibfnamefont {M.}~\bibnamefont {Fechner}},\
  and\ \bibinfo {author} {\bibfnamefont {D.~M.}\ \bibnamefont {Juraschek}},\
  }\bibfield  {title} {\bibinfo {title} {Dynamically induced multiferroic
  polarization},\ }\href
  {https://journals.aps.org/prl/abstract/10.1103/7lm1-wm3y} {\bibfield
  {journal} {\bibinfo  {journal} {Phys. Rev. Lett.}\ }\textbf {\bibinfo
  {volume} {135}},\ \bibinfo {pages} {066702} (\bibinfo {year}
  {2025})}\BibitemShut {NoStop}%
\bibitem [{\citenamefont {Juraschek}\ \emph {et~al.}(2017)\citenamefont
  {Juraschek}, \citenamefont {Fechner},\ and\ \citenamefont
  {Spaldin}}]{D.M.Juraschek_2017}%
  \BibitemOpen
  \bibfield  {author} {\bibinfo {author} {\bibfnamefont {D.~M.}\ \bibnamefont
  {Juraschek}}, \bibinfo {author} {\bibfnamefont {M.}~\bibnamefont {Fechner}},\
  and\ \bibinfo {author} {\bibfnamefont {N.~A.}\ \bibnamefont {Spaldin}},\
  }\bibfield  {title} {\bibinfo {title} {Ultrafast structure switching through
  nonlinear phononics},\ }\href
  {http://dx.doi.org/10.1103/PhysRevLett.118.054101} {\bibfield  {journal}
  {\bibinfo  {journal} {Phys. Rev. Lett.}\ }\textbf {\bibinfo {volume} {118}}
  (\bibinfo {year} {2017})}\BibitemShut {NoStop}%
\bibitem [{\citenamefont {Gu}\ and\ \citenamefont
  {Rondinelli}(2018{\natexlab{a}})}]{G.Mingqiang_2018}%
  \BibitemOpen
  \bibfield  {author} {\bibinfo {author} {\bibfnamefont {M.}~\bibnamefont
  {Gu}}\ and\ \bibinfo {author} {\bibfnamefont {J.~M.}\ \bibnamefont
  {Rondinelli}},\ }\bibfield  {title} {\bibinfo {title} {Nonlinear phononic
  control and emergent magnetism in mott insulating titanates},\ }\href
  {https://doi.org/10.1103/PhysRevB.98.024102} {\bibfield  {journal} {\bibinfo
  {journal} {Phys. Rev. B}\ }\textbf {\bibinfo {volume} {98}},\ \bibinfo
  {pages} {024102} (\bibinfo {year} {2018}{\natexlab{a}})}\BibitemShut
  {NoStop}%
\bibitem [{\citenamefont {Disa}\ \emph {et~al.}(2020)\citenamefont {Disa},
  \citenamefont {Fechner}, \citenamefont {Nova}, \citenamefont {Liu},
  \citenamefont {F{\"o}rst}, \citenamefont {Prabhakaran}, \citenamefont
  {Radaelli},\ and\ \citenamefont {Cavalleri}}]{A.S.Disa_2020}%
  \BibitemOpen
  \bibfield  {author} {\bibinfo {author} {\bibfnamefont {A.~S.}\ \bibnamefont
  {Disa}}, \bibinfo {author} {\bibfnamefont {M.}~\bibnamefont {Fechner}},
  \bibinfo {author} {\bibfnamefont {T.~F.}\ \bibnamefont {Nova}}, \bibinfo
  {author} {\bibfnamefont {B.}~\bibnamefont {Liu}}, \bibinfo {author}
  {\bibfnamefont {M.}~\bibnamefont {F{\"o}rst}}, \bibinfo {author}
  {\bibfnamefont {D.}~\bibnamefont {Prabhakaran}}, \bibinfo {author}
  {\bibfnamefont {P.~G.}\ \bibnamefont {Radaelli}},\ and\ \bibinfo {author}
  {\bibfnamefont {A.}~\bibnamefont {Cavalleri}},\ }\bibfield  {title} {\bibinfo
  {title} {Polarizing an antiferromagnet by optical engineering of the crystal
  field},\ }\href {https://doi.org/10.1038/s41567-020-0936-3} {\bibfield
  {journal} {\bibinfo  {journal} {Nat. Phys.}\ }\textbf {\bibinfo {volume}
  {16}},\ \bibinfo {pages} {937} (\bibinfo {year} {2020})}\BibitemShut
  {NoStop}%
\bibitem [{\citenamefont {Zeng}(2023)}]{Y.Zeng_2023}%
  \BibitemOpen
  \bibfield  {author} {\bibinfo {author} {\bibfnamefont {Y.}~\bibnamefont
  {Zeng}},\ }\bibfield  {title} {\bibinfo {title} {Nonlinear phononic coupling
  in antiferromagnetic monolayer \ce{MnPS3}},\ }\href
  {https://doi.org/10.1103/PhysRevB.108.054308} {\bibfield  {journal} {\bibinfo
   {journal} {Phys. Rev. B}\ }\textbf {\bibinfo {volume} {108}},\ \bibinfo
  {pages} {054308} (\bibinfo {year} {2023})}\BibitemShut {NoStop}%
\bibitem [{\citenamefont {Tang}\ \emph {et~al.}(2023)\citenamefont {Tang},
  \citenamefont {Boi},\ and\ \citenamefont {Cheng}}]{R.Tang_2023}%
  \BibitemOpen
  \bibfield  {author} {\bibinfo {author} {\bibfnamefont {R.}~\bibnamefont
  {Tang}}, \bibinfo {author} {\bibfnamefont {F.}~\bibnamefont {Boi}},\ and\
  \bibinfo {author} {\bibfnamefont {Y.-H.}\ \bibnamefont {Cheng}},\ }\bibfield
  {title} {\bibinfo {title} {Light-induced topological phase transition via
  nonlinear phononics in superconductor \ce{CsV3Sb5}},\ }\href
  {https://doi.org/10.1038/s41535-023-00609-z} {\bibfield  {journal} {\bibinfo
  {journal} {npj Quantum Mater.}\ }\textbf {\bibinfo {volume} {8}},\ \bibinfo
  {pages} {78} (\bibinfo {year} {2023})}\BibitemShut {NoStop}%
\bibitem [{\citenamefont {Blank}\ \emph {et~al.}(2023)\citenamefont {Blank},
  \citenamefont {Grishunin}, \citenamefont {Zvezdin}, \citenamefont {Hai},
  \citenamefont {Wu}, \citenamefont {Su}, \citenamefont {Huang}, \citenamefont
  {Zvezdin},\ and\ \citenamefont {Kimel}}]{T.G.Blank_2023}%
  \BibitemOpen
  \bibfield  {author} {\bibinfo {author} {\bibfnamefont {T.~G.~H.}\
  \bibnamefont {Blank}}, \bibinfo {author} {\bibfnamefont {K.~A.}\ \bibnamefont
  {Grishunin}}, \bibinfo {author} {\bibfnamefont {K.~A.}\ \bibnamefont
  {Zvezdin}}, \bibinfo {author} {\bibfnamefont {N.~T.}\ \bibnamefont {Hai}},
  \bibinfo {author} {\bibfnamefont {J.~C.}\ \bibnamefont {Wu}}, \bibinfo
  {author} {\bibfnamefont {S.-H.}\ \bibnamefont {Su}}, \bibinfo {author}
  {\bibfnamefont {J.-C.~A.}\ \bibnamefont {Huang}}, \bibinfo {author}
  {\bibfnamefont {A.~K.}\ \bibnamefont {Zvezdin}},\ and\ \bibinfo {author}
  {\bibfnamefont {A.~V.}\ \bibnamefont {Kimel}},\ }\bibfield  {title} {\bibinfo
  {title} {Two-dimensional terahertz spectroscopy of nonlinear phononics in the
  topological insulator \ce{MnBi2Te4}},\ }\href
  {https://doi.org/10.1103/PhysRevLett.131.026902} {\bibfield  {journal}
  {\bibinfo  {journal} {Phys. Rev. Lett.}\ }\textbf {\bibinfo {volume} {131}},\
  \bibinfo {pages} {026902} (\bibinfo {year} {2023})}\BibitemShut {NoStop}%
\bibitem [{\citenamefont {Juraschek}\ \emph {et~al.}(2021)\citenamefont
  {Juraschek}, \citenamefont {Neuman}, \citenamefont {Flick},\ and\
  \citenamefont {Narang}}]{phono_cavity}%
  \BibitemOpen
  \bibfield  {author} {\bibinfo {author} {\bibfnamefont {D.~M.}\ \bibnamefont
  {Juraschek}}, \bibinfo {author} {\bibfnamefont {T.}~\bibnamefont {Neuman}},
  \bibinfo {author} {\bibfnamefont {J.}~\bibnamefont {Flick}},\ and\ \bibinfo
  {author} {\bibfnamefont {P.}~\bibnamefont {Narang}},\ }\bibfield  {title}
  {\bibinfo {title} {Cavity control of nonlinear phononics},\ }\href
  {https://doi.org/10.1103/PhysRevResearch.3.L032046} {\bibfield  {journal}
  {\bibinfo  {journal} {Phys. Rev. Res.}\ }\textbf {\bibinfo {volume} {3}},\
  \bibinfo {pages} {L032046} (\bibinfo {year} {2021})}\BibitemShut {NoStop}%
\bibitem [{\citenamefont {Disa}\ \emph {et~al.}(2021)\citenamefont {Disa},
  \citenamefont {Nova},\ and\ \citenamefont {Cavalleri}}]{phono_review_disa}%
  \BibitemOpen
  \bibfield  {author} {\bibinfo {author} {\bibfnamefont {A.~S.}\ \bibnamefont
  {Disa}}, \bibinfo {author} {\bibfnamefont {T.~F.}\ \bibnamefont {Nova}},\
  and\ \bibinfo {author} {\bibfnamefont {A.}~\bibnamefont {Cavalleri}},\
  }\bibfield  {title} {\bibinfo {title} {Engineering crystal structures with
  light},\ }\href {https://doi.org/10.1038/s41567-021-01366-1} {\bibfield
  {journal} {\bibinfo  {journal} {Nat. Phys.}\ }\textbf {\bibinfo {volume}
  {17}},\ \bibinfo {pages} {1087} (\bibinfo {year} {2021})}\BibitemShut
  {NoStop}%
\bibitem [{\citenamefont {Subedi}(2021)}]{phono_review_comp}%
  \BibitemOpen
  \bibfield  {author} {\bibinfo {author} {\bibfnamefont {A.}~\bibnamefont
  {Subedi}},\ }\bibfield  {title} {\bibinfo {title} {Light-control of materials
  via nonlinear phononics},\ }\href {https://doi.org/10.5802/crphys.44}
  {\bibfield  {journal} {\bibinfo  {journal} {C. R. Phys.}\ }\textbf {\bibinfo
  {volume} {22}},\ \bibinfo {pages} {161} (\bibinfo {year} {2021})}\BibitemShut
  {NoStop}%
\bibitem [{\citenamefont {Perdew}\ \emph {et~al.}(2008)\citenamefont {Perdew},
  \citenamefont {Ruzsinszky}, \citenamefont {Csonka}, \citenamefont {Vydrov},
  \citenamefont {Scuseria}, \citenamefont {Constantin}, \citenamefont {Zhou},\
  and\ \citenamefont {Burke}}]{PBEsol_1}%
  \BibitemOpen
  \bibfield  {author} {\bibinfo {author} {\bibfnamefont {J.~P.}\ \bibnamefont
  {Perdew}}, \bibinfo {author} {\bibfnamefont {A.}~\bibnamefont {Ruzsinszky}},
  \bibinfo {author} {\bibfnamefont {G.~I.}\ \bibnamefont {Csonka}}, \bibinfo
  {author} {\bibfnamefont {O.~A.}\ \bibnamefont {Vydrov}}, \bibinfo {author}
  {\bibfnamefont {G.~E.}\ \bibnamefont {Scuseria}}, \bibinfo {author}
  {\bibfnamefont {L.~A.}\ \bibnamefont {Constantin}}, \bibinfo {author}
  {\bibfnamefont {X.}~\bibnamefont {Zhou}},\ and\ \bibinfo {author}
  {\bibfnamefont {K.}~\bibnamefont {Burke}},\ }\bibfield  {title} {\bibinfo
  {title} {Restoring the density-gradient expansion for exchange in solids and
  surfaces},\ }\href {https://doi.org/10.1103/PhysRevLett.100.136406}
  {\bibfield  {journal} {\bibinfo  {journal} {Phys. Rev. Lett.}\ }\textbf
  {\bibinfo {volume} {100}},\ \bibinfo {pages} {136406} (\bibinfo {year}
  {2008})}\BibitemShut {NoStop}%
\bibitem [{\citenamefont {Perdew}\ \emph {et~al.}(2009)\citenamefont {Perdew},
  \citenamefont {Ruzsinszky}, \citenamefont {Csonka}, \citenamefont {Vydrov},
  \citenamefont {Scuseria}, \citenamefont {Constantin}, \citenamefont {Zhou},\
  and\ \citenamefont {Burke}}]{PBEsol_2}%
  \BibitemOpen
  \bibfield  {author} {\bibinfo {author} {\bibfnamefont {J.~P.}\ \bibnamefont
  {Perdew}}, \bibinfo {author} {\bibfnamefont {A.}~\bibnamefont {Ruzsinszky}},
  \bibinfo {author} {\bibfnamefont {G.~I.}\ \bibnamefont {Csonka}}, \bibinfo
  {author} {\bibfnamefont {O.~A.}\ \bibnamefont {Vydrov}}, \bibinfo {author}
  {\bibfnamefont {G.~E.}\ \bibnamefont {Scuseria}}, \bibinfo {author}
  {\bibfnamefont {L.~A.}\ \bibnamefont {Constantin}}, \bibinfo {author}
  {\bibfnamefont {X.}~\bibnamefont {Zhou}},\ and\ \bibinfo {author}
  {\bibfnamefont {K.}~\bibnamefont {Burke}},\ }\bibfield  {title} {\bibinfo
  {title} {Erratum: Restoring the density-gradient expansion for exchange in
  solids and surfaces},\ }\href
  {https://doi.org/10.1103/PhysRevLett.102.039902} {\bibfield  {journal}
  {\bibinfo  {journal} {Phys. Rev. Lett.}\ }\textbf {\bibinfo {volume} {102}},\
  \bibinfo {pages} {039902} (\bibinfo {year} {2009})}\BibitemShut {NoStop}%
\bibitem [{\citenamefont {Kresse}\ and\ \citenamefont {Hafner}(1993)}]{vasp1}%
  \BibitemOpen
  \bibfield  {author} {\bibinfo {author} {\bibfnamefont {G.}~\bibnamefont
  {Kresse}}\ and\ \bibinfo {author} {\bibfnamefont {J.}~\bibnamefont
  {Hafner}},\ }\bibfield  {title} {\bibinfo {title} {Ab initio molecular
  dynamics for liquid metals},\ }\href
  {https://doi.org/10.1103/PhysRevB.47.558} {\bibfield  {journal} {\bibinfo
  {journal} {Phys. Rev. B}\ }\textbf {\bibinfo {volume} {47}},\ \bibinfo
  {pages} {558} (\bibinfo {year} {1993})}\BibitemShut {NoStop}%
\bibitem [{\citenamefont {Kresse}\ and\ \citenamefont {Hafner}(1994)}]{vasp2}%
  \BibitemOpen
  \bibfield  {author} {\bibinfo {author} {\bibfnamefont {G.}~\bibnamefont
  {Kresse}}\ and\ \bibinfo {author} {\bibfnamefont {J.}~\bibnamefont
  {Hafner}},\ }\bibfield  {title} {\bibinfo {title} {Ab initio
  molecular-dynamics simulation of the liquid-metal--amorphous-semiconductor
  transition in germanium},\ }\href {https://doi.org/10.1103/PhysRevB.49.14251}
  {\bibfield  {journal} {\bibinfo  {journal} {Phys. Rev. B}\ }\textbf {\bibinfo
  {volume} {49}},\ \bibinfo {pages} {14251} (\bibinfo {year}
  {1994})}\BibitemShut {NoStop}%
\bibitem [{\citenamefont {Kresse}\ and\ \citenamefont
  {Furthm{\"u}ller}(1996)}]{vasp3}%
  \BibitemOpen
  \bibfield  {author} {\bibinfo {author} {\bibfnamefont {G.}~\bibnamefont
  {Kresse}}\ and\ \bibinfo {author} {\bibfnamefont {J.}~\bibnamefont
  {Furthm{\"u}ller}},\ }\bibfield  {title} {\bibinfo {title} {Efficiency of
  ab-initio total energy calculations for metals and semiconductors using a
  plane-wave basis set},\ }\href
  {https://doi.org/https://doi.org/10.1016/0927-0256(96)00008-0} {\bibfield
  {journal} {\bibinfo  {journal} {Comput. Mater. Sci.}\ }\textbf {\bibinfo
  {volume} {6}},\ \bibinfo {pages} {15} (\bibinfo {year} {1996})}\BibitemShut
  {NoStop}%
\bibitem [{\citenamefont {Kresse}\ and\ \citenamefont
  {Furthm\"uller}(1996)}]{vasp}%
  \BibitemOpen
  \bibfield  {author} {\bibinfo {author} {\bibfnamefont {G.}~\bibnamefont
  {Kresse}}\ and\ \bibinfo {author} {\bibfnamefont {J.}~\bibnamefont
  {Furthm\"uller}},\ }\bibfield  {title} {\bibinfo {title} {Efficient iterative
  schemes for ab initio total-energy calculations using a plane-wave basis
  set},\ }\href {https://doi.org/10.1103/PhysRevB.54.11169} {\bibfield
  {journal} {\bibinfo  {journal} {Phys. Rev. B}\ }\textbf {\bibinfo {volume}
  {54}},\ \bibinfo {pages} {11169} (\bibinfo {year} {1996})}\BibitemShut
  {NoStop}%
\bibitem [{\citenamefont {Kresse}\ and\ \citenamefont {Joubert}(1999)}]{vasp4}%
  \BibitemOpen
  \bibfield  {author} {\bibinfo {author} {\bibfnamefont {G.}~\bibnamefont
  {Kresse}}\ and\ \bibinfo {author} {\bibfnamefont {D.}~\bibnamefont
  {Joubert}},\ }\bibfield  {title} {\bibinfo {title} {From ultrasoft
  pseudopotentials to the projector augmented-wave method},\ }\href
  {https://doi.org/10.1103/PhysRevB.59.1758} {\bibfield  {journal} {\bibinfo
  {journal} {Phys. Rev. B}\ }\textbf {\bibinfo {volume} {59}},\ \bibinfo
  {pages} {1758} (\bibinfo {year} {1999})}\BibitemShut {NoStop}%
\bibitem [{Note1()}]{Note1}%
  \BibitemOpen
  \bibinfo {note} {We confirm that the optimized interlayer Ni-O-Ni angle
  obtained by PBEsol, $170.4^{\circ }$, also agrees well with that obtained by
  PBE+$U$ ($U=4$ eV)~\cite {ref_PBE,ref_U}, $169.9^{\circ }$.}\BibitemShut
  {Stop}%
\bibitem [{\citenamefont {Togo}\ \emph {et~al.}(2023)\citenamefont {Togo},
  \citenamefont {Chaput}, \citenamefont {Tadano},\ and\ \citenamefont
  {Tanaka}}]{phonopy_1}%
  \BibitemOpen
  \bibfield  {author} {\bibinfo {author} {\bibfnamefont {A.}~\bibnamefont
  {Togo}}, \bibinfo {author} {\bibfnamefont {L.}~\bibnamefont {Chaput}},
  \bibinfo {author} {\bibfnamefont {T.}~\bibnamefont {Tadano}},\ and\ \bibinfo
  {author} {\bibfnamefont {I.}~\bibnamefont {Tanaka}},\ }\bibfield  {title}
  {\bibinfo {title} {Implementation strategies in phonopy and phono3py},\
  }\href {https://doi.org/10.1088/1361-648X/acd831} {\bibfield  {journal}
  {\bibinfo  {journal} {J. Phys.: Condens. Matter}\ }\textbf {\bibinfo {volume}
  {35}},\ \bibinfo {pages} {353001} (\bibinfo {year} {2023})}\BibitemShut
  {NoStop}%
\bibitem [{\citenamefont {Togo}(2023)}]{phonopy_2}%
  \BibitemOpen
  \bibfield  {author} {\bibinfo {author} {\bibfnamefont {A.}~\bibnamefont
  {Togo}},\ }\bibfield  {title} {\bibinfo {title} {First-principles phonon
  calculations with phonopy and phono3py},\ }\href
  {https://doi.org/10.7566/JPSJ.92.012001} {\bibfield  {journal} {\bibinfo
  {journal} {J. Phys. Soc. Jpn.}\ }\textbf {\bibinfo {volume} {92}},\ \bibinfo
  {pages} {012001} (\bibinfo {year} {2023})}\BibitemShut {NoStop}%
\bibitem [{\citenamefont {Momma}\ and\ \citenamefont {Izumi}(2011)}]{vesta}%
  \BibitemOpen
  \bibfield  {author} {\bibinfo {author} {\bibfnamefont {K.}~\bibnamefont
  {Momma}}\ and\ \bibinfo {author} {\bibfnamefont {F.}~\bibnamefont {Izumi}},\
  }\bibfield  {title} {\bibinfo {title} {{{\it VESTA3} for three-dimensional
  visualization of crystal, volumetric and morphology data}},\ }\href
  {https://doi.org/10.1107/S0021889811038970} {\bibfield  {journal} {\bibinfo
  {journal} {J. Appl. Cryst.}\ }\textbf {\bibinfo {volume} {44}},\ \bibinfo
  {pages} {1272} (\bibinfo {year} {2011})}\BibitemShut {NoStop}%
\bibitem [{Note2()}]{Note2}%
  \BibitemOpen
  \bibinfo {note} {We evaluate a time average of $\theta (t)$ after light
  irradiation in the time domain $[t_{\protect \mathrm {min}}, t_{\protect
  \mathrm {max}}]$ as follows: \begin {equation*} \protect \bar {\theta } =
  \protect \frac {1}{t_{\protect \mathrm {max}}-t_{\protect \mathrm
  {min}}}\DOTSI \intop \ilimits@ _{t_{\protect \mathrm {min}}}^{t_{\protect
  \mathrm {max}}} dt' \DOTSI \intop \ilimits@ _{t_{\protect \mathrm
  {min}}}^{t'}dt\ \protect \frac {\theta (t)}{t'-t_{\protect \mathrm {min}}},
  \end {equation*} where the time average over $t\in [t_{\protect \mathrm
  {min}}, t']$ is again time-averaged over $t'\in [t_{\protect \mathrm {min}},
  t_{\protect \mathrm {max}}]$. While the first time average, $\DOTSI \intop
  \ilimits@ _{t_{\protect \mathrm {min}}}^{t'}dt\ \protect \frac {\theta
  (t)}{t'-t_{\protect \mathrm {min}}}$, exhibits a decaying oscillation against
  $t'$, a center of that oscillation is efficiently obtained by the second
  average. We set $t_{\protect \mathrm {min}}=0.8\protect \,\protect \rm {ps}$,
  which is sufficiently larger than $\tau =0.3\protect \,\protect \rm {ps}$ so
  that $F(t_{\protect \mathrm {min}})\simeq 0$, and $t_{\protect \mathrm
  {max}}=25\protect \,\protect \rm {ps}$, which is sufficiently large to get a
  converged value for the time average.}\BibitemShut {Stop}%
\bibitem [{\citenamefont {Gu}\ and\ \citenamefont
  {Rondinelli}(2018{\natexlab{b}})}]{Raman-Raman}%
  \BibitemOpen
  \bibfield  {author} {\bibinfo {author} {\bibfnamefont {M.}~\bibnamefont
  {Gu}}\ and\ \bibinfo {author} {\bibfnamefont {J.~M.}\ \bibnamefont
  {Rondinelli}},\ }\bibfield  {title} {\bibinfo {title} {Coupled raman-raman
  modes in the ionic raman scattering process},\ }\href
  {https://doi.org/10.1063/1.5048037} {\bibfield  {journal} {\bibinfo
  {journal} {Appl. Phys. Lett.}\ }\textbf {\bibinfo {volume} {113}},\ \bibinfo
  {pages} {112903} (\bibinfo {year} {2018}{\natexlab{b}})}\BibitemShut
  {NoStop}%
\bibitem [{\citenamefont {Klemens}(1966)}]{P.G.Klemens_1966}%
  \BibitemOpen
  \bibfield  {author} {\bibinfo {author} {\bibfnamefont {P.~G.}\ \bibnamefont
  {Klemens}},\ }\bibfield  {title} {\bibinfo {title} {Anharmonic decay of
  optical phonons},\ }\href {https://doi.org/10.1103/PhysRev.148.845}
  {\bibfield  {journal} {\bibinfo  {journal} {Phys. Rev.}\ }\textbf {\bibinfo
  {volume} {148}},\ \bibinfo {pages} {845} (\bibinfo {year}
  {1966})}\BibitemShut {NoStop}%
\bibitem [{\citenamefont {Teitelbaum}\ \emph {et~al.}(2018)\citenamefont
  {Teitelbaum}, \citenamefont {Henighan}, \citenamefont {Huang}, \citenamefont
  {Liu}, \citenamefont {Jiang}, \citenamefont {Zhu}, \citenamefont {Chollet},
  \citenamefont {Sato}, \citenamefont {Murray}, \citenamefont {Fahy},
  \citenamefont {O'Mahony}, \citenamefont {Bailey}, \citenamefont {Uher},
  \citenamefont {Trigo},\ and\ \citenamefont {Reis}}]{T.Samuel_2018}%
  \BibitemOpen
  \bibfield  {author} {\bibinfo {author} {\bibfnamefont {S.~W.}\ \bibnamefont
  {Teitelbaum}}, \bibinfo {author} {\bibfnamefont {T.}~\bibnamefont
  {Henighan}}, \bibinfo {author} {\bibfnamefont {Y.}~\bibnamefont {Huang}},
  \bibinfo {author} {\bibfnamefont {H.}~\bibnamefont {Liu}}, \bibinfo {author}
  {\bibfnamefont {M.~P.}\ \bibnamefont {Jiang}}, \bibinfo {author}
  {\bibfnamefont {D.}~\bibnamefont {Zhu}}, \bibinfo {author} {\bibfnamefont
  {M.}~\bibnamefont {Chollet}}, \bibinfo {author} {\bibfnamefont
  {T.}~\bibnamefont {Sato}}, \bibinfo {author} {\bibfnamefont {E.~D.}\
  \bibnamefont {Murray}}, \bibinfo {author} {\bibfnamefont {S.}~\bibnamefont
  {Fahy}}, \bibinfo {author} {\bibfnamefont {S.}~\bibnamefont {O'Mahony}},
  \bibinfo {author} {\bibfnamefont {T.~P.}\ \bibnamefont {Bailey}}, \bibinfo
  {author} {\bibfnamefont {C.}~\bibnamefont {Uher}}, \bibinfo {author}
  {\bibfnamefont {M.}~\bibnamefont {Trigo}},\ and\ \bibinfo {author}
  {\bibfnamefont {D.~A.}\ \bibnamefont {Reis}},\ }\bibfield  {title} {\bibinfo
  {title} {Direct measurement of anharmonic decay channels of a coherent
  phonon},\ }\href {https://doi.org/10.1103/PhysRevLett.121.125901} {\bibfield
  {journal} {\bibinfo  {journal} {Phys. Rev. Lett.}\ }\textbf {\bibinfo
  {volume} {121}},\ \bibinfo {pages} {125901} (\bibinfo {year}
  {2018})}\BibitemShut {NoStop}%
\bibitem [{\citenamefont {Feng}\ and\ \citenamefont {Ruan}(2016)}]{T.Feng_16}%
  \BibitemOpen
  \bibfield  {author} {\bibinfo {author} {\bibfnamefont {T.}~\bibnamefont
  {Feng}}\ and\ \bibinfo {author} {\bibfnamefont {X.}~\bibnamefont {Ruan}},\
  }\bibfield  {title} {\bibinfo {title} {Quantum mechanical prediction of
  four-phonon scattering rates and reduced thermal conductivity of solids},\
  }\href {https://doi.org/10.1103/PhysRevB.93.045202} {\bibfield  {journal}
  {\bibinfo  {journal} {Phys. Rev. B}\ }\textbf {\bibinfo {volume} {93}},\
  \bibinfo {pages} {045202} (\bibinfo {year} {2016})}\BibitemShut {NoStop}%
\bibitem [{\citenamefont {Maradudin}\ and\ \citenamefont
  {Fein}(1962)}]{A.Maradudin_62_1}%
  \BibitemOpen
  \bibfield  {author} {\bibinfo {author} {\bibfnamefont {A.~A.}\ \bibnamefont
  {Maradudin}}\ and\ \bibinfo {author} {\bibfnamefont {A.~E.}\ \bibnamefont
  {Fein}},\ }\bibfield  {title} {\bibinfo {title} {Scattering of neutrons by an
  anharmonic crystal},\ }\href {https://doi.org/10.1103/PhysRev.128.2589}
  {\bibfield  {journal} {\bibinfo  {journal} {Phys. Rev.}\ }\textbf {\bibinfo
  {volume} {128}},\ \bibinfo {pages} {2589} (\bibinfo {year}
  {1962})}\BibitemShut {NoStop}%
\bibitem [{\citenamefont {Maradudin}\ \emph {et~al.}(1962)\citenamefont
  {Maradudin}, \citenamefont {Fein},\ and\ \citenamefont
  {Vineyard}}]{A.Maradudin_62_2}%
  \BibitemOpen
  \bibfield  {author} {\bibinfo {author} {\bibfnamefont {A.~A.}\ \bibnamefont
  {Maradudin}}, \bibinfo {author} {\bibfnamefont {A.~E.}\ \bibnamefont
  {Fein}},\ and\ \bibinfo {author} {\bibfnamefont {G.~H.}\ \bibnamefont
  {Vineyard}},\ }\bibfield  {title} {\bibinfo {title} {On the evaluation of
  phonon widths and shifts},\ }\href
  {https://doi.org/https://doi.org/10.1002/pssb.19620021106} {\bibfield
  {journal} {\bibinfo  {journal} {Phys. Status Solidi B}\ }\textbf {\bibinfo
  {volume} {2}},\ \bibinfo {pages} {1479} (\bibinfo {year} {1962})}\BibitemShut
  {NoStop}%
\bibitem [{\citenamefont {Feng}\ and\ \citenamefont {Ruan}(2014)}]{T.Feng_14}%
  \BibitemOpen
  \bibfield  {author} {\bibinfo {author} {\bibfnamefont {T.}~\bibnamefont
  {Feng}}\ and\ \bibinfo {author} {\bibfnamefont {X.}~\bibnamefont {Ruan}},\
  }\bibfield  {title} {\bibinfo {title} {Prediction of spectral phonon mean
  free path and thermal conductivity with applications to thermoelectrics and
  thermal management: A review},\ }\href
  {https://doi.org/https://doi.org/10.1155/2014/206370} {\bibfield  {journal}
  {\bibinfo  {journal} {J. Nanomater.}\ }\textbf {\bibinfo {volume} {2014}},\
  \bibinfo {pages} {206370} (\bibinfo {year} {2014})}\BibitemShut {NoStop}%
\bibitem [{\citenamefont {Monacelli}\ and\ \citenamefont
  {Mauri}(2021)}]{L.Monacelli_21}%
  \BibitemOpen
  \bibfield  {author} {\bibinfo {author} {\bibfnamefont {L.}~\bibnamefont
  {Monacelli}}\ and\ \bibinfo {author} {\bibfnamefont {F.}~\bibnamefont
  {Mauri}},\ }\bibfield  {title} {\bibinfo {title} {Time-dependent
  self-consistent harmonic approximation: Anharmonic nuclear quantum dynamics
  and time correlation functions},\ }\href
  {https://doi.org/10.1103/PhysRevB.103.104305} {\bibfield  {journal} {\bibinfo
   {journal} {Phys. Rev. B}\ }\textbf {\bibinfo {volume} {103}},\ \bibinfo
  {pages} {104305} (\bibinfo {year} {2021})}\BibitemShut {NoStop}%
\bibitem [{\citenamefont {Siciliano}\ \emph {et~al.}(2023)\citenamefont
  {Siciliano}, \citenamefont {Monacelli}, \citenamefont {Caldarelli},\ and\
  \citenamefont {Mauri}}]{A.Siciliano_23}%
  \BibitemOpen
  \bibfield  {author} {\bibinfo {author} {\bibfnamefont {A.}~\bibnamefont
  {Siciliano}}, \bibinfo {author} {\bibfnamefont {L.}~\bibnamefont
  {Monacelli}}, \bibinfo {author} {\bibfnamefont {G.}~\bibnamefont
  {Caldarelli}},\ and\ \bibinfo {author} {\bibfnamefont {F.}~\bibnamefont
  {Mauri}},\ }\bibfield  {title} {\bibinfo {title} {Wigner gaussian dynamics:
  Simulating the anharmonic and quantum ionic motion},\ }\href
  {https://doi.org/10.1103/PhysRevB.107.174307} {\bibfield  {journal} {\bibinfo
   {journal} {Phys. Rev. B}\ }\textbf {\bibinfo {volume} {107}},\ \bibinfo
  {pages} {174307} (\bibinfo {year} {2023})}\BibitemShut {NoStop}%
\bibitem [{\citenamefont {Lihm}\ and\ \citenamefont
  {Park}(2021)}]{J.M.Lihm_21}%
  \BibitemOpen
  \bibfield  {author} {\bibinfo {author} {\bibfnamefont {J.-M.}\ \bibnamefont
  {Lihm}}\ and\ \bibinfo {author} {\bibfnamefont {C.-H.}\ \bibnamefont
  {Park}},\ }\bibfield  {title} {\bibinfo {title} {Gaussian time-dependent
  variational principle for the finite-temperature anharmonic lattice
  dynamics},\ }\href {https://doi.org/10.1103/PhysRevResearch.3.L032017}
  {\bibfield  {journal} {\bibinfo  {journal} {Phys. Rev. Res.}\ }\textbf
  {\bibinfo {volume} {3}},\ \bibinfo {pages} {L032017} (\bibinfo {year}
  {2021})}\BibitemShut {NoStop}%
\bibitem [{\citenamefont {Liu}\ \emph {et~al.}(2023{\natexlab{b}})\citenamefont
  {Liu}, \citenamefont {Sun}, \citenamefont {Huo}, \citenamefont {Ma},
  \citenamefont {Ji}, \citenamefont {Yi}, \citenamefont {Li}, \citenamefont
  {Liu}, \citenamefont {Yu}, \citenamefont {Zhang}, \citenamefont {Chen},
  \citenamefont {Liang}, \citenamefont {Dong}, \citenamefont {Guo},
  \citenamefont {Zhong}, \citenamefont {Shen}, \citenamefont {Li},\ and\
  \citenamefont {Wang}}]{density_wave_327_326}%
  \BibitemOpen
  \bibfield  {author} {\bibinfo {author} {\bibfnamefont {Z.}~\bibnamefont
  {Liu}}, \bibinfo {author} {\bibfnamefont {H.}~\bibnamefont {Sun}}, \bibinfo
  {author} {\bibfnamefont {M.}~\bibnamefont {Huo}}, \bibinfo {author}
  {\bibfnamefont {X.}~\bibnamefont {Ma}}, \bibinfo {author} {\bibfnamefont
  {Y.}~\bibnamefont {Ji}}, \bibinfo {author} {\bibfnamefont {E.}~\bibnamefont
  {Yi}}, \bibinfo {author} {\bibfnamefont {L.}~\bibnamefont {Li}}, \bibinfo
  {author} {\bibfnamefont {H.}~\bibnamefont {Liu}}, \bibinfo {author}
  {\bibfnamefont {J.}~\bibnamefont {Yu}}, \bibinfo {author} {\bibfnamefont
  {Z.}~\bibnamefont {Zhang}}, \bibinfo {author} {\bibfnamefont
  {Z.}~\bibnamefont {Chen}}, \bibinfo {author} {\bibfnamefont {F.}~\bibnamefont
  {Liang}}, \bibinfo {author} {\bibfnamefont {H.}~\bibnamefont {Dong}},
  \bibinfo {author} {\bibfnamefont {H.}~\bibnamefont {Guo}}, \bibinfo {author}
  {\bibfnamefont {D.}~\bibnamefont {Zhong}}, \bibinfo {author} {\bibfnamefont
  {B.}~\bibnamefont {Shen}}, \bibinfo {author} {\bibfnamefont {S.}~\bibnamefont
  {Li}},\ and\ \bibinfo {author} {\bibfnamefont {M.}~\bibnamefont {Wang}},\
  }\bibfield  {title} {\bibinfo {title} {Evidence for charge and spin density
  waves in single crystals of \ce{La3Ni2O7} and \ce{La3Ni2O6}},\ }\href
  {https://doi.org/10.1007/s11433-022-1962-4} {\bibfield  {journal} {\bibinfo
  {journal} {Sci. China Phys. Mech. Astron.}\ }\textbf {\bibinfo {volume}
  {66}},\ \bibinfo {pages} {217411} (\bibinfo {year}
  {2023}{\natexlab{b}})}\BibitemShut {NoStop}%
\bibitem [{\citenamefont {Chen}\ \emph
  {et~al.}(2024{\natexlab{a}})\citenamefont {Chen}, \citenamefont {Liu},
  \citenamefont {Jiao}, \citenamefont {Zou}, \citenamefont {Jiang},
  \citenamefont {Li}, \citenamefont {Luo}, \citenamefont {Wu}, \citenamefont
  {Zhang}, \citenamefont {Guo},\ and\ \citenamefont {Shu}}]{Chen_Liu}%
  \BibitemOpen
  \bibfield  {author} {\bibinfo {author} {\bibfnamefont {K.}~\bibnamefont
  {Chen}}, \bibinfo {author} {\bibfnamefont {X.}~\bibnamefont {Liu}}, \bibinfo
  {author} {\bibfnamefont {J.}~\bibnamefont {Jiao}}, \bibinfo {author}
  {\bibfnamefont {M.}~\bibnamefont {Zou}}, \bibinfo {author} {\bibfnamefont
  {C.}~\bibnamefont {Jiang}}, \bibinfo {author} {\bibfnamefont
  {X.}~\bibnamefont {Li}}, \bibinfo {author} {\bibfnamefont {Y.}~\bibnamefont
  {Luo}}, \bibinfo {author} {\bibfnamefont {Q.}~\bibnamefont {Wu}}, \bibinfo
  {author} {\bibfnamefont {N.}~\bibnamefont {Zhang}}, \bibinfo {author}
  {\bibfnamefont {Y.}~\bibnamefont {Guo}},\ and\ \bibinfo {author}
  {\bibfnamefont {L.}~\bibnamefont {Shu}},\ }\bibfield  {title} {\bibinfo
  {title} {Evidence of spin density waves in
  {La}$_{3}${Ni}$_{2}${O}$_{7-\delta}$},\ }\href
  {https://doi.org/10.1103/PhysRevLett.132.256503} {\bibfield  {journal}
  {\bibinfo  {journal} {Phys. Rev. Lett.}\ }\textbf {\bibinfo {volume} {132}},\
  \bibinfo {pages} {256503} (\bibinfo {year} {2024}{\natexlab{a}})}\BibitemShut
  {NoStop}%
\bibitem [{\citenamefont {Kakoi}\ \emph
  {et~al.}(2024{\natexlab{b}})\citenamefont {Kakoi}, \citenamefont {Oi},
  \citenamefont {Ohshita}, \citenamefont {Yashima}, \citenamefont {Kuroki},
  \citenamefont {Kato}, \citenamefont {Takahashi}, \citenamefont {Ishiwata},
  \citenamefont {Adachi}, \citenamefont {Hatada}, \citenamefont {Uda},\ and\
  \citenamefont {Mukuda}}]{Kakoi_Oi}%
  \BibitemOpen
  \bibfield  {author} {\bibinfo {author} {\bibfnamefont {M.}~\bibnamefont
  {Kakoi}}, \bibinfo {author} {\bibfnamefont {T.}~\bibnamefont {Oi}}, \bibinfo
  {author} {\bibfnamefont {Y.}~\bibnamefont {Ohshita}}, \bibinfo {author}
  {\bibfnamefont {M.}~\bibnamefont {Yashima}}, \bibinfo {author} {\bibfnamefont
  {K.}~\bibnamefont {Kuroki}}, \bibinfo {author} {\bibfnamefont
  {T.}~\bibnamefont {Kato}}, \bibinfo {author} {\bibfnamefont {H.}~\bibnamefont
  {Takahashi}}, \bibinfo {author} {\bibfnamefont {S.}~\bibnamefont {Ishiwata}},
  \bibinfo {author} {\bibfnamefont {Y.}~\bibnamefont {Adachi}}, \bibinfo
  {author} {\bibfnamefont {N.}~\bibnamefont {Hatada}}, \bibinfo {author}
  {\bibfnamefont {T.}~\bibnamefont {Uda}},\ and\ \bibinfo {author}
  {\bibfnamefont {H.}~\bibnamefont {Mukuda}},\ }\bibfield  {title} {\bibinfo
  {title} {Multiband metallic ground state in multilayered nickelates
  {La}$_{3}${Ni}$_2${O}$_7$ and {La}$_{4}${Ni}$_3${O}$_{10}$ probed by
  $^{139}${La}-{NMR} at ambient pressure},\ }\href
  {https://doi.org/10.7566/JPSJ.93.053702} {\bibfield  {journal} {\bibinfo
  {journal} {J. Phys. Soc. Jpn.}\ }\textbf {\bibinfo {volume} {93}},\ \bibinfo
  {pages} {053702} (\bibinfo {year} {2024}{\natexlab{b}})}\BibitemShut
  {NoStop}%
\bibitem [{\citenamefont {Xie}\ \emph {et~al.}(2024)\citenamefont {Xie},
  \citenamefont {Huo}, \citenamefont {Ni}, \citenamefont {Shen}, \citenamefont
  {Huang}, \citenamefont {Sun}, \citenamefont {Walker}, \citenamefont {Adroja},
  \citenamefont {Yu}, \citenamefont {Shen}, \citenamefont {He}, \citenamefont
  {Cao},\ and\ \citenamefont {Wang}}]{Xie_Huo}%
  \BibitemOpen
  \bibfield  {author} {\bibinfo {author} {\bibfnamefont {T.}~\bibnamefont
  {Xie}}, \bibinfo {author} {\bibfnamefont {M.}~\bibnamefont {Huo}}, \bibinfo
  {author} {\bibfnamefont {X.}~\bibnamefont {Ni}}, \bibinfo {author}
  {\bibfnamefont {F.}~\bibnamefont {Shen}}, \bibinfo {author} {\bibfnamefont
  {X.}~\bibnamefont {Huang}}, \bibinfo {author} {\bibfnamefont
  {H.}~\bibnamefont {Sun}}, \bibinfo {author} {\bibfnamefont {H.~C.}\
  \bibnamefont {Walker}}, \bibinfo {author} {\bibfnamefont {D.}~\bibnamefont
  {Adroja}}, \bibinfo {author} {\bibfnamefont {D.}~\bibnamefont {Yu}}, \bibinfo
  {author} {\bibfnamefont {B.}~\bibnamefont {Shen}}, \bibinfo {author}
  {\bibfnamefont {L.}~\bibnamefont {He}}, \bibinfo {author} {\bibfnamefont
  {K.}~\bibnamefont {Cao}},\ and\ \bibinfo {author} {\bibfnamefont
  {M.}~\bibnamefont {Wang}},\ }\bibfield  {title} {\bibinfo {title} {Strong
  interlayer magnetic exchange coupling in \ce{La3Ni2O_{7-$\delta$}} revealed
  by inelastic neutron scattering},\ }\href
  {https://doi.org/https://doi.org/10.1016/j.scib.2024.07.030} {\bibfield
  {journal} {\bibinfo  {journal} {Science Bulletin}\ }\textbf {\bibinfo
  {volume} {69}},\ \bibinfo {pages} {3221} (\bibinfo {year}
  {2024})}\BibitemShut {NoStop}%
\bibitem [{\citenamefont {Chen}\ \emph
  {et~al.}(2024{\natexlab{b}})\citenamefont {Chen}, \citenamefont {Choi},
  \citenamefont {Jiang}, \citenamefont {Mei}, \citenamefont {Jiang},
  \citenamefont {Li}, \citenamefont {Agrestini}, \citenamefont
  {Garcia-Fernandez}, \citenamefont {Sun}, \citenamefont {Huang}, \citenamefont
  {Shen}, \citenamefont {Wang}, \citenamefont {Hu}, \citenamefont {Lu},
  \citenamefont {Zhou},\ and\ \citenamefont {Feng}}]{Chen_Choi}%
  \BibitemOpen
  \bibfield  {author} {\bibinfo {author} {\bibfnamefont {X.}~\bibnamefont
  {Chen}}, \bibinfo {author} {\bibfnamefont {J.}~\bibnamefont {Choi}}, \bibinfo
  {author} {\bibfnamefont {Z.}~\bibnamefont {Jiang}}, \bibinfo {author}
  {\bibfnamefont {J.}~\bibnamefont {Mei}}, \bibinfo {author} {\bibfnamefont
  {K.}~\bibnamefont {Jiang}}, \bibinfo {author} {\bibfnamefont
  {J.}~\bibnamefont {Li}}, \bibinfo {author} {\bibfnamefont {S.}~\bibnamefont
  {Agrestini}}, \bibinfo {author} {\bibfnamefont {M.}~\bibnamefont
  {Garcia-Fernandez}}, \bibinfo {author} {\bibfnamefont {H.}~\bibnamefont
  {Sun}}, \bibinfo {author} {\bibfnamefont {X.}~\bibnamefont {Huang}}, \bibinfo
  {author} {\bibfnamefont {D.}~\bibnamefont {Shen}}, \bibinfo {author}
  {\bibfnamefont {M.}~\bibnamefont {Wang}}, \bibinfo {author} {\bibfnamefont
  {J.}~\bibnamefont {Hu}}, \bibinfo {author} {\bibfnamefont {Y.}~\bibnamefont
  {Lu}}, \bibinfo {author} {\bibfnamefont {K.-J.}\ \bibnamefont {Zhou}},\ and\
  \bibinfo {author} {\bibfnamefont {D.}~\bibnamefont {Feng}},\ }\bibfield
  {title} {\bibinfo {title} {Electronic and magnetic excitations in
  {La}$_3${Ni}$_2${O}$_7$},\ }\href
  {https://doi.org/https://doi.org/10.1038/s41467-024-53863-5} {\bibfield
  {journal} {\bibinfo  {journal} {Nat. Commun.}\ }\textbf {\bibinfo {volume}
  {15}},\ \bibinfo {pages} {9597} (\bibinfo {year}
  {2024}{\natexlab{b}})}\BibitemShut {NoStop}%
\bibitem [{\citenamefont {Wang}\ \emph
  {et~al.}(2024{\natexlab{d}})\citenamefont {Wang}, \citenamefont {Jiang},
  \citenamefont {Wang}, \citenamefont {Zhang},\ and\ \citenamefont
  {Hu}}]{Wang_Jiang}%
  \BibitemOpen
  \bibfield  {author} {\bibinfo {author} {\bibfnamefont {Y.}~\bibnamefont
  {Wang}}, \bibinfo {author} {\bibfnamefont {K.}~\bibnamefont {Jiang}},
  \bibinfo {author} {\bibfnamefont {Z.}~\bibnamefont {Wang}}, \bibinfo {author}
  {\bibfnamefont {F.-C.}\ \bibnamefont {Zhang}},\ and\ \bibinfo {author}
  {\bibfnamefont {J.}~\bibnamefont {Hu}},\ }\bibfield  {title} {\bibinfo
  {title} {Electronic and magnetic structures of bilayer
  {La}$_{3}${Ni}$_2${O}$_{7}$ at ambient pressure},\ }\href
  {https://doi.org/10.1103/PhysRevB.110.205122} {\bibfield  {journal} {\bibinfo
   {journal} {Phys. Rev. B}\ }\textbf {\bibinfo {volume} {110}},\ \bibinfo
  {pages} {205122} (\bibinfo {year} {2024}{\natexlab{d}})}\BibitemShut
  {NoStop}%
\bibitem [{\citenamefont {Zhao}\ \emph {et~al.}(2025)\citenamefont {Zhao},
  \citenamefont {Zhou}, \citenamefont {Huo}, \citenamefont {Wang},
  \citenamefont {Nie}, \citenamefont {Yang}, \citenamefont {Ying},
  \citenamefont {Wang}, \citenamefont {Wu},\ and\ \citenamefont
  {Chen}}]{Dan_Zhou}%
  \BibitemOpen
  \bibfield  {author} {\bibinfo {author} {\bibfnamefont {D.}~\bibnamefont
  {Zhao}}, \bibinfo {author} {\bibfnamefont {Y.}~\bibnamefont {Zhou}}, \bibinfo
  {author} {\bibfnamefont {M.}~\bibnamefont {Huo}}, \bibinfo {author}
  {\bibfnamefont {Y.}~\bibnamefont {Wang}}, \bibinfo {author} {\bibfnamefont
  {L.}~\bibnamefont {Nie}}, \bibinfo {author} {\bibfnamefont {Y.}~\bibnamefont
  {Yang}}, \bibinfo {author} {\bibfnamefont {J.}~\bibnamefont {Ying}}, \bibinfo
  {author} {\bibfnamefont {M.}~\bibnamefont {Wang}}, \bibinfo {author}
  {\bibfnamefont {T.}~\bibnamefont {Wu}},\ and\ \bibinfo {author}
  {\bibfnamefont {X.}~\bibnamefont {Chen}},\ }\bibfield  {title} {\bibinfo
  {title} {Pressure-enhanced spin-density-wave transition in double-layer
  nickelate \ce{La3Ni2O_{7-$\delta$}}},\ }\href
  {https://www.sciencedirect.com/science/article/pii/S2095927325001811}
  {\bibfield  {journal} {\bibinfo  {journal} {Science Bulletin}\ }\textbf
  {\bibinfo {volume} {70}},\ \bibinfo {pages} {1239} (\bibinfo {year}
  {2025})}\BibitemShut {NoStop}%
\bibitem [{\citenamefont {Khasanov}\ \emph {et~al.}(2025)\citenamefont
  {Khasanov}, \citenamefont {Hicken}, \citenamefont {Gawryluk}, \citenamefont
  {Sorel}, \citenamefont {B\"{o}tzel}, \citenamefont {Lechermann},
  \citenamefont {Eremin}, \citenamefont {Luetkens},\ and\ \citenamefont
  {Guguchia}}]{Khasanov_Hicken}%
  \BibitemOpen
  \bibfield  {author} {\bibinfo {author} {\bibfnamefont {R.}~\bibnamefont
  {Khasanov}}, \bibinfo {author} {\bibfnamefont {T.~J.}\ \bibnamefont
  {Hicken}}, \bibinfo {author} {\bibfnamefont {D.~J.}\ \bibnamefont
  {Gawryluk}}, \bibinfo {author} {\bibfnamefont {L.~P.}\ \bibnamefont {Sorel}},
  \bibinfo {author} {\bibfnamefont {S.}~\bibnamefont {B\"{o}tzel}}, \bibinfo
  {author} {\bibfnamefont {F.}~\bibnamefont {Lechermann}}, \bibinfo {author}
  {\bibfnamefont {I.~M.}\ \bibnamefont {Eremin}}, \bibinfo {author}
  {\bibfnamefont {H.}~\bibnamefont {Luetkens}},\ and\ \bibinfo {author}
  {\bibfnamefont {Z.}~\bibnamefont {Guguchia}},\ }\bibfield  {title} {\bibinfo
  {title} {Pressure-induced split of the density wave transitions in
  \ce{La3Ni2O_{7-$\delta$}}},\ }\href
  {https://doi.org/10.1038/s41567-024-02754-z} {\bibfield  {journal} {\bibinfo
  {journal} {Nature}\ }\textbf {\bibinfo {volume} {21}},\ \bibinfo {pages}
  {430} (\bibinfo {year} {2025})}\BibitemShut {NoStop}%
\bibitem [{\citenamefont {Meng}\ \emph {et~al.}(2024)\citenamefont {Meng},
  \citenamefont {Yang}, \citenamefont {Sun}, \citenamefont {Zhang},
  \citenamefont {Luo}, \citenamefont {Chen}, \citenamefont {Ma}, \citenamefont
  {Wang}, \citenamefont {Hong}, \citenamefont {Wang},\ and\ \citenamefont
  {Yu}}]{Meng_Yang}%
  \BibitemOpen
  \bibfield  {author} {\bibinfo {author} {\bibfnamefont {Y.}~\bibnamefont
  {Meng}}, \bibinfo {author} {\bibfnamefont {Y.}~\bibnamefont {Yang}}, \bibinfo
  {author} {\bibfnamefont {H.}~\bibnamefont {Sun}}, \bibinfo {author}
  {\bibfnamefont {S.}~\bibnamefont {Zhang}}, \bibinfo {author} {\bibfnamefont
  {J.}~\bibnamefont {Luo}}, \bibinfo {author} {\bibfnamefont {L.}~\bibnamefont
  {Chen}}, \bibinfo {author} {\bibfnamefont {X.}~\bibnamefont {Ma}}, \bibinfo
  {author} {\bibfnamefont {M.}~\bibnamefont {Wang}}, \bibinfo {author}
  {\bibfnamefont {F.}~\bibnamefont {Hong}}, \bibinfo {author} {\bibfnamefont
  {X.}~\bibnamefont {Wang}},\ and\ \bibinfo {author} {\bibfnamefont
  {X.}~\bibnamefont {Yu}},\ }\bibfield  {title} {\bibinfo {title}
  {Density-wave-like gap evolution in \ce{La3Ni2O7} under high pressure
  revealed by ultrafast optical spectroscopy},\ }\href
  {https://doi.org/10.1038/s41467-024-54518-1} {\bibfield  {journal} {\bibinfo
  {journal} {Nat. Commun.}\ }\textbf {\bibinfo {volume} {15}},\ \bibinfo
  {pages} {10408} (\bibinfo {year} {2024})}\BibitemShut {NoStop}%
\bibitem [{\citenamefont {LaBollita}\ \emph {et~al.}(2024)\citenamefont
  {LaBollita}, \citenamefont {Pardo}, \citenamefont {Norman},\ and\
  \citenamefont {Botana}}]{LaBollita_Pardo_116}%
  \BibitemOpen
  \bibfield  {author} {\bibinfo {author} {\bibfnamefont {H.}~\bibnamefont
  {LaBollita}}, \bibinfo {author} {\bibfnamefont {V.}~\bibnamefont {Pardo}},
  \bibinfo {author} {\bibfnamefont {M.~R.}\ \bibnamefont {Norman}},\ and\
  \bibinfo {author} {\bibfnamefont {A.~S.}\ \bibnamefont {Botana}},\ }\bibfield
   {title} {\bibinfo {title} {Assessing the formation of spin and charge
  stripes in \ce{La3Ni2O7} from first-principles},\ }\href
  {https://doi.org/10.1103/PhysRevMaterials.8.L111801} {\bibfield  {journal}
  {\bibinfo  {journal} {Phys. Rev. Mater.}\ }\textbf {\bibinfo {volume} {8}},\
  \bibinfo {pages} {L111801} (\bibinfo {year} {2024})}\BibitemShut {NoStop}%
\bibitem [{\citenamefont {Zhang}\ \emph
  {et~al.}(2024{\natexlab{d}})\citenamefont {Zhang}, \citenamefont {Xu},\ and\
  \citenamefont {Xiang}}]{Zhang_Xu}%
  \BibitemOpen
  \bibfield  {author} {\bibinfo {author} {\bibfnamefont {B.}~\bibnamefont
  {Zhang}}, \bibinfo {author} {\bibfnamefont {C.}~\bibnamefont {Xu}},\ and\
  \bibinfo {author} {\bibfnamefont {H.}~\bibnamefont {Xiang}},\ }\href@noop {}
  {\bibinfo {title} {Emergent spin-charge-orbital order in superconductor
  \ce{La3Ni2O7}}} (\bibinfo {year} {2024}{\natexlab{d}}),\ \Eprint
  {https://arxiv.org/abs/2407.18473} {arXiv:2407.18473} \BibitemShut {NoStop}%
\bibitem [{\citenamefont {Ni}\ \emph {et~al.}(2025)\citenamefont {Ni},
  \citenamefont {Ji}, \citenamefont {He}, \citenamefont {Xie}, \citenamefont
  {Yao}, \citenamefont {Wang},\ and\ \citenamefont {Cao}}]{Ni_Ji}%
  \BibitemOpen
  \bibfield  {author} {\bibinfo {author} {\bibfnamefont {X.-S.}\ \bibnamefont
  {Ni}}, \bibinfo {author} {\bibfnamefont {Y.}~\bibnamefont {Ji}}, \bibinfo
  {author} {\bibfnamefont {L.}~\bibnamefont {He}}, \bibinfo {author}
  {\bibfnamefont {T.}~\bibnamefont {Xie}}, \bibinfo {author} {\bibfnamefont
  {D.-X.}\ \bibnamefont {Yao}}, \bibinfo {author} {\bibfnamefont
  {M.}~\bibnamefont {Wang}},\ and\ \bibinfo {author} {\bibfnamefont
  {K.}~\bibnamefont {Cao}},\ }\bibfield  {title} {\bibinfo {title} {Spin
  density wave in the bilayered nickelate \ce{La3Ni2O_{7-$\delta$}} at ambient
  pressure},\ }\href {https://doi.org/10.1038/s41535-025-00740-z} {\bibfield
  {journal} {\bibinfo  {journal} {npj Quantum Mater.}\ }\textbf {\bibinfo
  {volume} {10}} (\bibinfo {year} {2025})}\BibitemShut {NoStop}%
\bibitem [{\citenamefont {Lin}\ \emph {et~al.}(2024)\citenamefont {Lin},
  \citenamefont {Zhang}, \citenamefont {Kaushal}, \citenamefont {Alvarez},
  \citenamefont {Maier}, \citenamefont {Moreo},\ and\ \citenamefont
  {Dagotto}}]{Lin_Zhang}%
  \BibitemOpen
  \bibfield  {author} {\bibinfo {author} {\bibfnamefont {L.-F.}\ \bibnamefont
  {Lin}}, \bibinfo {author} {\bibfnamefont {Y.}~\bibnamefont {Zhang}}, \bibinfo
  {author} {\bibfnamefont {N.}~\bibnamefont {Kaushal}}, \bibinfo {author}
  {\bibfnamefont {G.}~\bibnamefont {Alvarez}}, \bibinfo {author} {\bibfnamefont
  {T.~A.}\ \bibnamefont {Maier}}, \bibinfo {author} {\bibfnamefont
  {A.}~\bibnamefont {Moreo}},\ and\ \bibinfo {author} {\bibfnamefont
  {E.}~\bibnamefont {Dagotto}},\ }\bibfield  {title} {\bibinfo {title}
  {Magnetic phase diagram of a two-orbital model for bilayer nickelates with
  varying doping},\ }\href {https://doi.org/10.1103/PhysRevB.110.195135}
  {\bibfield  {journal} {\bibinfo  {journal} {Phys. Rev. B}\ }\textbf {\bibinfo
  {volume} {110}},\ \bibinfo {pages} {195135} (\bibinfo {year}
  {2024})}\BibitemShut {NoStop}%
\bibitem [{\citenamefont {Bistoni}\ \emph {et~al.}(2019)\citenamefont
  {Bistoni}, \citenamefont {Barone}, \citenamefont {Cappelluti}, \citenamefont
  {Benfatto},\ and\ \citenamefont {Mauri}}]{dynamicalZ1}%
  \BibitemOpen
  \bibfield  {author} {\bibinfo {author} {\bibfnamefont {O.}~\bibnamefont
  {Bistoni}}, \bibinfo {author} {\bibfnamefont {P.}~\bibnamefont {Barone}},
  \bibinfo {author} {\bibfnamefont {E.}~\bibnamefont {Cappelluti}}, \bibinfo
  {author} {\bibfnamefont {L.}~\bibnamefont {Benfatto}},\ and\ \bibinfo
  {author} {\bibfnamefont {F.}~\bibnamefont {Mauri}},\ }\bibfield  {title}
  {\bibinfo {title} {Giant effective charges and piezoelectricity in gapped
  graphene},\ }\href {https://doi.org/10.1088/2053-1583/ab2ce0} {\bibfield
  {journal} {\bibinfo  {journal} {2D Mater.}\ }\textbf {\bibinfo {volume}
  {6}},\ \bibinfo {pages} {045015} (\bibinfo {year} {2019})}\BibitemShut
  {NoStop}%
\bibitem [{\citenamefont {Binci}\ \emph {et~al.}(2021)\citenamefont {Binci},
  \citenamefont {Barone},\ and\ \citenamefont {Mauri}}]{dynamicalZ2}%
  \BibitemOpen
  \bibfield  {author} {\bibinfo {author} {\bibfnamefont {L.}~\bibnamefont
  {Binci}}, \bibinfo {author} {\bibfnamefont {P.}~\bibnamefont {Barone}},\ and\
  \bibinfo {author} {\bibfnamefont {F.}~\bibnamefont {Mauri}},\ }\bibfield
  {title} {\bibinfo {title} {First-principles theory of infrared vibrational
  spectroscopy of metals and semimetals: Application to graphite},\ }\href
  {https://doi.org/10.1103/PhysRevB.103.134304} {\bibfield  {journal} {\bibinfo
   {journal} {Phys. Rev. B}\ }\textbf {\bibinfo {volume} {103}},\ \bibinfo
  {pages} {134304} (\bibinfo {year} {2021})}\BibitemShut {NoStop}%
\bibitem [{\citenamefont {Dreyer}\ \emph {et~al.}(2022)\citenamefont {Dreyer},
  \citenamefont {Coh},\ and\ \citenamefont {Stengel}}]{dynamicalZ3}%
  \BibitemOpen
  \bibfield  {author} {\bibinfo {author} {\bibfnamefont {C.~E.}\ \bibnamefont
  {Dreyer}}, \bibinfo {author} {\bibfnamefont {S.}~\bibnamefont {Coh}},\ and\
  \bibinfo {author} {\bibfnamefont {M.}~\bibnamefont {Stengel}},\ }\bibfield
  {title} {\bibinfo {title} {Nonadiabatic born effective charges in metals and
  the drude weight},\ }\href {https://doi.org/10.1103/PhysRevLett.128.095901}
  {\bibfield  {journal} {\bibinfo  {journal} {Phys. Rev. Lett.}\ }\textbf
  {\bibinfo {volume} {128}},\ \bibinfo {pages} {095901} (\bibinfo {year}
  {2022})}\BibitemShut {NoStop}%
\bibitem [{\citenamefont {Wang}\ \emph {et~al.}(2022)\citenamefont {Wang},
  \citenamefont {Sharma}, \citenamefont {Gross},\ and\ \citenamefont
  {Dewhurst}}]{dynamicalZ4}%
  \BibitemOpen
  \bibfield  {author} {\bibinfo {author} {\bibfnamefont {C.-Y.}\ \bibnamefont
  {Wang}}, \bibinfo {author} {\bibfnamefont {S.}~\bibnamefont {Sharma}},
  \bibinfo {author} {\bibfnamefont {E.~K.~U.}\ \bibnamefont {Gross}},\ and\
  \bibinfo {author} {\bibfnamefont {J.~K.}\ \bibnamefont {Dewhurst}},\
  }\bibfield  {title} {\bibinfo {title} {Dynamical born effective charges},\
  }\href {https://doi.org/10.1103/PhysRevB.106.L180303} {\bibfield  {journal}
  {\bibinfo  {journal} {Phys. Rev. B}\ }\textbf {\bibinfo {volume} {106}},\
  \bibinfo {pages} {L180303} (\bibinfo {year} {2022})}\BibitemShut {NoStop}%
\bibitem [{Note3()}]{Note3}%
  \BibitemOpen
  \bibinfo {note} {\protect \url
  {https://hdl.handle.net/11094/102582}}\BibitemShut {NoStop}%
\bibitem [{\citenamefont {Bradley}\ and\ \citenamefont
  {Cracknell}(1972)}]{group1}%
  \BibitemOpen
  \bibfield  {author} {\bibinfo {author} {\bibfnamefont {C.~J.}\ \bibnamefont
  {Bradley}}\ and\ \bibinfo {author} {\bibfnamefont {A.~P.}\ \bibnamefont
  {Cracknell}},\ }\href@noop {} {\emph {\bibinfo {title} {The mathematical
  theory of symmetry in solids : representation theory for point groups and
  space groups}}}\ (\bibinfo  {publisher} {Clarendon Press},\ \bibinfo {year}
  {1972})\BibitemShut {NoStop}%
\bibitem [{\citenamefont {Altmann}\ and\ \citenamefont
  {Herzig}(1994)}]{group2}%
  \BibitemOpen
  \bibfield  {author} {\bibinfo {author} {\bibfnamefont {S.~L.}\ \bibnamefont
  {Altmann}}\ and\ \bibinfo {author} {\bibfnamefont {P.}~\bibnamefont
  {Herzig}},\ }\href@noop {} {\emph {\bibinfo {title} {Point-group Theory
  Tables}}}\ (\bibinfo  {publisher} {Clarendon Press},\ \bibinfo {year}
  {1994})\BibitemShut {NoStop}%
\bibitem [{\citenamefont {Perdew}\ \emph {et~al.}(1996)\citenamefont {Perdew},
  \citenamefont {Burke},\ and\ \citenamefont {Ernzerhof}}]{ref_PBE}%
  \BibitemOpen
  \bibfield  {author} {\bibinfo {author} {\bibfnamefont {J.~P.}\ \bibnamefont
  {Perdew}}, \bibinfo {author} {\bibfnamefont {K.}~\bibnamefont {Burke}},\ and\
  \bibinfo {author} {\bibfnamefont {M.}~\bibnamefont {Ernzerhof}},\ }\bibfield
  {title} {\bibinfo {title} {Generalized gradient approximation made simple},\
  }\href {https://doi.org/10.1103/PhysRevLett.77.3865} {\bibfield  {journal}
  {\bibinfo  {journal} {Phys. Rev. Lett.}\ }\textbf {\bibinfo {volume} {77}},\
  \bibinfo {pages} {3865} (\bibinfo {year} {1996})}\BibitemShut {NoStop}%
\bibitem [{\citenamefont {Dudarev}\ \emph {et~al.}(1998)\citenamefont
  {Dudarev}, \citenamefont {Botton}, \citenamefont {Savrasov}, \citenamefont
  {Humphreys},\ and\ \citenamefont {Sutton}}]{ref_U}%
  \BibitemOpen
  \bibfield  {author} {\bibinfo {author} {\bibfnamefont {S.~L.}\ \bibnamefont
  {Dudarev}}, \bibinfo {author} {\bibfnamefont {G.~A.}\ \bibnamefont {Botton}},
  \bibinfo {author} {\bibfnamefont {S.~Y.}\ \bibnamefont {Savrasov}}, \bibinfo
  {author} {\bibfnamefont {C.~J.}\ \bibnamefont {Humphreys}},\ and\ \bibinfo
  {author} {\bibfnamefont {A.~P.}\ \bibnamefont {Sutton}},\ }\bibfield  {title}
  {\bibinfo {title} {Electron-energy-loss spectra and the structural stability
  of nickel oxide: An {LSDA+U} study},\ }\href
  {https://doi.org/10.1103/PhysRevB.57.1505} {\bibfield  {journal} {\bibinfo
  {journal} {Phys. Rev. B}\ }\textbf {\bibinfo {volume} {57}},\ \bibinfo
  {pages} {1505} (\bibinfo {year} {1998})}\BibitemShut {NoStop}%
\end{thebibliography}%

\end{document}